\documentclass[useAMS, usenatbib]{mn2e}
\usepackage{times}
\usepackage{amsmath}
\usepackage{amssymb}
\usepackage{graphicx}
\usepackage[tight,hang]{subfigure}
\usepackage[letterpaper, margin=0.65in]{geometry}
\usepackage{natbib}
\usepackage{hyperref}
\usepackage{enumitem}
\usepackage[usenames]{color}

\title[Ram Pressure Stripping of X-ray Coronae]{Ram Pressure Stripping of Hot Coronal Gas from Group and Cluster Galaxies and the Detectability of Surviving X-ray Coronae}
\author[R. Vijayaraghavan \& P. M. Ricker]{Rukmani Vijayaraghavan$^1$\thanks{E-mail: vijayar2@illinois.edu} \& Paul M. Ricker$^1$\\
        $^1$Department of Astronomy, University of Illinois at Urbana-Champaign, 1002 W. Green Street, Urbana, IL 61801, USA}

\begin{document}

\date{\today}
\maketitle

\begin{abstract}
  Ram pressure stripping can remove hot and cold gas from galaxies in the intracluster medium (ICM), as shown by observations of X-ray and HI galaxy wakes in nearby clusters of galaxies. However, ram pressure stripping, including pre-processing in group environments, does not remove all the hot coronal gas from cluster galaxies. Recent high-resolution \textit{Chandra} observations have shown that $\sim 1 - 4$\ kpc extended, hot galactic coronae are ubiquitous in group and cluster galaxies. To better understand this result, we simulate ram pressure stripping of a cosmologically motivated population of galaxies in isolated group and cluster environments. The galaxies and the host group and cluster are composed of collisionless dark matter and hot gas initially in hydrostatic equilibrium with the galaxy and host potentials. We show that the rate at which gas is lost depends on the galactic and host halo mass.  Using synthetic X-ray observations, we evaluate the detectability of stripped galactic coronae in real observations by stacking images on the known galaxy centers. We find that coronal emission should be detected within $\sim 10 \arcsec$, or $\sim 5$ kpc up to $\sim 2.3$ Gyr in the lowest (0.1 -- 1.2 keV) energy band. Thus the presence of observed coronae in cluster galaxies significantly smaller than the hot X-ray halos of field galaxies indicates that at least some gas removal occurs within cluster environments for recently accreted galaxies. Finally, we evaluate the possibility that existing and future X-ray cluster catalogs can be used in combination with optical galaxy positions to detect galactic coronal emission via stacking analysis. We briefly discuss the effects of additional physical processes on coronal survival, and will address them in detail in future papers in this series.
\end{abstract}


\section{Introduction}
\label{sec:intro}
Galaxy groups and clusters are hostile environments for their galaxies. The hot intracluster medium (ICM\footnote{For brevity and to avoid confusion with the intergalactic medium, we refer to the intragroup medium as the ICM in this paper.}) comprises most of the baryonic mass and about 10\% of the total mass in these systems. Through ram pressure stripping, the ICM can efficiently strip galaxies of their hot and cold interstellar medium (ISM) gas (\citealt{Gunn72}, \citealt{Quilis00}). In addition to ram pressure stripping, galaxies lose their ISM gas due to thermal conduction between the ICM and ISM (\citealt{Sarazin86}), as well as tidal stripping (\citealt{Gnedin03b}) and galaxy harassment (\citealt{Moore96}, \citealt{Gnedin03a}). The loss of gas suppresses star formation in group and cluster galaxies, making them appear `red and dead' compared to field galaxies.

Removal of the cold disk component of the ISM shuts off ongoing star formation. Theoretical (e.g.\ \citealt{Quilis00}, \citealt{Vollmer01}, \citealt{Schulz01}, \citealt{Roediger06}, \citealt{Tonnesen09}, \citealt{Kapferer09}) and observational (e.g.\ \citealt{Kenney99}, \citealt{Oosterloo05}, \citealt{Chung07}, \citealt{Sun07b}, \citealt{Abramson11}) studies show that the stripped gas trails galaxies in the form of atomic gas (HI) or H$\alpha$ tails. Ram pressure can also compress the cold ISM gas and induce star formation, both in the galactic disk and in stripped wakes and tails. In stripped tails, the absence of ionizing galactic radiation favors a scenario where the stripped gas cools and forms stars, seen observationally as intracluster star formation (e.g., \citealt{Cortese07}, \citealt{Sun10}).

Ram pressure also removes the hot coronal (or halo) component of galactic ISM gas (\citealt{Larson80}, \citealt{Kawata08}, \citealt{McCarthy08}). This process, while not responsible for the immediate suppression of star formation, results in the loss of long-term star formation fuel by removing gas that can radiatively cool and eventually form stars.  The stripping of hot coronal gas by ram pressure is referred to as `strangulation' or `starvation'. Ram pressure-stripped, X-ray emitting wakes and tails are observed trailing their galaxies in both early- (\citealt{Forman79}, \citealt{Irwin96},  \citealt{Sivakoff04}, \citealt{Machacek05}, \citealt{Machacek06}, \citealt{Randall08}, \citealt{Kim08}, \citealt{Kraft11}) and late-type (\citealt{Wang04}, \citealt{Machacek04}, \citealt{Sun05b}, \citealt{Sun10}, \citealt{Zhang13}) group and cluster galaxies. In general, the term `wake' refers to the density enhanced, gravitationally focused ICM trailing a galaxy, while the term `tail' refers to stripped galactic gas originally bound to a galaxy. These terms have been used interchangeably in the literature.

In the presence of efficient ram pressure stripping and evaporation due to thermal conduction, galaxies are not expected to retain their hot coronae. However, recent observations of galaxies in dense cluster environments show that $\sim 40 - 80\%$ of galaxies in group and cluster environments have extended X-ray coronae, suggesting that these coronae survive on timescales comparable to the lifetimes of clusters. \citet{Vikhlinin01}, using \textit{Chandra} observations, reported the first detection of hot X-ray coronae centered on NGC 4874 and NGC 4889, the two central Coma cluster galaxies. These $\sim 1 - 2$ keV coronae are remnants of hot galactic ISM gas and are confined by the 9 keV Coma ICM. Additionally, these coronae are much smaller ($\sim 3$ kpc) than those of typical field galaxies ($\sim 100$ kpc). While ram pressure is not expected to strip these coronae given that these are central galaxies that do not move significantly with respect to their surrounding ICM, their survival depends upon a balance between thermal conduction and radiative cooling, influenced by the ICM and ISM magnetic fields. \citet{Yamasaki02}, using \textit{Chandra} observations of Abell 1060, showed that its two central giant elliptical galaxies have $2 - 3$ kpc, $0.7 - 0.9$ keV coronae that do not appear to be undergoing stripping. \citet{Sun05} observed the Abell 1367 galaxy cluster with \textit{Chandra} and found that four of its galaxies have $0.4 - 1$ keV thermal coronae. \citet{Sun05} also show that the coronae of the two more massive galaxies in their sample are relaxed and symmetric, while the smaller galaxies appear to be in the process of being stripped. \citet{Sun05c} show that the NGC 1265 radio galaxy in the Perseus cluster has a 0.6 keV X-ray corona, and its asymmetric structure indicates that the galaxy is currently subject to ram pressure stripping.

More recent systematic studies have shown that galactic coronae in clusters are ubiquitous and that their properties depend on their environment. \citet{Sun07} studied 179 galaxies in 25 nearby ($z < 0.05$) galaxy clusters using \textit{Chandra} observations. Excluding cD galaxies, they found that more than 60\% of early-type galaxies with 2MASS $K_s$-band luminosities $L_{K_s} > 2L_{*}$, 40\% of $L_{*} < L_{K_s} < 2L_{*}$ galaxies, and 15\% of $L_{K_s} < L_{*}$ galaxies host $1.4 - 4$ kpc embedded X-ray coronae. They also found that $\sim 30\%$ of the late-type galaxies in their sample host observable coronae. \citet{Jeltema08}, using \textit{Chandra} observations of 13 nearby galaxy groups, found that $\sim 80\%$ of $L_{K_s} > L_{*}$ early-type group galaxies and 4 of 11 late-type galaxies host hot coronae. They also show that $\sim 5\%$ of the galaxies in their sample have wakes consistent with tidal and ram pressure stripping. Taken together with the \citet{Sun07} study, these results indicate that less massive group environments can strip galactic halos but are less efficient than massive groups. 

Theoretical studies of hot galactic coronae have primarily focused on the rate of mass loss due to ram pressure in individual galaxies and the observable properties of galaxy wakes and tails. The earliest of these studies were by \citet{Gisler76} and \citet{Lea76}, who showed using analytic calculations and numerical simulations that ram pressure can remove most of a galaxy's gas within a cluster environment. \citet{Nulsen82} showed that transport processes like viscosity and thermal conduction can enhance gas stripping in galaxies in addition to ram pressure stripping. \citet{Takeda84} showed that a galaxy on a radial cluster-centric orbit can lose almost all of its gas due to the drastic rise in ram pressure during core passage. \citet{Stevens99} performed a series of hydrodynamical simulations and showed that galaxies in the process of being ram pressure stripped by ICM gas display bow shocks and prominent stripped tails. \citet{Stevens99} also showed that galaxies in cooler, less massive systems, galaxies with active stellar mass loss, and galaxies in the outer regions of clusters were more likely to have significant X-ray tails. 

\citet{Toniazzo01} performed three-dimensional simulations of elliptical galaxies in cluster orbits and showed that their X-ray luminosities varied significantly during their orbital evolution. They also showed that the initial post-infall stripping of their model galaxies were consistent with X-ray observations of M86 in the Virgo cluster. \citet{Acreman03}, using simulations of a range of galaxies being ram pressure stripped, showed that the observed X-ray luminosities of these galaxies varied with galactic mass injection and replenishment rates, and that observed X-ray wakes were most prominent during the first passages of galaxies through clusters. \citet{McCarthy08}, using 3D simulations of spherically symmetric galaxies with hot gas halos, showed that these galaxies can retain up to $30\%$ of their initial gas after 10 Gyr, and that the amount of gas retained can be reproduced by analytic models of ram pressure stripping. \citet{Tonnesen11} simulated ram pressure stripping of a cold disk gas by the ICM and showed that stripped cold gas, compressed by the ICM to high pressures, can emit X-rays before being mixed in with the ICM. \citet{Roediger14a} and \citet{Roediger14b} performed simulations of an M89-like isolated elliptical galaxy subject to an ICM wind, with varying ICM viscosity, to investigate the detailed dynamics of the stripped galactic atmosphere. \citet{Roediger14a} disentangle the flow of the ICM around a galaxy and the flow of the stripped galaxies' gas. \citet{Roediger14b} show that a viscous ICM plasma suppresses Kelvin-Helmholtz instabilities and the mixing of stripped gas with the ICM.

The above theoretical studies of ram pressure stripped galaxies and their coronae have primarily been `wind-tunnel' simulations that include a single model galaxy in a box whose fluid parameters mimic those of a realistic ICM. Realistic groups and clusters, however, have a population of galaxies with a range of masses. These galaxies also have a range of radial and circular cluster-centric orbits and therefore experience strong and weak ram pressure at various locations. In a previous paper (\citet{VR13}), we used a test particle model within isolated and merging dark matter plus hot gas groups and clusters to calculate the effect of tidal and ram pressure stripping on galaxies with realistic orbits. We showed that on average, galaxies at larger group- and cluster-centric radii are significantly less stripped than galaxies that are closer to the center. We also showed that group environments in group-cluster mergers can efficiently `pre-process' their galaxies by removing at least $\sim 85\%$ of their galaxies' gas before cluster infall. In this paper, we extend this study of galaxies on realistic orbits by simulating a group and cluster environment with realistic galaxy populations. Each galaxy consists of a dark matter halo and hot gas initially in hydrostatic equilibrium with the galaxy potential. We focus on the evolution of galaxies in isolated group and cluster environments and defer study of group-cluster mergers to a future paper.

The survival of unstripped coronae in groups and clusters is a complex problem, involving the interplay among  various physical processes in the ICM and ISM that remove and replenish coronae. Tidal stripping, ram pressure stripping, and thermal conduction between the ICM and ISM contribute to removal and evaporation of these coronae, while magnetic fields can shield the coronal gas by suppressing conduction and the growth of shear instabilities. Galactic coronae can be replenished by stellar outflows and AGN feedback. In the absence of cold gas fuel, particularly in cluster environments, star formation and AGN activity are likely suppressed, so they may not play a significant role in these environments.  A systematic theoretical study that models all these processes is needed to disentangle the relative importance of the various mechanisms that influence the survival or destruction of galactic coronae. This is the first in a series of papers in which we progressively model the above mechanisms. In this paper, we describe two $N$-body + adiabatic hydrodynamics simulations of galaxies evolving in realistic group and cluster simulations. We study the formation of hot tails and wakes as a result of stripping, as well as the detectability of surviving coronae as a function of time spent by galaxies within group and cluster environments. 

This paper is structured as follows: in \S~\ref{sec:methods} we describe our simulation initial conditions and parameters together with convergence tests that illustrate the effect of varying spatial resolution. In \S~\ref{sec:results}, we describe the results of our simulations, including a qualitative overview of ram pressure-stripped galaxies in \S~\ref{sec:tmap}. We quantify the amount of gas stripped in group and cluster environments and the variation in relative mass loss with galaxy mass in \S~\ref{sec:massloss} and correlate the properties of `confinement surfaces' and stripped tails with ICM properties in \S~\ref{sec:xraywakes}. In \S~\ref{sec:xraycoronae}, we generate synthetic X-ray observations of the group and cluster, including their galaxies at various timesteps, and evaluate the detectability of surviving X-ray coronae by stacking X-ray observations centered on galaxies. In \S~\ref{sec:discussion}, we discuss our results in the context of prior theoretical and observational studies of galactic coronae in groups and clusters and discuss the limitations of our results. We discuss in \S~\ref{sec:disc_xray} the prospect of using existing and future X-ray cluster catalogs to detect stacked galactic coronal emission and potential systematic studies that can be performed with such analyses. We summarize our results in \S~\ref{sec:conclusions}.

\section{Methods}
\label{sec:methods}

The simulations in this paper were performed using \textsc{FLASH 4} (\citealt{Fryxell00}, \citealt{Dubey08}), a parallel $N$-body plus adaptive mesh refinement (AMR) Eulerian hydrodynamics code. In  \textsc{FLASH}, particles are mapped to the mesh using cloud-in-cell (CIC) mapping, and a direct multigrid solver (\citealt{Ricker08}) is used to calculate the gravitational potential on the mesh. To solve Euler's equations, we use the directionally split piecewise parabolic method (\citealt{Colella84}). AMR is implemented using \textsc{PARAMESH 4} (\citealt{MacNeice00}). We perform two idealized simulations of an isolated group and cluster with galaxies. The group and cluster as well as their galaxies initially consist of spherical dark matter halos and hot gas in hydrostatic equilibrium with the overall potential. 

\subsection{Initial conditions}
\label{sec:ic}

\begin{table*}
\begin{center}
  \begin{tabular}{c c c c c c c c c c c}
    \hline 
    Halo & $M_{200} (\mbox{M}_{\odot}$) & $R_{200}$ (kpc) & $r_{\rm s}$ (kpc) & $f_{\rm g}$ & $S_0$ (keV cm$^2$) & $S_1$ (keV cm$^2$) &$N_{\rm sat}$ & $M_{\rm sub, tot} (\mbox{M}_{\odot})$ & $M_{\rm BCG} (\mbox{M}_{\odot})$\\
    \hline
    Cluster & $1.2 \times 10^{14}$ & $687$ & 186  & 0.096 & 4.8 & 90.0 & 152 & $4.1 \times 10^{13} $ & $1.4 \times 10^{12} $\\
    Group & $3.2 \times 10^{13}$ & $446$ & 108 & 0.066 & 2.0 & 40.0 & 26 & $6.7 \times 10^{12} $ & $1.3 \times 10^{12} $\\
    \hline 
  \end{tabular}
  \caption{Group and cluster parameters. 
  \label{table1}} 
\end{center}
\end{table*}

To initialize the dark matter and hot gas in our group and cluster and their subhalos, we use the cluster initialization technique developed in \citet{ZuHone11} and used in \citet{VR13}. We assume standard cosmological parameter values of $H_0 = 71$ km s$^{-1}$ Mpc$^{-1}$, $\Omega_{\rm m} = 0.3$, and $\Omega_{\Lambda} = 0.7$ to calculate the critical density of the Universe and the redshift-dependent halo concentrations. The group and cluster correspond to isolated systems that evolve quiescently from a redshift $z = 1$. The aim of this paper is to quantify the effects of the group and cluster environments alone on galaxy evolution; therefore, we study the evolution of the group and cluster in isolated boxes under the assumption that they are collapsed systems whose evolution is unaffected by large-scale cosmic velocity fields. The parameters of the group and cluster are summarized in Table~\ref{table1}. All halos and subhalos are initially assumed to be spherically symmetric, with total density profiles (including subhalo contribution for the halos) specified using a Navarro-Frenk-White profile (NFW, \citealt{Navarro97}):
\begin{equation}
  \rho_{\rm tot}(r \leq R_{200}) = \frac{\rho_{\rm s}}{r/r_{\rm s} (1 + r/r_{\rm s})^2}. 
\end{equation}
The subhalos in the group and cluster are truncated at a distance $R_{200}$\footnote{$R_{200}$ is the radius within which the mean density of the halo, $\overline{\rho} = 3M_{200}/4 \pi R_{200}^3$, is given by $\overline{\rho} = 200 \rho_{\rm crit}$, and $\rho_{\rm crit}$ is the critical density of the universe at $z = 1$.} from their centers, while densities of the group and cluster halos are assumed to fall off exponentially at $r >  R_{200}$:
\begin{equation}
\rho_{\rm tot}(r > R_{200}) =
\frac{\rho_{\rm s}}{c_{200}(1 + c_{200})^2}\left(\frac{r}{R_{200}}\right)^{\kappa} \exp\left(-\frac{r-R_{200}}{r_{\rm decay}}\right).   
\end{equation}
Here $c_{200}$, the concentration parameter, is determined using the redshift-dependent concentration-mass relationship in \citet{Duffy08} at $z = 1$. $r_{\rm s}$ is the NFW scale radius, and $\rho_{\rm s}$ is the NFW scale density. We assume $r_{\rm decay} = 0.1R_{200}$, and $\kappa$ is chosen such that the magnitude and slope of the density profile are continuous at $R_{200}$. The relationships among these parameters are
\begin{align}
  r_{\rm s} &= \frac{R_{200}}{c_{200}} \\
  \rho_{\rm s} &= \frac{200}{3}\rho_{\rm crit} \frac{c_{200}^3} {\log(1 + c_{200}) - c_{200}/(1 + c_{200})}\\
  \kappa &= \frac{R_{200}}{r_{\rm decay}} - \frac{3c_{200}+1}{1+c_{200}} .
\end{align}

Using the observed conditional luminosity function (CLF) of \citet{Yang08}, we create 26 and 152 satellites more massive than $10^9\ \mbox{M}_{\odot}$ within the group and cluster respectively. We assume that the group and cluster galaxies have a constant dynamical mass-to-light ratio of $10\ \mbox{M}_{\odot} / \mbox{L}_{\odot}$, consistent with observations (\citealt{Gerhard01}, \citealt{Padmanabhan04}, \citealt{Humphrey06}). We also allow the group and cluster to have a central brightest cluster galaxy (BCG). The CLF and mass-to-light ratio determine the distribution of satellite galaxy masses.

The radial profiles of the gas distribution in the main halos and subhalos are first calculated. The gas fractions within the main group and cluster halos' $R_{200}$ radii are determined using the observed relation (\citealt{Vikhlinin09}):
\begin{equation}
  f_{\rm g}(h/0.72)^{1.5} = 0.125 + 0.037\log_{10} (M/10^{15} ~\mbox{M}_{\odot}).
\end{equation}
The ICM gas is constrained to be in hydrostatic equilibrium with the group and cluster halos' total gravitational potential (including the subhalo contribution) $\Phi$ using
\begin{equation}
  \frac{dP}{dr} = -\rho_{\rm gas}\frac{d\Phi}{dr}.
  \label{eqn:hse}
\end{equation}
The gas pressure, $P$, density, $\rho_{\rm gas}$, and temperature, $T$, are related in the usual ideal gas form,
\begin{equation}
  P = \frac{k_B}{\mu m_p}\rho_{\rm gas} T ,
\end{equation}
with $\mu\approx 0.59$ for a fully ionized hydrogen plus helium plasma with cosmic abundances. The corresponding adiabatic index is $\gamma = 5/3$. We impose the condition
\begin{equation}
  T(R_{200}) = \frac{1}{2}T_{200} ,
  \label{eqn:tvir}
\end{equation}
where $T_{200}$ is given by
\begin{equation}
  k_B T_{200} \equiv \frac{G M_{200} \mu m_p}{2 R_{200}} .
  \label{eqn:tvir2}
\end{equation}
The equation of hydrostatic equilibrium is solved to initialize the gas density profile, assuming that the cluster and group are relaxed, cool-core systems\footnote{The cool core assumption is justified in \citealt{ZuHone11} and references therein.}, with small core entropies and a given radial entropy profile $S(r) \equiv k_B T(r) n_e(r)^{-2/3}$, where $n_e$ is the electron number density. The entropy profile of each halo is based on observations by \citet{Cavagnolo09} and is of the form
\begin{equation}
  S(r) = S_0 + S_1\left(\frac{r}{R_{200}}\right)^{1.1}.
\end{equation}
We initially calculate $\rho_{\rm gas}(R_{200})$ from $T_{200}$ and $S(R_{200})$, and then numerically solve the equation of hydrostatic equilibrium along with the ideal gas law to calculate $\rho_{\rm gas} (r)$, $P (r)$, and $T(r)$.

We initialize the hot halo gas of galaxy subhalos by assuming that the gas mass is 10\% of the total mass and the gas density profile can be represented by a singular isothermal sphere, $\rho_{\rm gas}(r_{\rm gal}) = \rho_0 r_0^2/r_{\rm gal}^2$, where $r_{\rm gal}$ is the galaxy-centric radius. \footnote{This profile and mass fraction are somewhat {\it ad hoc} and not necessarily an accurate representation of all galactic coronae. However, as argued by \citet{McCarthy08}, non-gravitational processes like cooling and replenishment due to feedback can significantly modify the distribution of galactic gas. Modeling these processes is beyond the scope of this paper.} The temperature at $r_{\rm gal} = R_{\rm 200, gal}$ is determined using the virial temperature relation (Equation~\ref{eqn:tvir}), and the pressure is determined by constraining the subhalo gas to be in hydrostatic equilibrium with the individual subhalo potential. The densities and pressures of the satellite galaxies' gas are added to those of the parent halo. The BCG's hot gas halo is initialized in a similar fashion to the satellites', with the additional constraint that the density and pressure profiles continuously join onto those of the ICM.

We determine the positions of dark matter particles for the parent halos using the spherically averaged dark matter density profile, $\rho_{\rm DM} = \rho_{\rm tot} - \rho_{\rm gas} - \rho_{\rm subhalo}$. In this equation, the initial estimate for $\rho_{\rm subhalo}$ is calculated from $\rho_{\rm subhalo} = \rho_{\rm tot} \times M_{\rm subhalo, tot} / M_{\rm main, tot}$. The positions of the subhalo particles are determined in a similar fashion, from $\rho_{\rm DM, sub} = \rho_{\rm tot, sub} - \rho_{\rm gas, sub}$.  Given this profile for $\rho_{\rm DM}$, we use the procedure outlined in \citet{Kazantzidis04} to initialize the positions and velocities of dark matter particles, which each have mass $10^6\ \mbox{M}_{\odot}$. For each particle, we draw a uniform random deviate $u$ in $[0,1)$ and choose the particle's halo-centric radius, $r$, by inverting the function 
\begin{equation}
u = \frac{\int_0^r \rho_{\rm DM}(r) r^2 dr}{\int_0^\infty \rho_{\rm DM}(r) r^2 dr} .
\end{equation}
To calculate particle velocities, we use the Eddington formula for the distribution function (\citealt{Eddington16}, \citealt{Binney08}):
\begin{equation}
  f(\mathcal{E}) = \frac{1}{\sqrt{8}\pi^2}\left[\int_0^{\mathcal{E}} \frac{d^2\rho_{\rm DM}}{d\Psi^2}\frac{d\Psi}{\sqrt{\mathcal{E} - \Psi}} + \frac{1}{\sqrt{\mathcal{E}}}\left(\frac{d\rho_{\rm DM}}{d\Psi}\right)_{\Psi=0} \right].
\end{equation}
Here $\Psi = -\Phi$ is the relative potential of the particle, based on the total density $\rho_{\rm tot}$ or $ \rho_{\rm tot, sub}$, and $\mathcal{E} = \Psi - \frac{1}{2}v^2$ is the relative energy. Using an acceptance-rejection technique, we choose random particle speeds $v$ given $f(\mathcal{E})$, assuming an isotropic velocity distribution.
 
The positions and velocities of the satellites are drawn from the above distribution of dark matter particle positions and velocities. These are initialized to be non-overlapping, and the total density of the subhalo particles, $\rho_{\rm subhalo}$, is calculated. Introducing the subhalos within the main halo breaks the smoothness of the main halo's density profile. Therefore, to maintain a radially averaged smooth density profile, we re-initialize the main halo's particles, but now $\rho_{\rm subhalo}$ is the spherically averaged contribution of the subhalos to the total density profile. We neglect the effect of the breaking of spherical symmetry in the difference between the parent $\rho_{\rm tot}$ and the parent $\rho_{\rm gas}$ due to the subhalos. The BCG is initialized so that its center coincides with that of its parent halo and it has zero peculiar velocity with respect to its parent halo. The particles belonging to the subhalos are also initialized using the above technique, with the appropriate density profiles and distribution functions, and with the subhalo potentials.

The group simulation is performed in a cubic box of side $10^{25}~\mbox{cm}$ ($3.24~\mbox{Mpc}$, physical units), and the cluster simulation in a cubic box of side $2\times 10^{25}~ \mbox{cm}$ ($6.48~\mbox{Mpc}$). The group and cluster simulations have a maximum of 8 and 9 levels of refinement respectively, corresponding to a maximum resolution of 1.6 kpc. These are idealized isolated simulations, with the implicit assumption that the group and cluster are collapsed, gravitationally bound regions removed from the expansion of the Universe. The simulations were run for $7.61$ Gyr, corresponding to the lookback time at $z = 1$ for our chosen cosmological parameter values.

\subsection{Resolution and convergence tests}

\begin{figure*}
  \begin{center}
    \subfigure[L2 norm in $M(R_{200})$ and $M(r_{\rm s})$]
    {\includegraphics[width=3.4in]{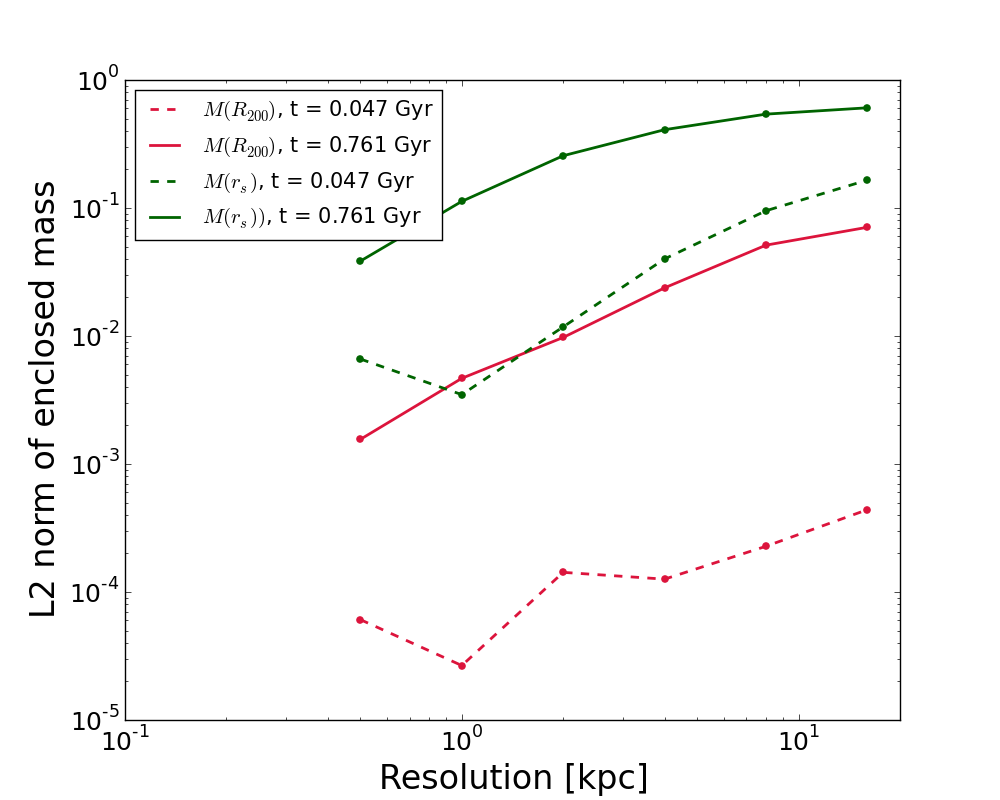}\label{fig:rvir_rs_l2norm}}
    \subfigure[L2 norm in $r(0.5 M_{200})$ and $r(0.1 M_{200})$]
    {\includegraphics[width=3.4in]{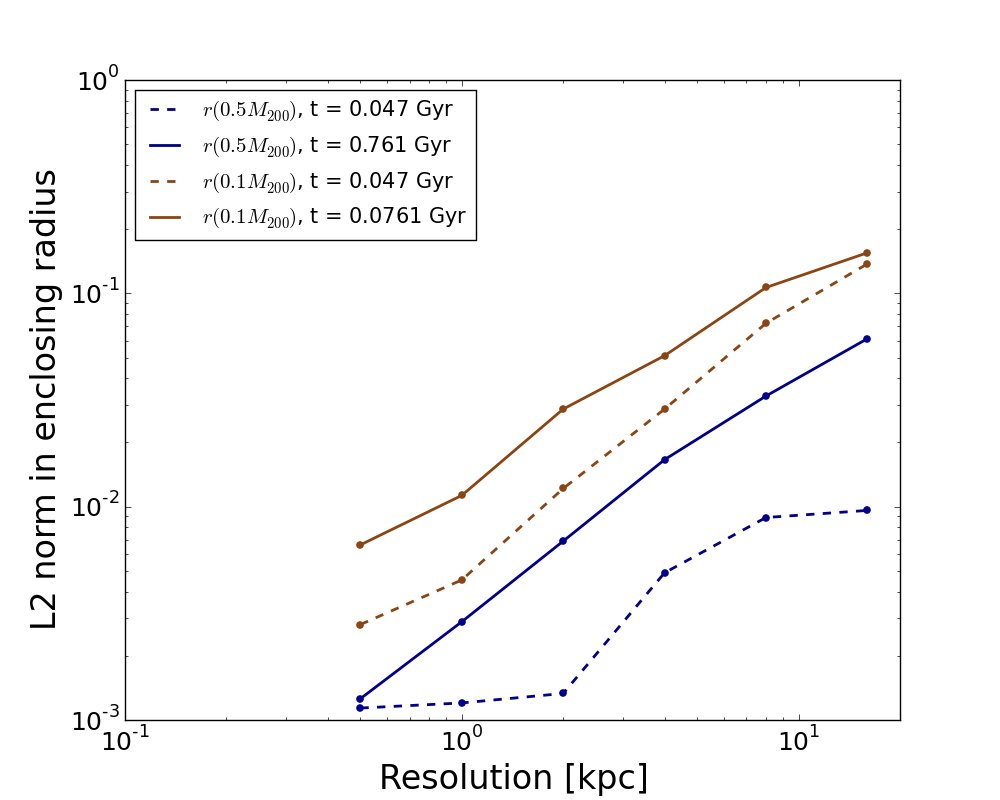}\label{fig:mhalf_mtenth_l2norm}}
    \caption{Left: The L2 norm in $M(r_{200})$ and $M(r_{\rm s})$, calculated and normalized with respect to the values of the 0.25 kpc simulation, at $t_{01} = 47.6$ Myr and $t_{16} = 761$ Myr.  Right:  The L2 norm in $r(0.5 M_{200})$ and $r(0.1 M_{200})$, calculated and normalized with respect to the values of the 0.25 kpc simulation, at $t_{01}$ and $t_{16}$. \label{fig:rfrac_mhalf_l2norm}}
  \end{center}  
\end{figure*}

\begin{figure*}
  \begin{center}
    {\includegraphics[width=3.6in]{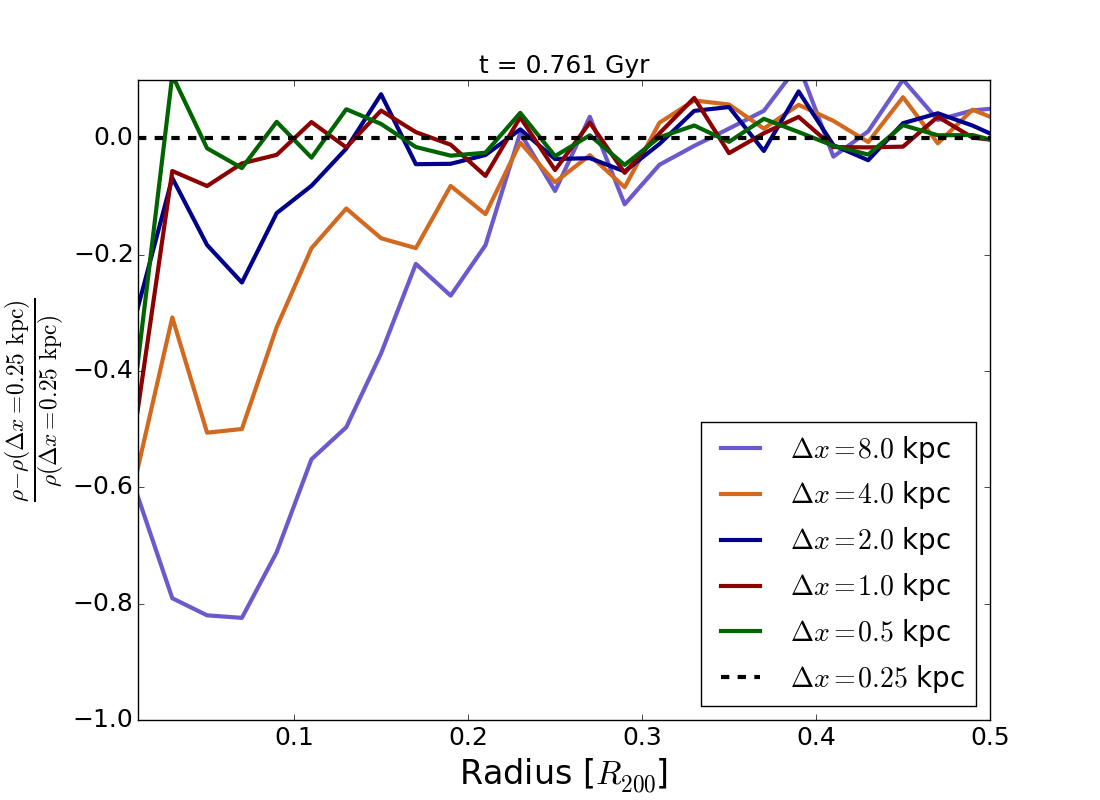}\label{fig:rhostackt16}}
    \caption{Normalized deviation in stacked galaxy density profiles from 0.25 kpc resolution run, with density normalized to each subhalo's scale density, $\rho_{\rm s}$, and radius normalized to $R_{200}$ at $t_{16} = 761$ Myr.\label{fig:rhostack_normdev}}
  \end{center}  
\end{figure*}

The simulations must have sufficient spatial resolution to prevent the artificial flattening of density profiles and avoid the rapid disruption of a galaxy's particles. To test the robustness of our simulations against such effects, we performed a series of simulations of the group and its subhalos with varying minimum spatial resolution (corresponding to the maximum refinement level) from 0.25 kpc to 16 kpc. These convergence tests used particle masses of $10^7\ \mbox{M}_{\odot}$ and lasted for $2.4 \times 10^{16}$ seconds (0.76 Gyr) each. 

We use four primary metrics to probe the evolution of subhalo structure within the group: the mass enclosed within the scale radius and virial radius of the original subhalo ($M(r_{\rm s})$ and $M(R_{200})$), and the radius enclosing half the mass (half-mass radius) and $10\%$ of the total mass of the original subhalo ($r(0.5 M_{200})$ and $r(0.1 M_{200})$). Overall, the mean mass within the subhalos' original $R_{200}$ decreases with time, and the half-mass radius increases with time. For a typical galaxy with a scale radius of $\sim$ 8 kpc, a minimum resolution of 2 kpc corresponds at least 8 zones per dimension across the central core of the subhalo. For resolutions of $\sim 4 - 8$ kpc, the central core is resolved with $2 - 4$ zones, so the subhalos in these simulations have poorly-resolved cores and are flattened out.

We quantify the dynamical effects of varying spatial resolution by comparing the relative error in the above quantities at a given simulation timestep. We define the L2 error norm in $M(R_{200})$ as
\begin{equation}
 {\rm L2} = \sqrt{\left \langle \left( \frac{M(R_{200}(\Delta x)) - M(R_{200}(\Delta x = 0.25 {\rm\ kpc}))}{M(R_{200}(\Delta x = 0.25 {\rm\ kpc}))} \right)^2 \right \rangle},
\end{equation}
where the average is taken over all the subhalos. We similarly define the L2 norms for the other three quantities and calculate them at $t_{01} = 0.0476$ Gyr and $t_{16} = 0.761$ Gyr. The results are plotted in Figure~\ref{fig:rfrac_mhalf_l2norm}. For the most part they are consistent with error growth ${\cal O}(\Delta x)$ with increasing minimum zone spacing $\Delta x$. Additionally, we see that the deviations with respect to the best-resolved simulation are larger at the later timestep ($t_{16}$, solid lines) compared to the deviations at the earlier timestep ($t_{01}$, dashed lines). This is consistent with a scenario in which poorly resolved subhalos are more susceptible over time to tidal disruption and smearing. Errors are also systematically larger for $M(r_{\rm s})$ and $r(0.1 M_{200})$, as expected since these scales are smaller than the virial radius.

To further illustrate the effects of varying spatial resolution on the internal structure of subhalos, we calculate the density profiles of individual subhalos and stack them at a given timestep. In this stacking process, the densities and radii are normalized to each subhalo's initial scale density, $\rho_{\rm s}$, and $R_{200}$ respectively. We then calculate the normalized deviation in stacked densities at each resolution level from the stacked profile for a resolution of 0.25 kpc using
\begin{equation}
 \rho_{\rm norm, dev} = \frac{\rho(\Delta x) - \rho(\Delta x = 0.25\ \mbox{kpc})}{\rho(\Delta x = 0.25\ \mbox{kpc})},
\end{equation}
where we have suppressed the radial coordinate in the profile for clarity.
Figure~\ref{fig:rhostack_normdev} illustrates the effect of spatial resolution on internal density. There is an obvious trend of decreasing central density with worsening spatial resolution, indicating that the subhalos are being smeared out. This deviation is $\sim 20\%$ for a resolution of $1 - 2$ kpc within $0.1 R_{200}$ and increases to $\sim 50\%$ for a resolution of $4 - 8$ kpc within $0.3 R_{200}$. $r_s$ is typically $0.15 - 0.2 R_{200}$, where the density error calculated in Figure~\ref{fig:rhostack_normdev} is less than $10 \%$. Given these results, we chose a minimum spatial resolution of 1.6 kpc, for which the L2 norms in $M(r_{\rm s})$ and $r(0.1 M_{200})$ compared to a minimum resolution of 0.25 kpc are $\sim 10^{-2} - 10^{-3}$. 

\section{Results}
\label{sec:results}

\subsection{Projected gas temperature maps}
\label{sec:tmap}

\begin{figure*}
  \begin{center}
    \subfigure
    {\includegraphics[width=3.2in]{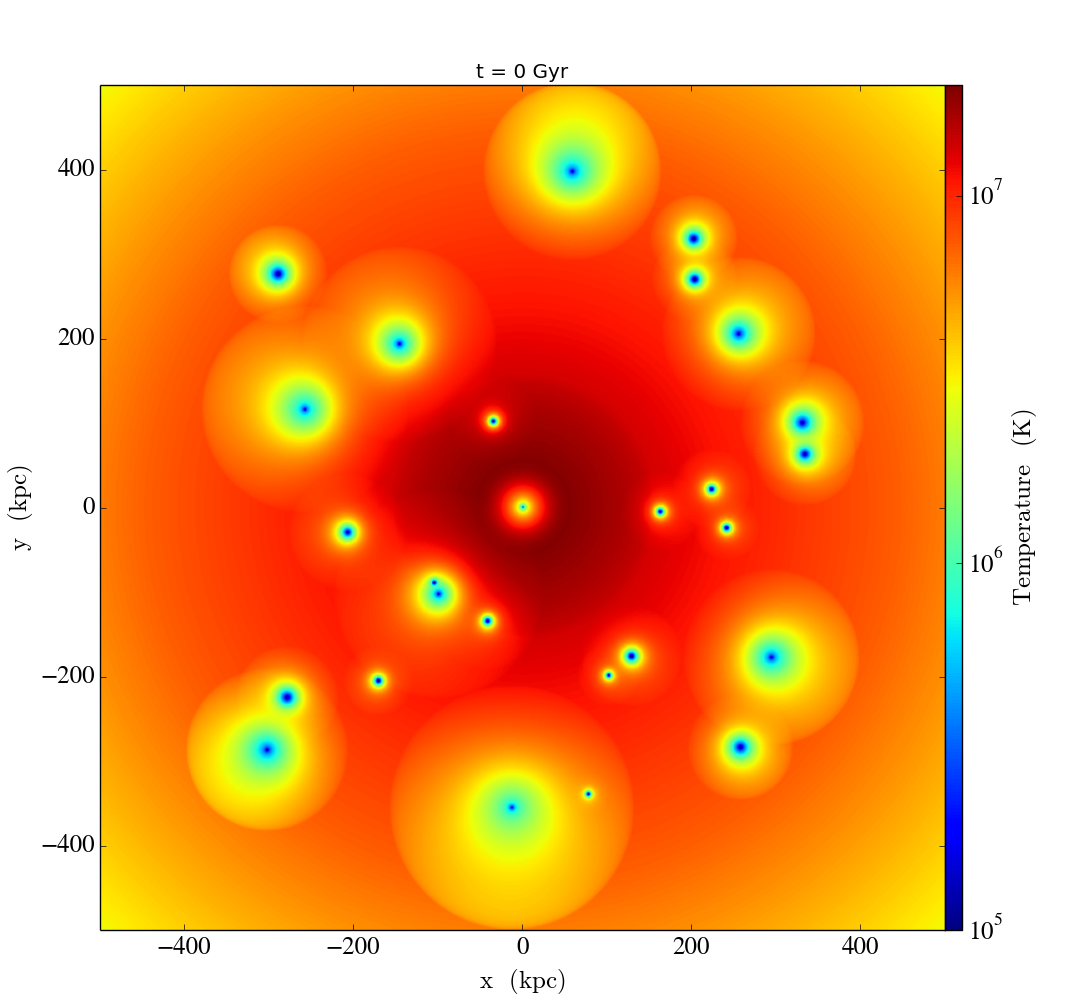}\label{fig:groupgasTEM0}}
    \subfigure
    {\includegraphics[width=3.2in]{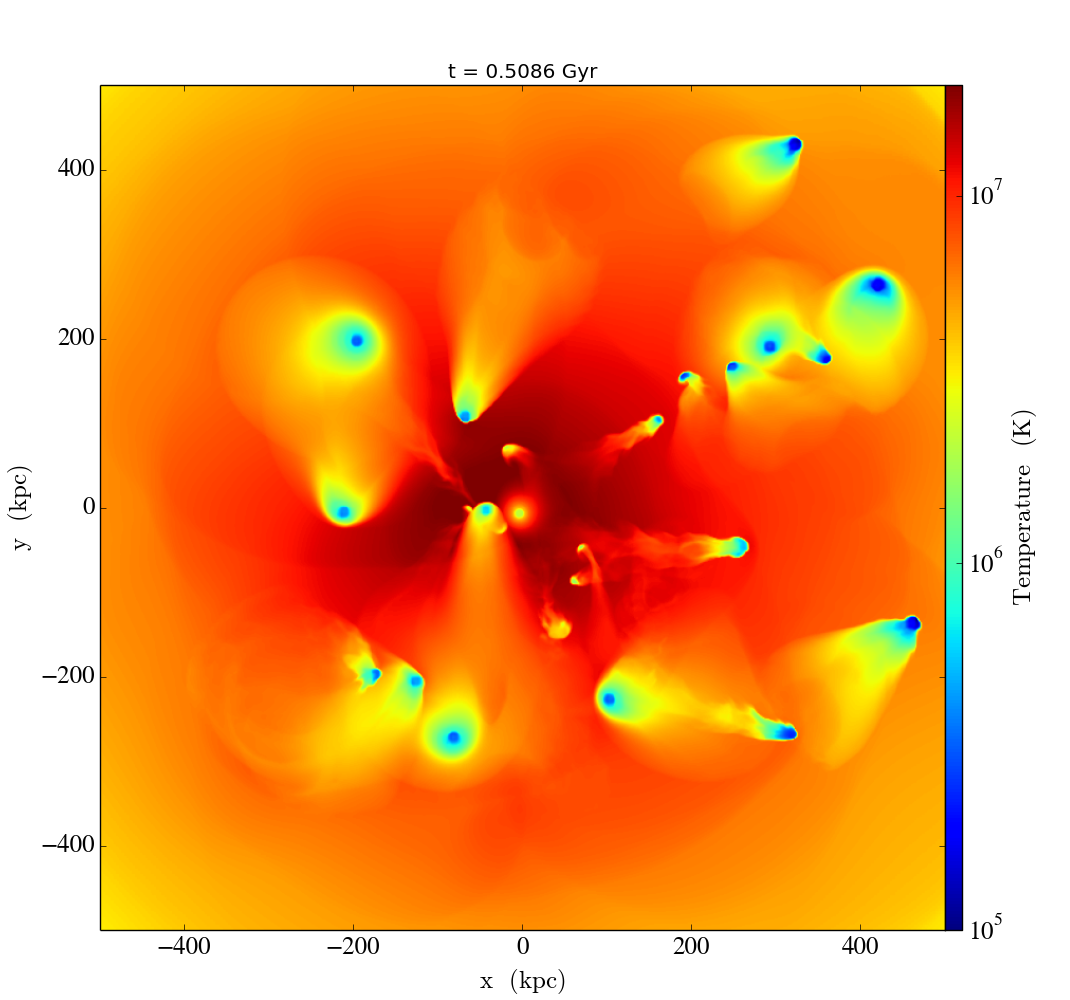}\label{fig:groupgasTEM32}}
    \subfigure
    {\includegraphics[width=3.2in]{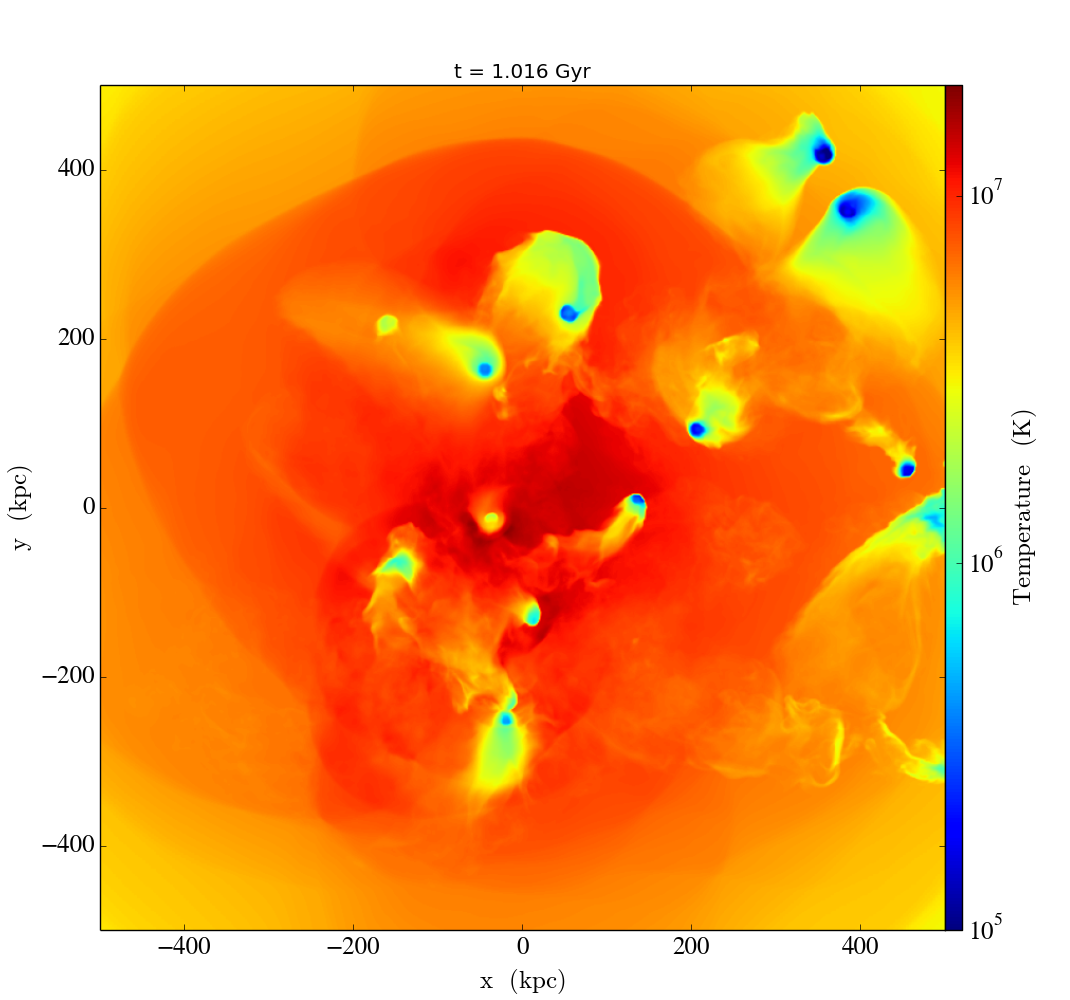}\label{fig:groupgasTEM64}}
    \subfigure
    {\includegraphics[width=3.2in]{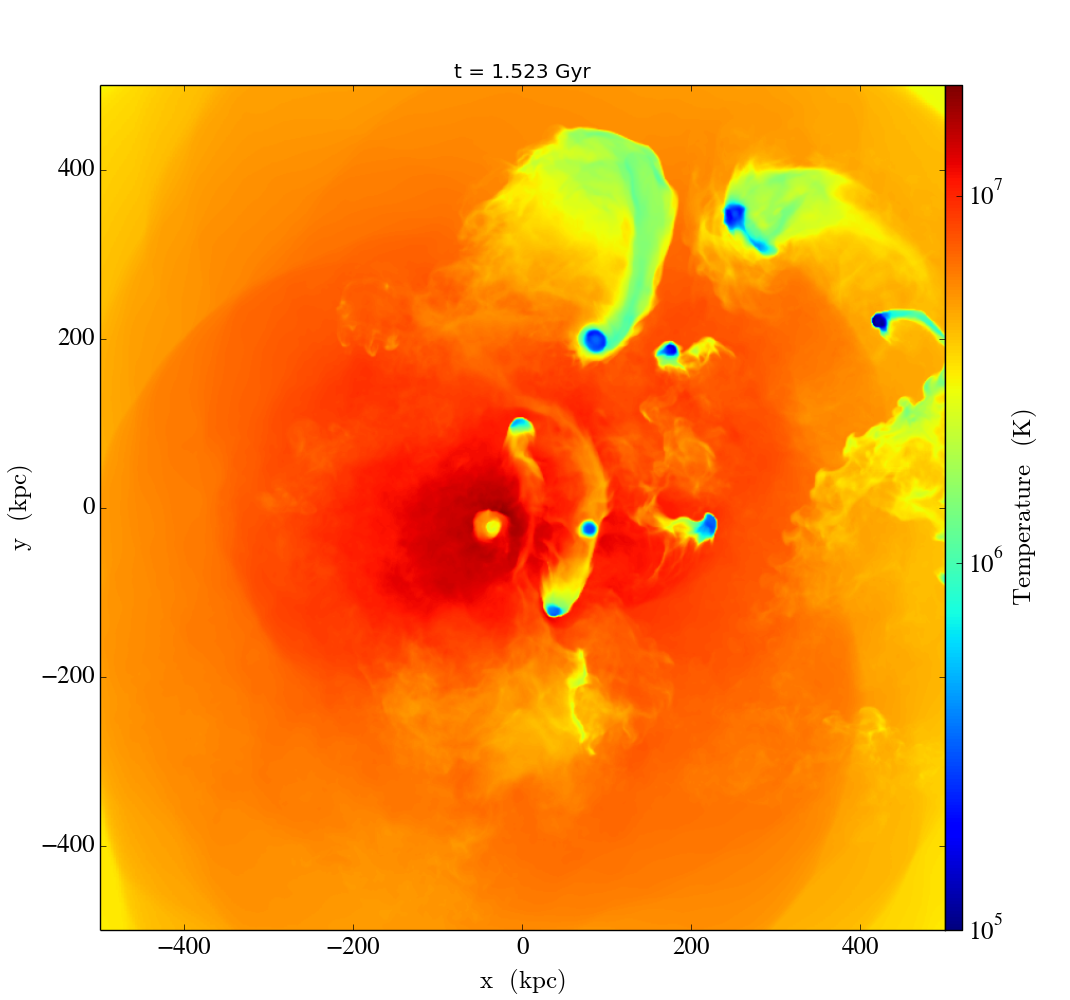}\label{fig:groupgasTEM96}}
    \subfigure
    {\includegraphics[width=3.2in]{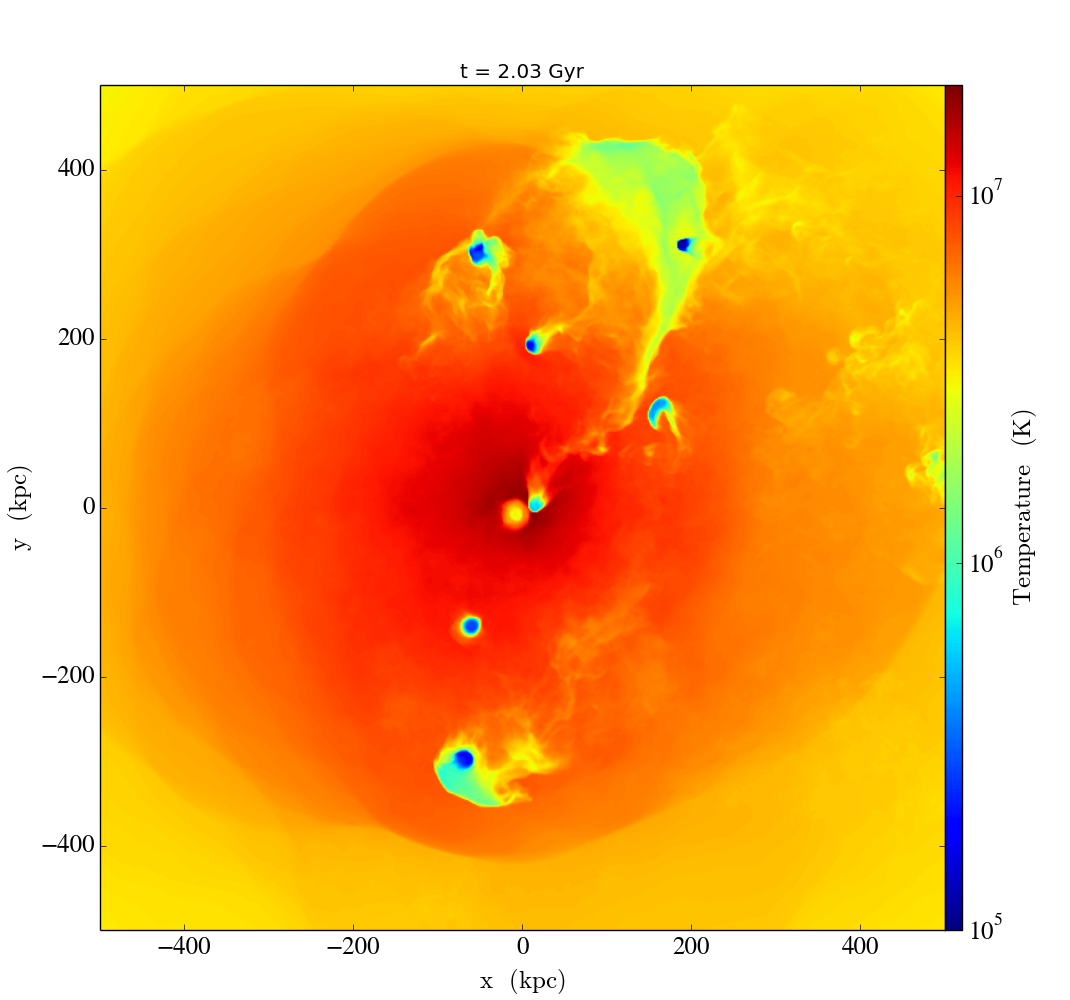}\label{fig:groupgasTEM128}}
    \subfigure
    {\includegraphics[width=3.2in]{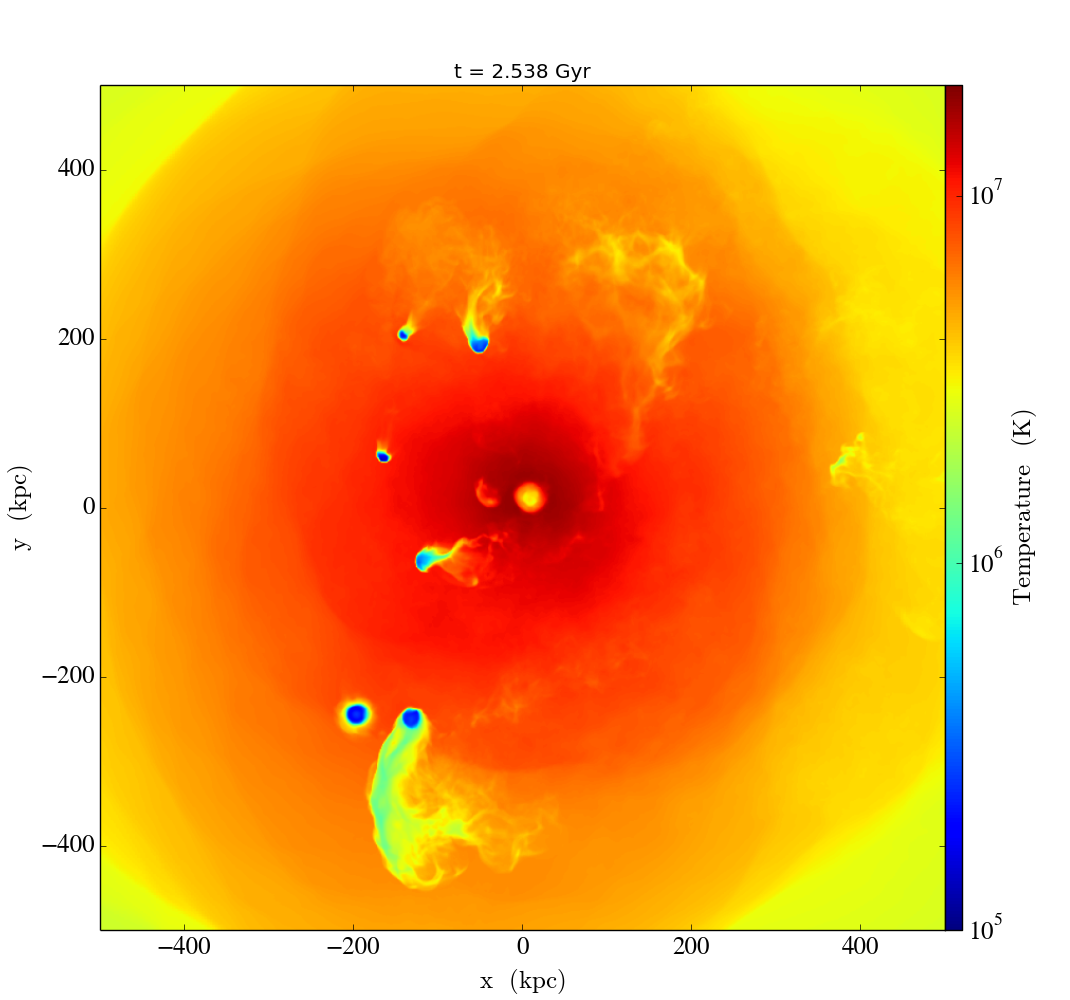}\label{fig:groupgasTEM160}}
     \caption{The evolution of gas in the isolated group and its galaxies, as seen in maps of emission measure-weighted temperature. Galaxies are stripped of their gas by the ICM, and the stripped gas trails galaxies in their orbits in the form of wakes before mixing with the ICM. \label{fig:groupgasT_EM}}
  \end{center}  
\end{figure*}

\begin{figure*}
  \begin{center}
    \subfigure
    {\includegraphics[width=3.2in]{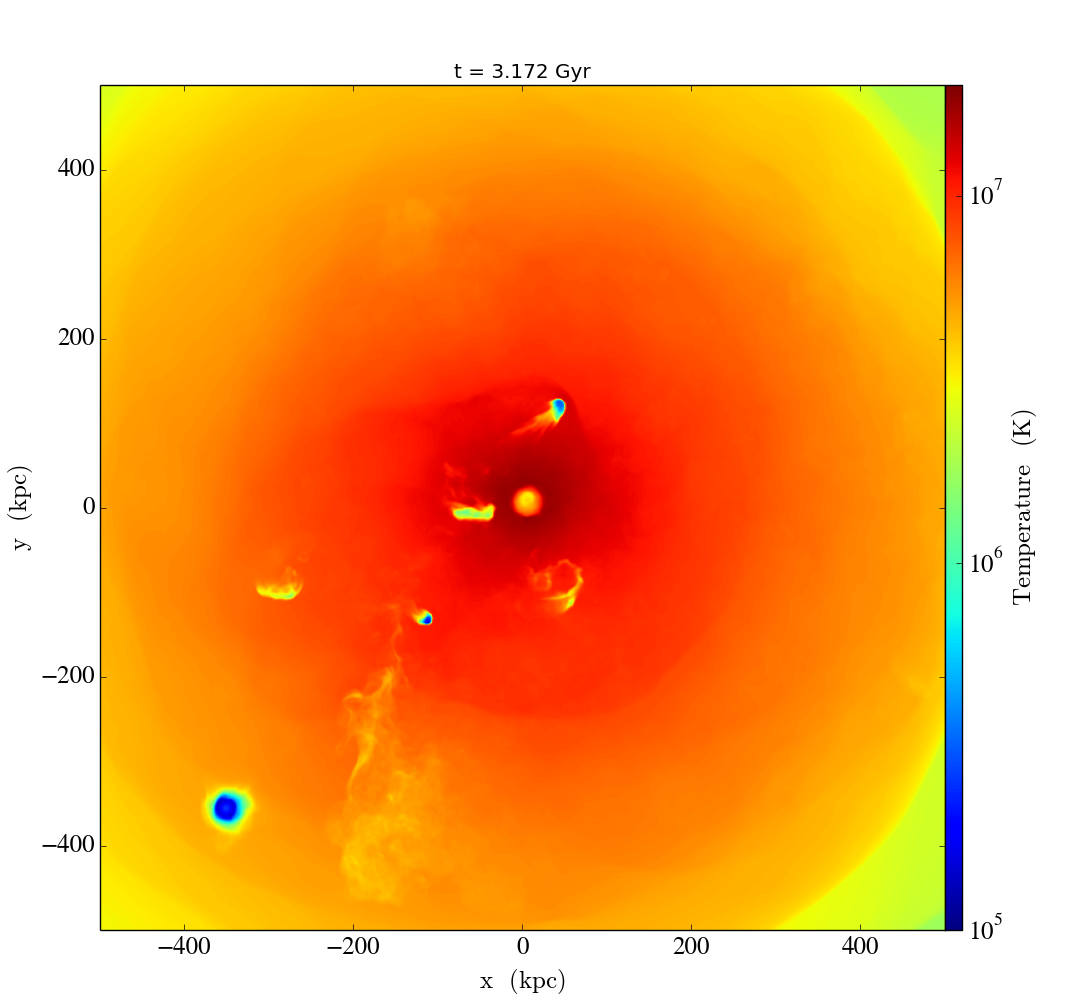}\label{fig:groupgasTEM200}}
    \subfigure
    {\includegraphics[width=3.2in]{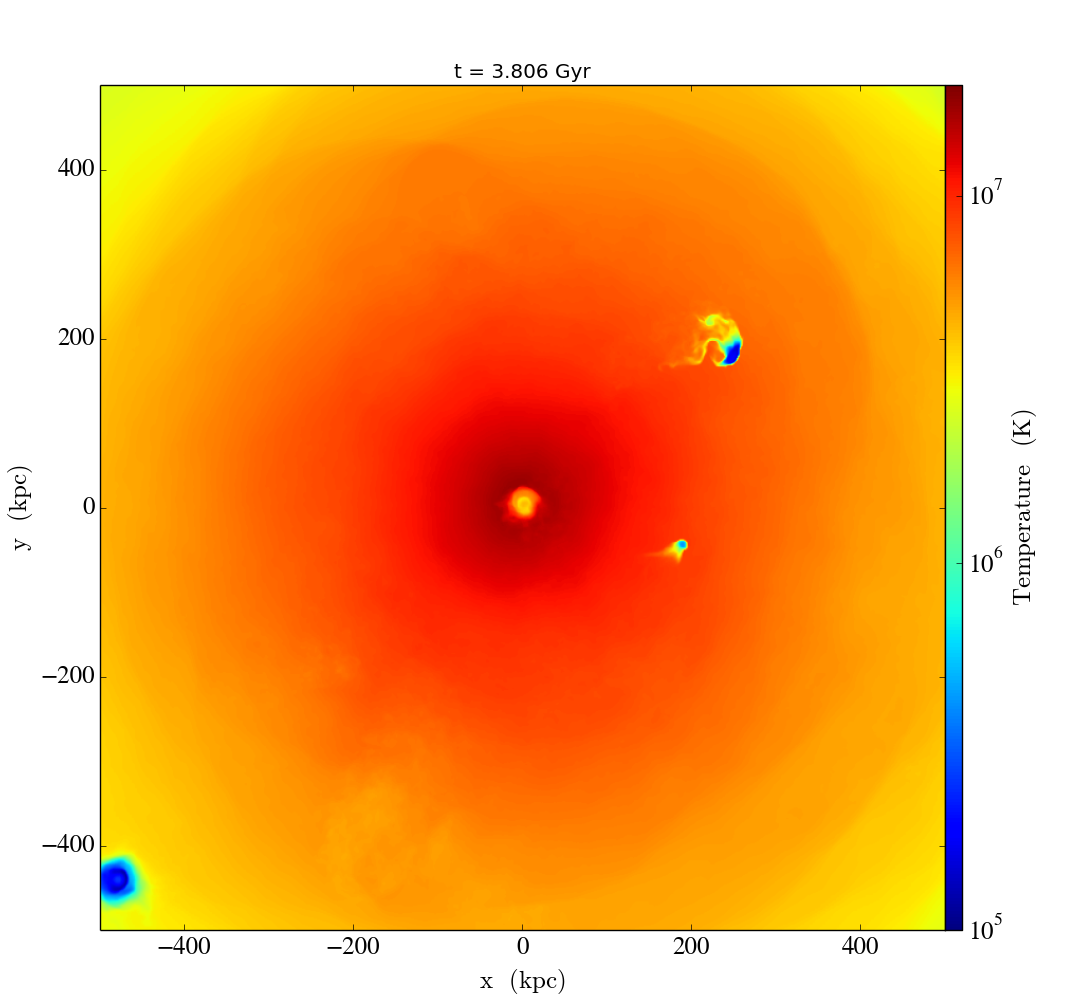}\label{fig:groupgasTEM240}}
    \subfigure
    {\includegraphics[width=3.2in]{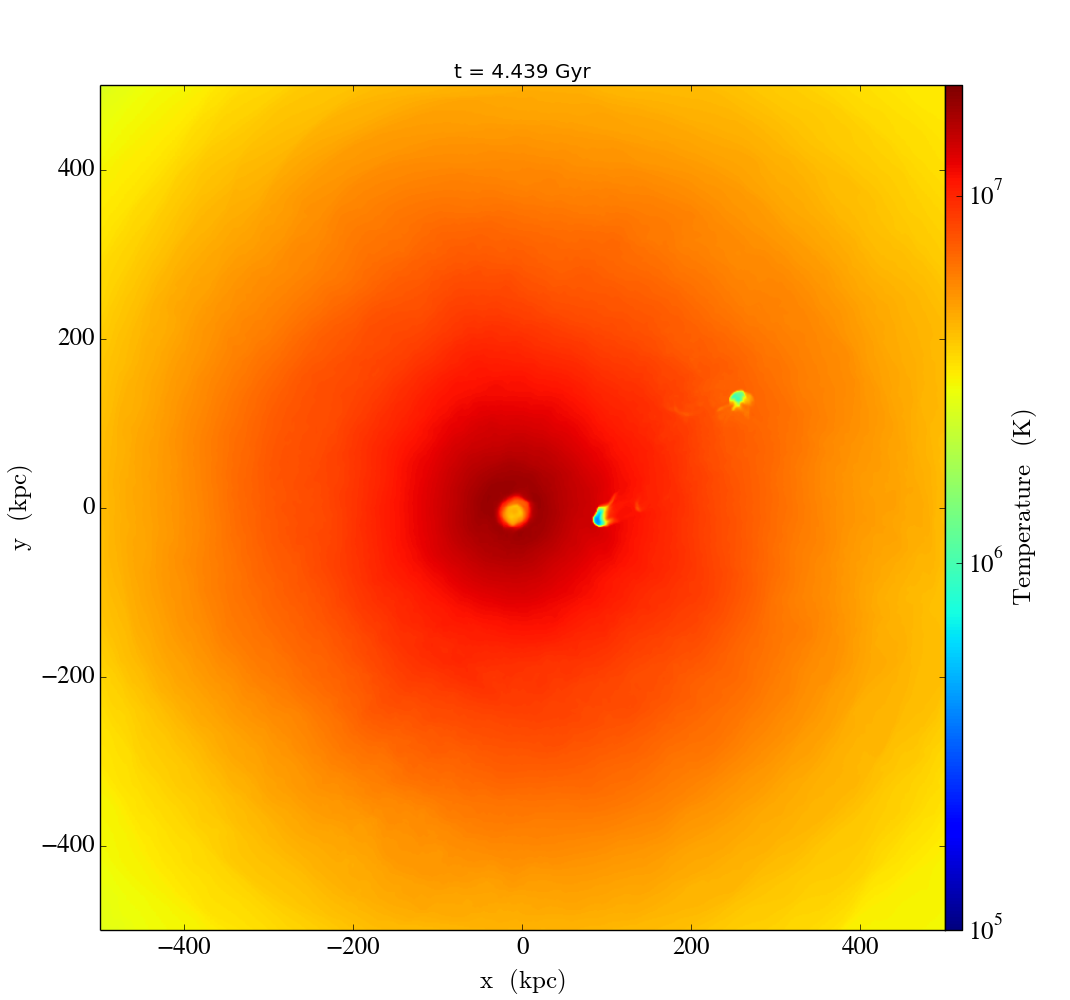}\label{fig:groupgasTEM280}}
    \subfigure
    {\includegraphics[width=3.2in]{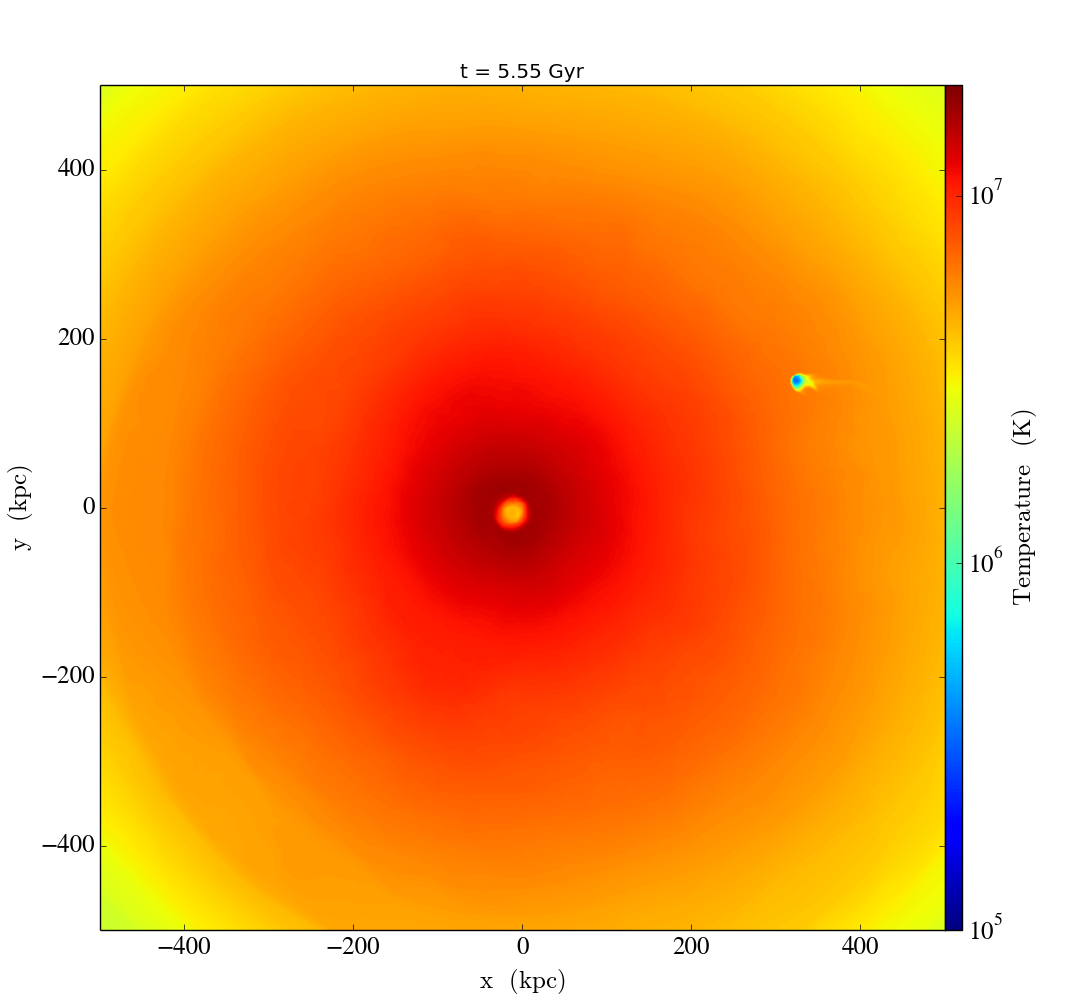}\label{fig:groupgasTEM350}}
    \subfigure
    {\includegraphics[width=3.2in]{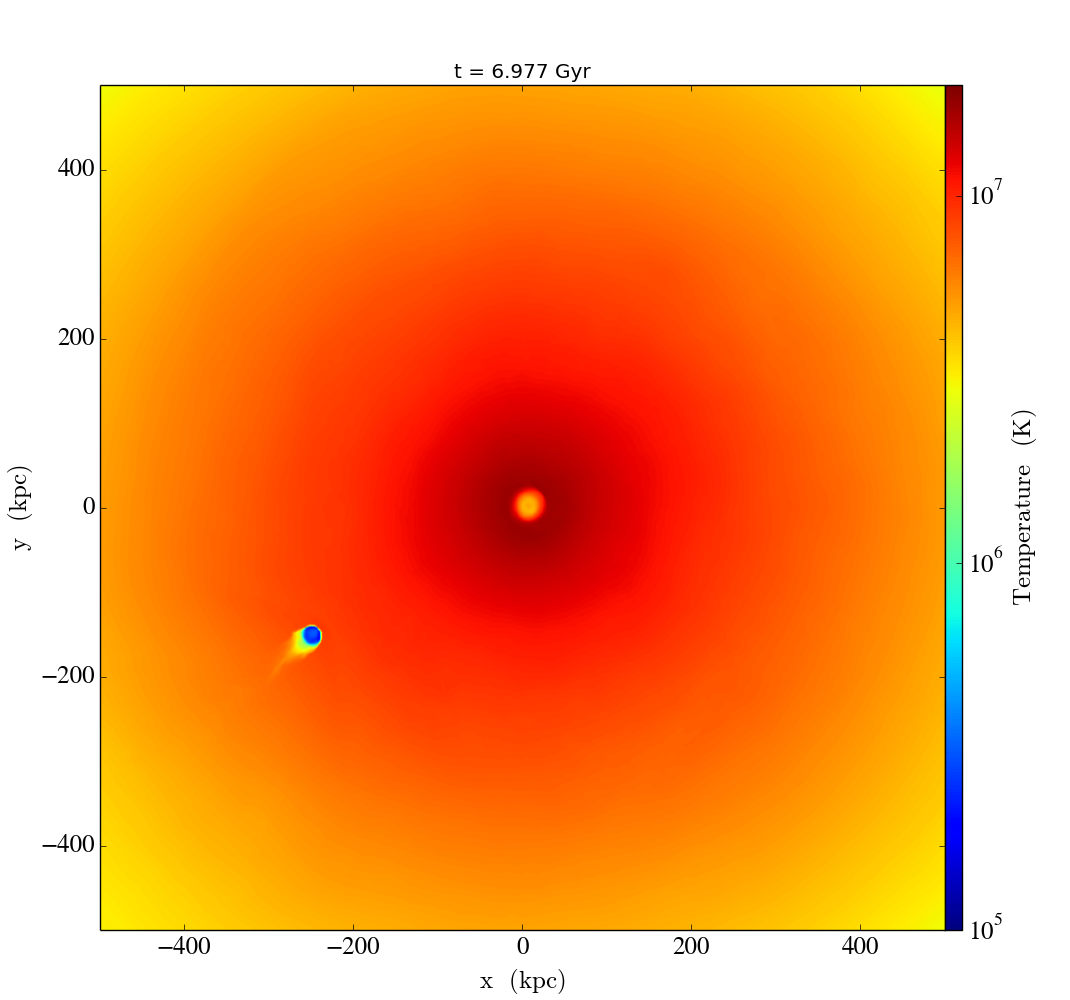}\label{fig:groupgasTEM440}}
    \subfigure
    {\includegraphics[width=3.2in]{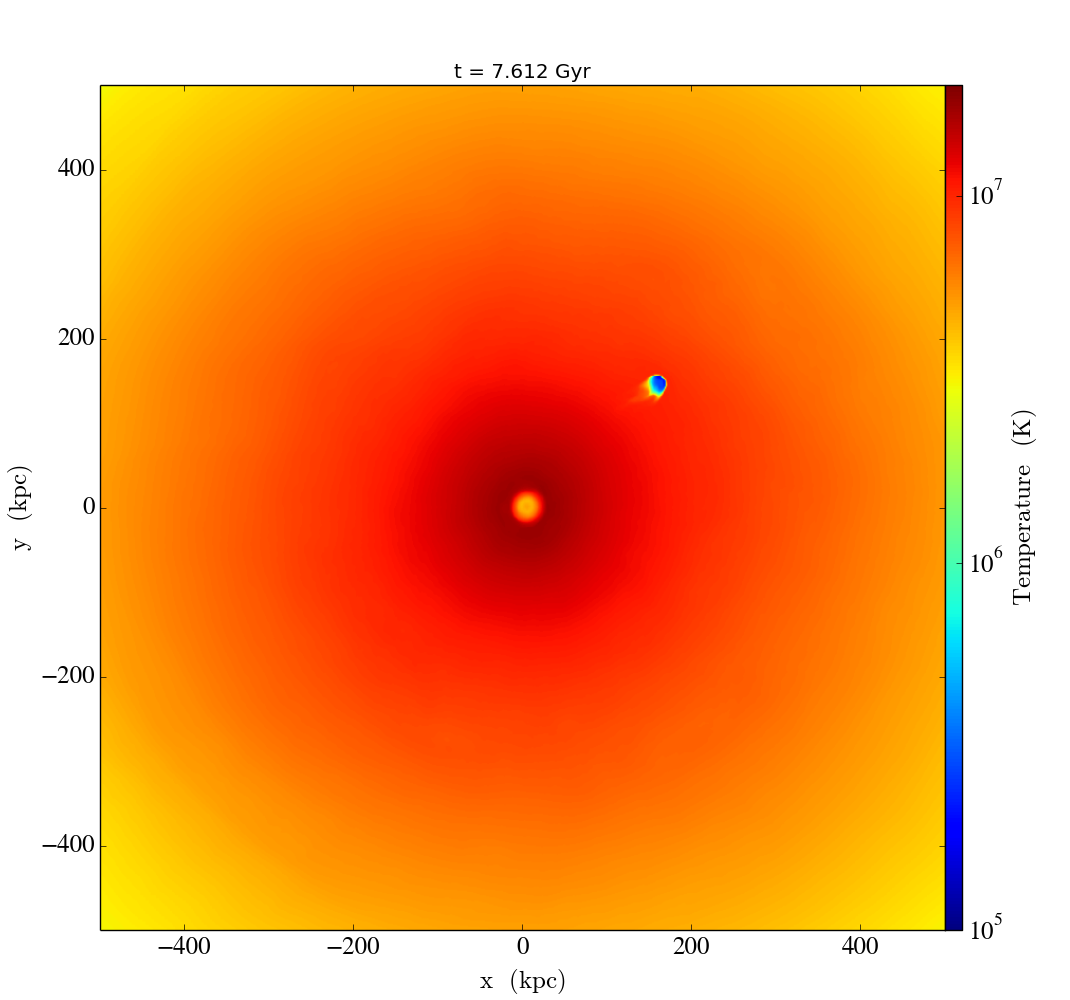}\label{fig:groupgasTEM480}}
     \caption{Emission measure-weighted temperature maps of the group and its galaxies at late times. Most galaxies have lost their gas by $t \gtrsim 3$ Gyr; a few coronae survive up to $\sim 4$ Gyr. The orbit of the galaxy at the bottom left corner of the first panel (at [$X, Y$] = [-350 kpc, -350 kpc], $t$ = 3.172 Gyr) is outside the inner $500 \times 500$ kpc region at $4.4 - 5.5$ Gyr and re-enters the central region at $\sim 6.5$ Gyr. It is the last surviving galactic corona.  The last panel corresponds to the end of the simulation at 7.6 Gyr, when the ICM has relaxed to equilibrium.   \label{fig:groupgasT_EM2}}
  \end{center}  
\end{figure*}

\begin{figure*}
  \begin{center}
    \subfigure
    {\includegraphics[width=3.2in]{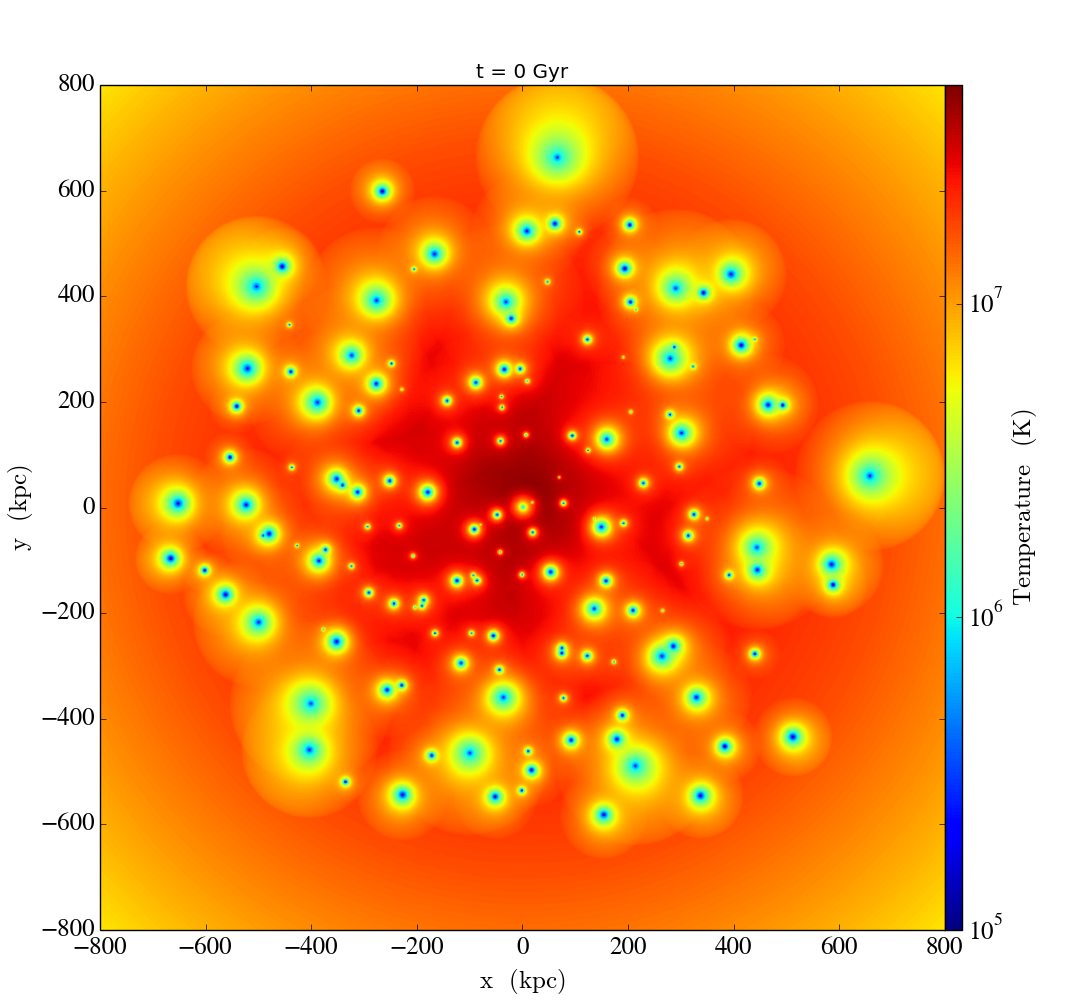}\label{fig:clustergasTEM0}}
    \subfigure
    {\includegraphics[width=3.2in]{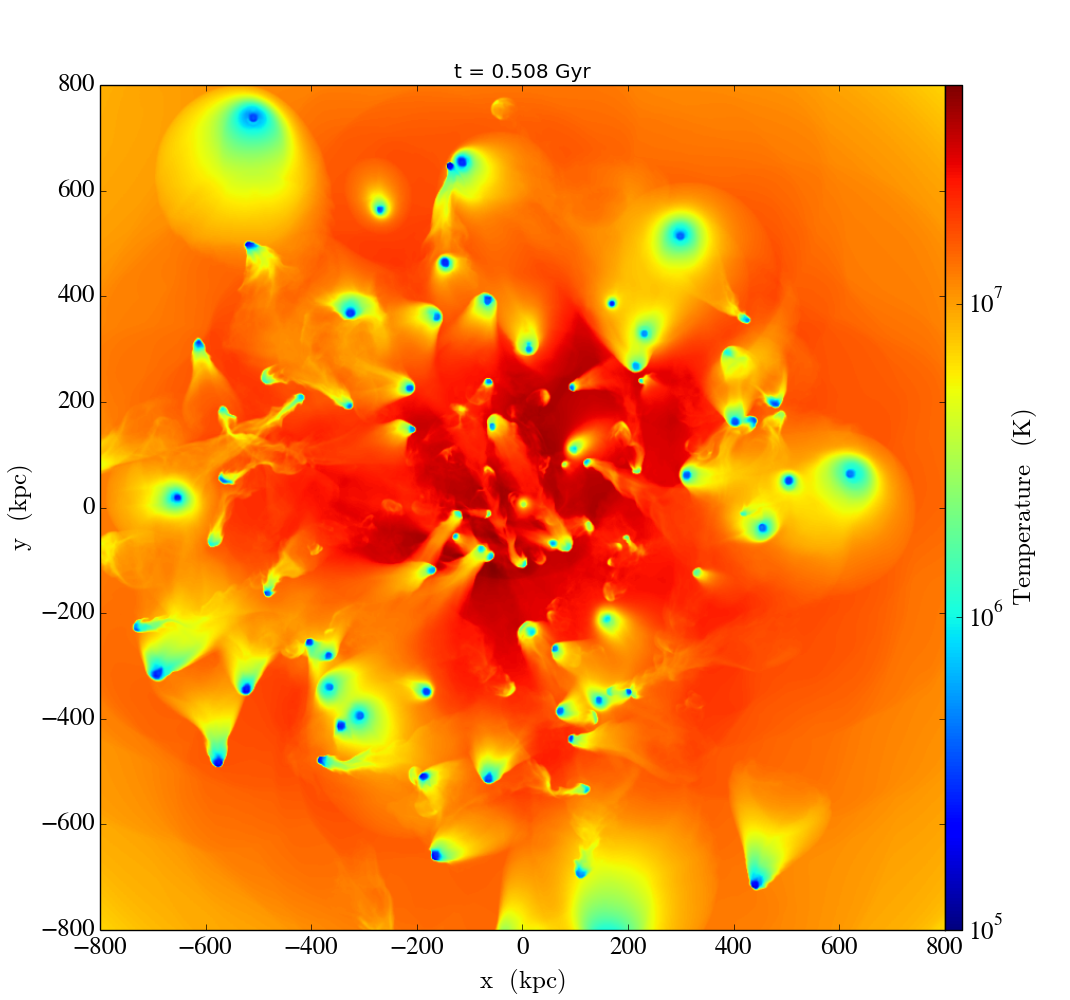}\label{fig:clustergasTEM32}}
    \subfigure
    {\includegraphics[width=3.2in]{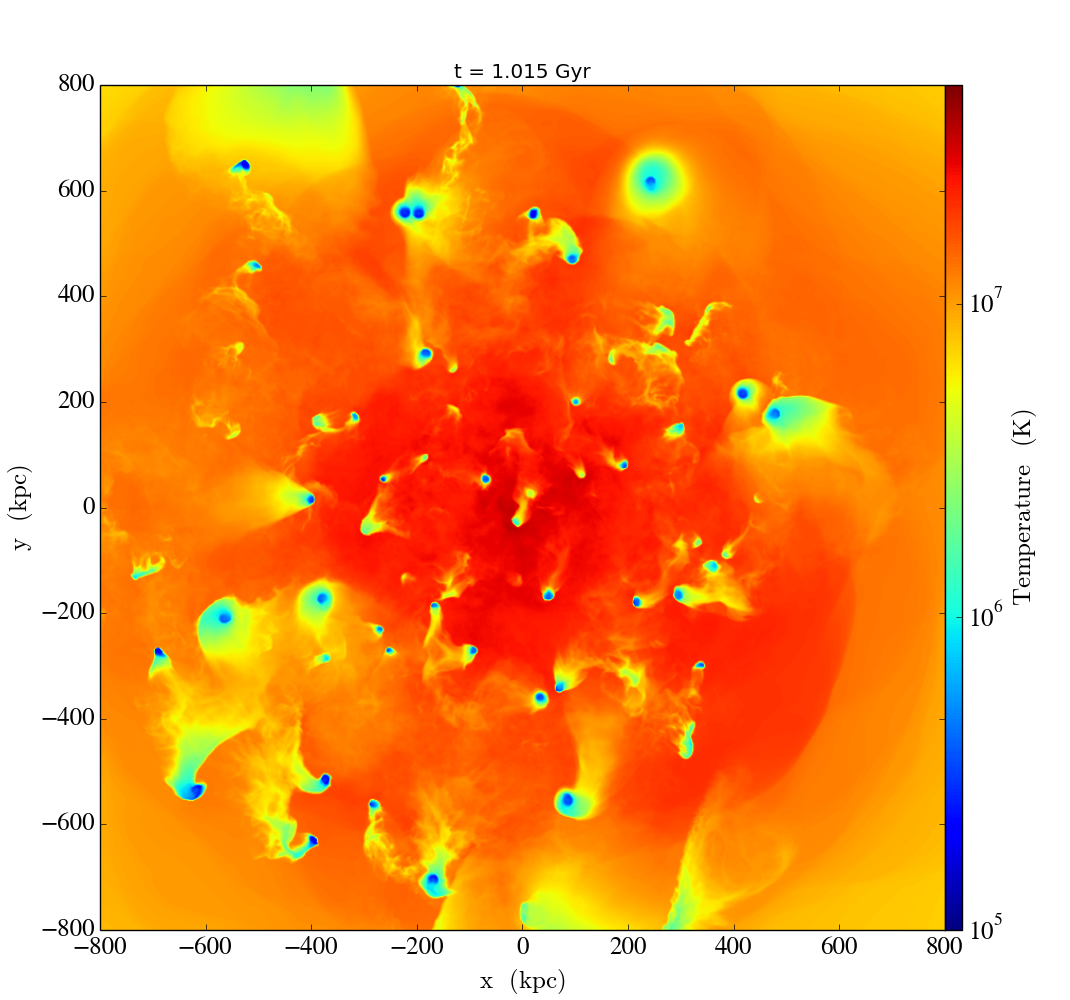}\label{fig:clustergasTEM64}}
    \subfigure
    {\includegraphics[width=3.2in]{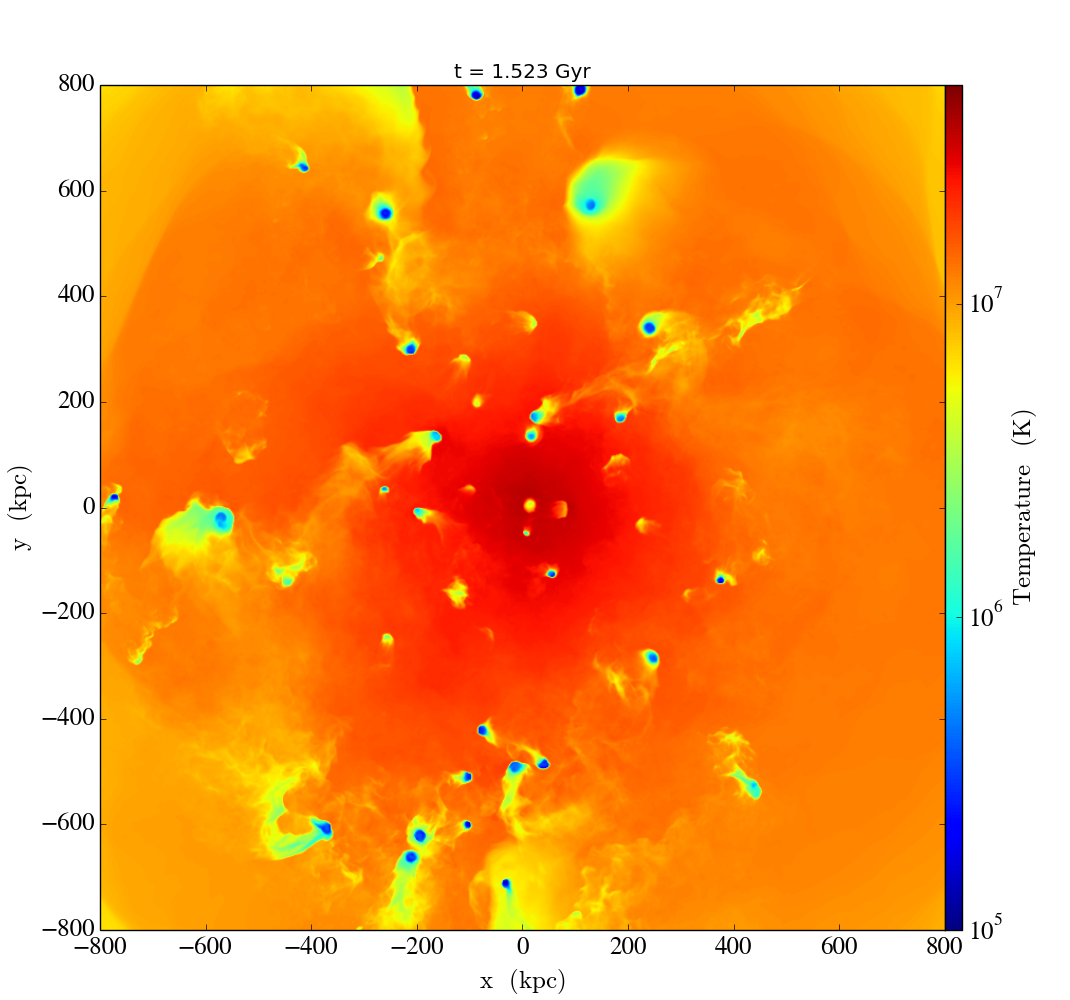}\label{fig:clustergasTEM96}}
    \subfigure
    {\includegraphics[width=3.2in]{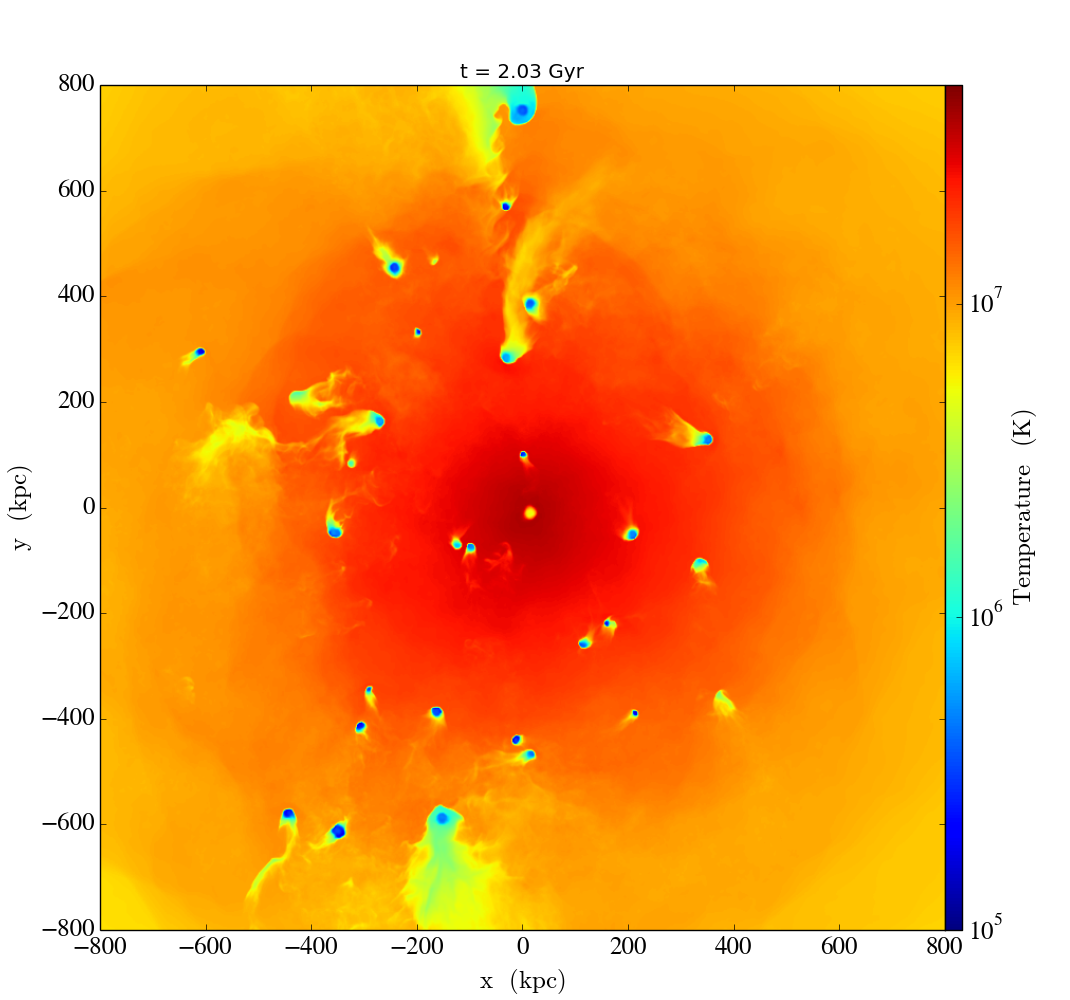}\label{fig:clustergasTEM128}}
    \subfigure
    {\includegraphics[width=3.2in]{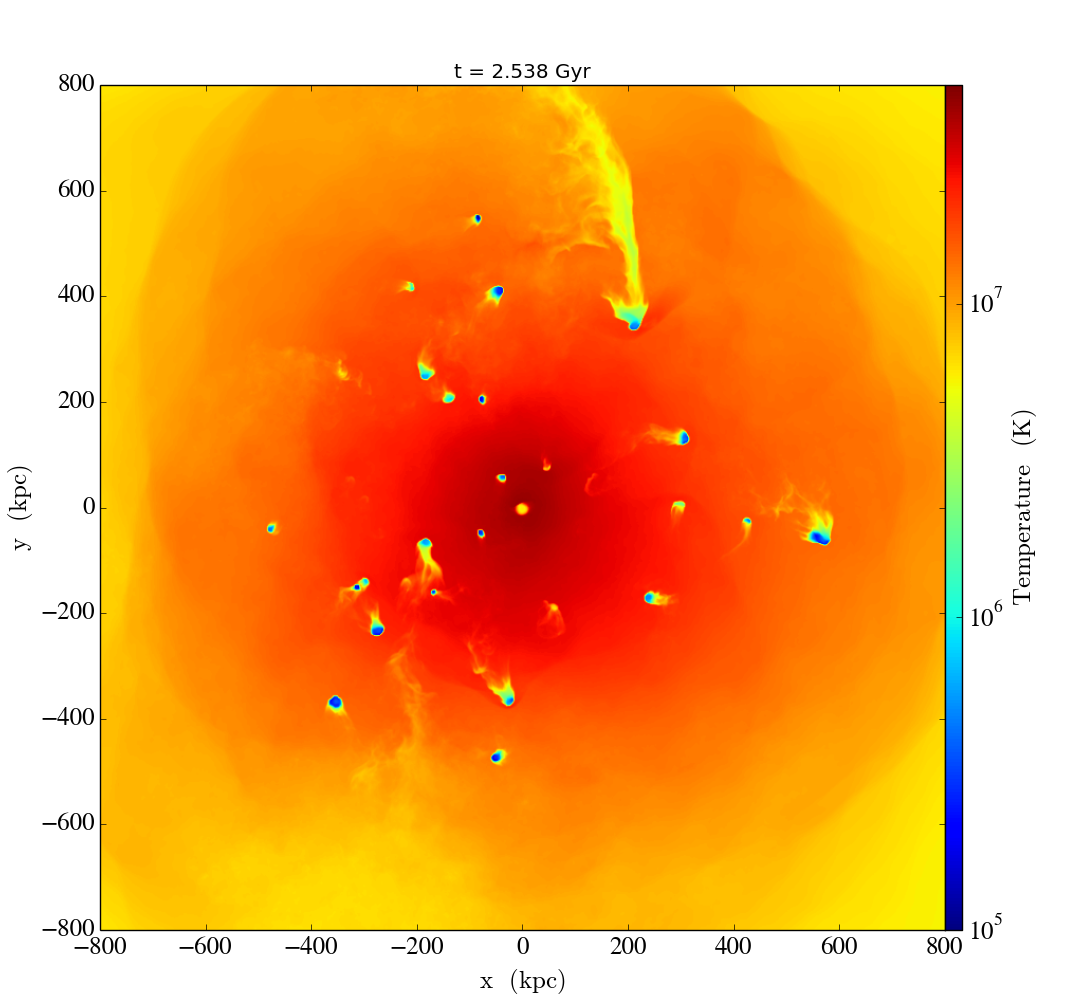}\label{fig:clustergasTEM160}}
     \caption{As for Figure~\ref{fig:groupgasT_EM}, but for the isolated cluster.   \label{fig:clustergasT_EM}}
  \end{center}  
\end{figure*}

In this section we present a qualitative overview of the evolution of the group and cluster galaxies' coronal gas. Figures~\ref{fig:groupgasT_EM}, ~\ref{fig:groupgasT_EM2}, and ~\ref{fig:clustergasT_EM} are emission measure-weighted temperature maps of the gas in the group and cluster, including the gas bound to their galaxies in the form of hot X-ray emitting coronae\footnote{All simulation snapshots were generated using the \texttt{yt} analysis package (\citealt{Turk11}, \url{http://yt-project.org/}) The emission measure is calculated as $\int n_e^2 dl$ along the line of sight through the group and cluster.}. The gas that is removed from galaxies trails them in their orbits in the form of wakes before dissipating within the ICM. A small fraction of the gas remains bound to the galaxies as dense coronae for $t \gtrsim 2 - 3$ Gyr. 

We note here that these snapshots are not at uniform time intervals and have been chosen to illustrate the various stages of ram pressure stripping and wake formation. Figures~\ref{fig:groupgasT_EM} and ~\ref{fig:clustergasT_EM} show the emission measure-weighted temperature maps for the first $\sim 2.54$ Gyr of evolution, when the stripped gas initially forms smooth wide wakes. These wakes narrow with time, and some wakes form shear instabilities at wake-ICM boundaries, seen in the form of characteristic Kelvin-Helmholtz rolls. As galaxies turn in their orbits, trailing wakes appear bent in projection. Wakes at larger radii are longer-lived than those of galaxies in the inner regions of the group and cluster.

Figure~\ref{fig:groupgasT_EM2} shows the late time evolution of the group's gas at $t > 3$ Gyr. Two group galaxies' orbits are outside the inner $1 \, \mbox{Mpc} \, \times \, 1 \, \mbox{Mpc} $ region shown in these maps. The galaxy at the bottom left corner of the first panel in Figure~\ref{fig:groupgasT_EM2} leaves this central region at $t \simeq 3.8$ Gyr and therefore retains its corona in the low-density outer ICM, where it is subject to weak ram pressure. It re-enters the inner $1 \, \mbox{Mpc} \, \times \, 1 \, \mbox{Mpc} $ region at $\sim 6.35 $ Gyr and is the last surviving galactic corona at the end of the simulation ($t = 7.612$ Gyr). Most galactic wakes have dissipated by $t > 3$ Gyr, but a few coronae survive before being almost completely stripped by $t \simeq 5.5$ Gyr. The stripped gas drives shock waves in the ICM, and the resulting inhomogeneities in the ICM are smoothed by $t \sim 6$ Gyr. 

\subsection{Mass loss due to ram pressure stripping}
\label{sec:massloss}
\begin{figure*}
  \begin{center}
    \subfigure[Group]
    {\includegraphics[width=3.5in]{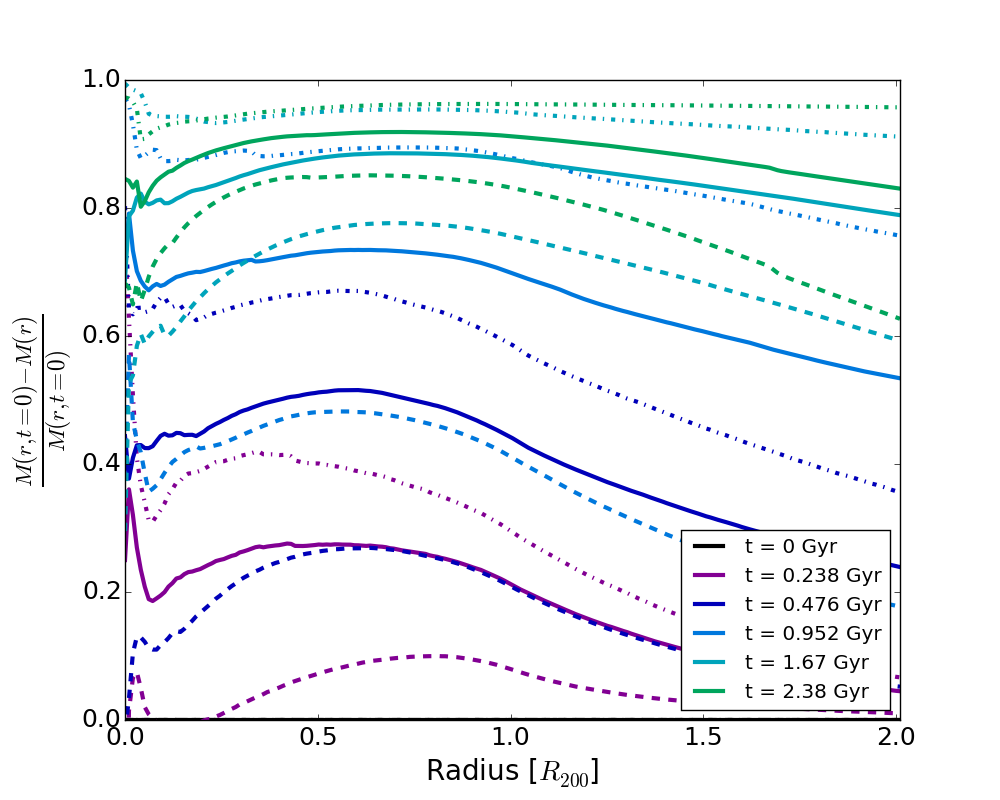} \label{fig:groupgaslossrate}}
    \subfigure[Cluster]
    {\includegraphics[width=3.5in]{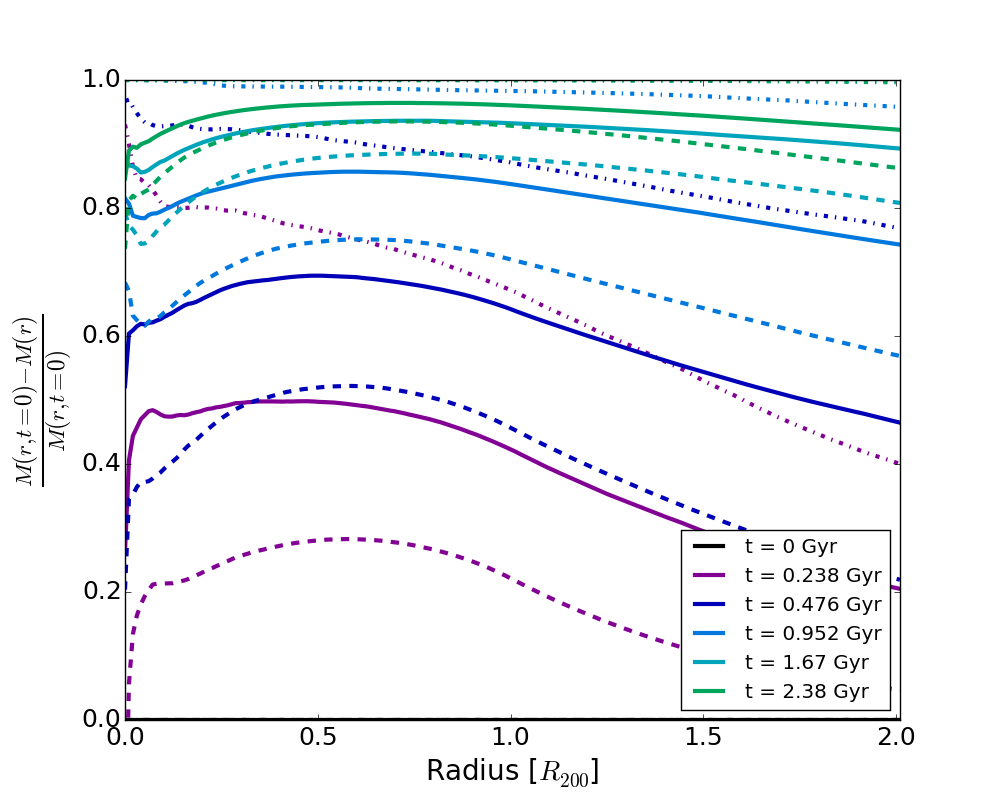} \label{fig:clustergaslossrate}}    
    \caption{Stacked differential mass profiles as a function of time for group and cluster galaxies. The solid lines correspond to all group and cluster galaxies. The dashed lines are for galaxies that have initial masses $M > 10^{11}\ \mbox{M}_{\odot}$, and the dotted lines are for galaxies with initial masses $M < 10^{11}\ \mbox{M}_{\odot}$ \label{fig:gaslossrate}}
  \end{center}  
\end{figure*}

We quantify the rate at which group and cluster galaxies are stripped of their gas using their differential gas mass loss profiles, defined as
\begin{equation}
\Delta M (r) = \frac{M(r, t = 0) - M(r)}{M(r, t = 0)}, 
\end{equation} 
where $M(r)$ is the gas mass enclosed within a galaxy-centric radius $r$ for a galaxy. We calculate the stacked differential gas mass profiles for different samples of galaxies: all galaxies initialized in the group and cluster, galaxies more massive than $10^{11}\ \mbox{M}_{\odot}$, and galaxies less massive than $10^{11}\ \mbox{M}_{\odot}$. To stack galaxies, we calculate the mean radial gas mass profile in linearly spaced radial bins for each galaxy, normalize these mass profiles to the initial gas mass for that galaxy, calculate the average radial mass profile for each of the three galaxy samples, and then calculate the differential mass loss compared to the stacked profiles at $t = 0$. 

When comparing the group and cluster, we expect cluster galaxies to experience stronger ram pressure and consequently lose their gas faster compared to group galaxies. This is because the ram pressure, $P_{\rm ram} = \rho_{\rm ICM} v_{\rm gal}^2$, that galaxies experience depends on the host system's velocity dispersion, which increases with increasing halo mass. In Figure~\ref{fig:gaslossrate}, we plot $\Delta M (r)$ for stacked samples of group and cluster  galaxies at various times up to $t = 2.38$ Gyr. $\Delta M (r)$ is the fraction of gas lost; for instance, $\Delta M (r) = 1$ corresponds to all of the gas within that radius being stripped. We see on comparing the overall stacked differential mass loss profiles (solid lines) that cluster galaxies indeed lose their gas faster than group galaxies: at $t = 0.238$ Gyr, cluster galaxies have on average lost $\sim 40\%$ of the initial gas within $R_{\rm 200, gal}$ while group galaxies on average have lost $\sim 20\%$ of their initial gas. Group galaxies lose $\sim 80\%$ of their initial gas by $1.6 - 1.7$ Gyr, while cluster galaxies lose the same amount of gas within $1$ Gyr. 

Galaxy gas loss rates also depend on the mass of the host galaxy: more massive galaxies exert larger gravitational restoring forces that can better withstand stripping. Comparing the low-mass (dashed lines) sample to the high-mass (dotted lines) sample in Figure~\ref{fig:gaslossrate}, we see that lower-mass galaxies are stripped at a significantly higher rate. For instance, in Figure~\ref{fig:clustergaslossrate}, at $t = 0.238$ Gyr, the lower mass sample of galaxies loses on average $65\%$ of the gas mass within $R_{\rm 200, gal}$, while the more massive sample loses only about $20\%$. We see the same trend in group galaxies: at the same timestep, massive group galaxies lose less than $10\%$ of their gas, while lower-mass group galaxies lose $\sim 30\%$ of their gas, on average.

\subsection{Properties of X-ray tails and wakes}
\label{sec:xraywakes}

\subsubsection{Confinement surface}
\label{sec:confsurface}

\begin{figure*}
  \begin{center}
    \subfigure[Tracer particle positions]
    {\includegraphics[width=3.5in]{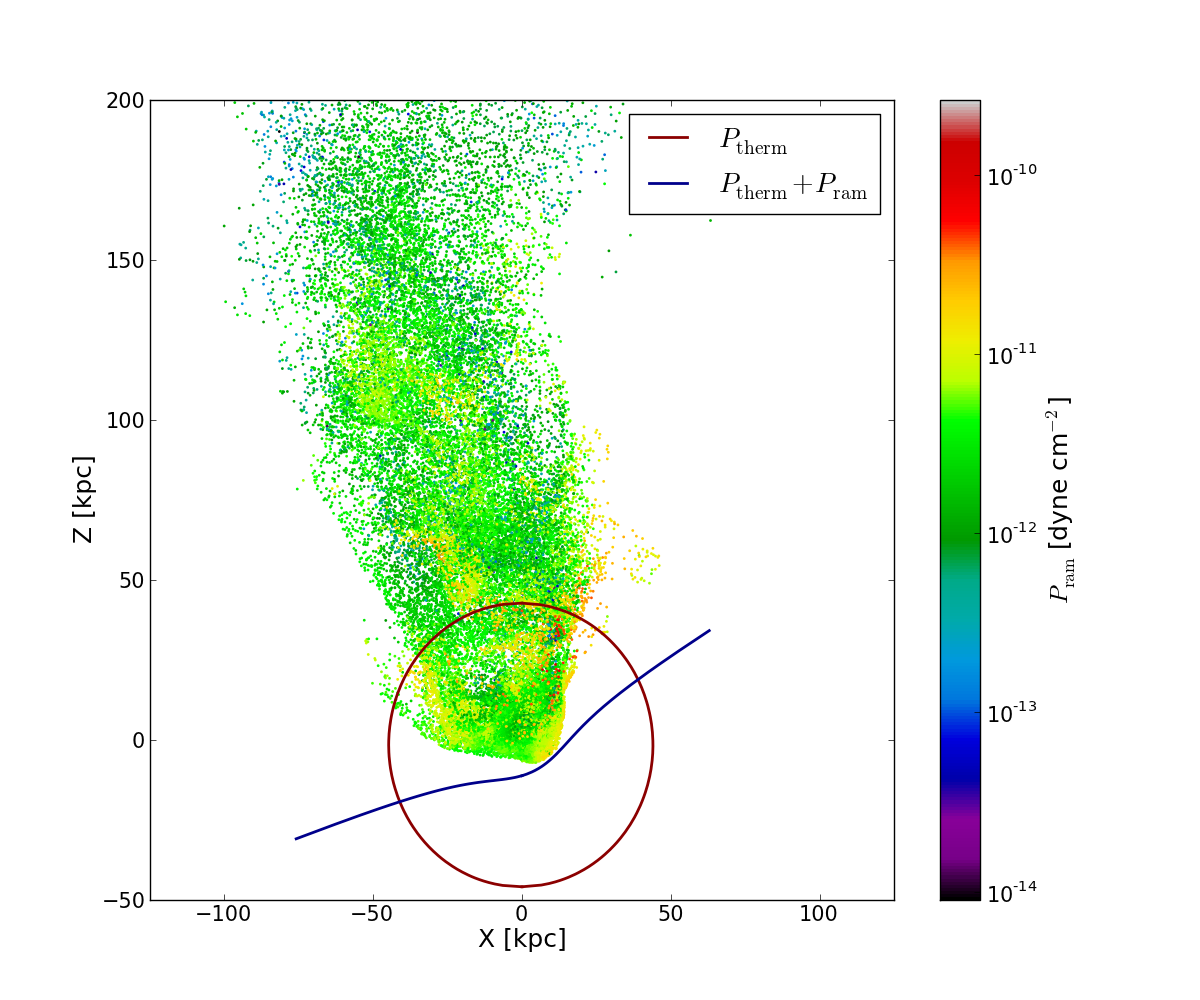} \label{fig:prampthermillus}}   
    \subfigure[Galaxy contours]
    {\includegraphics[width=3in]{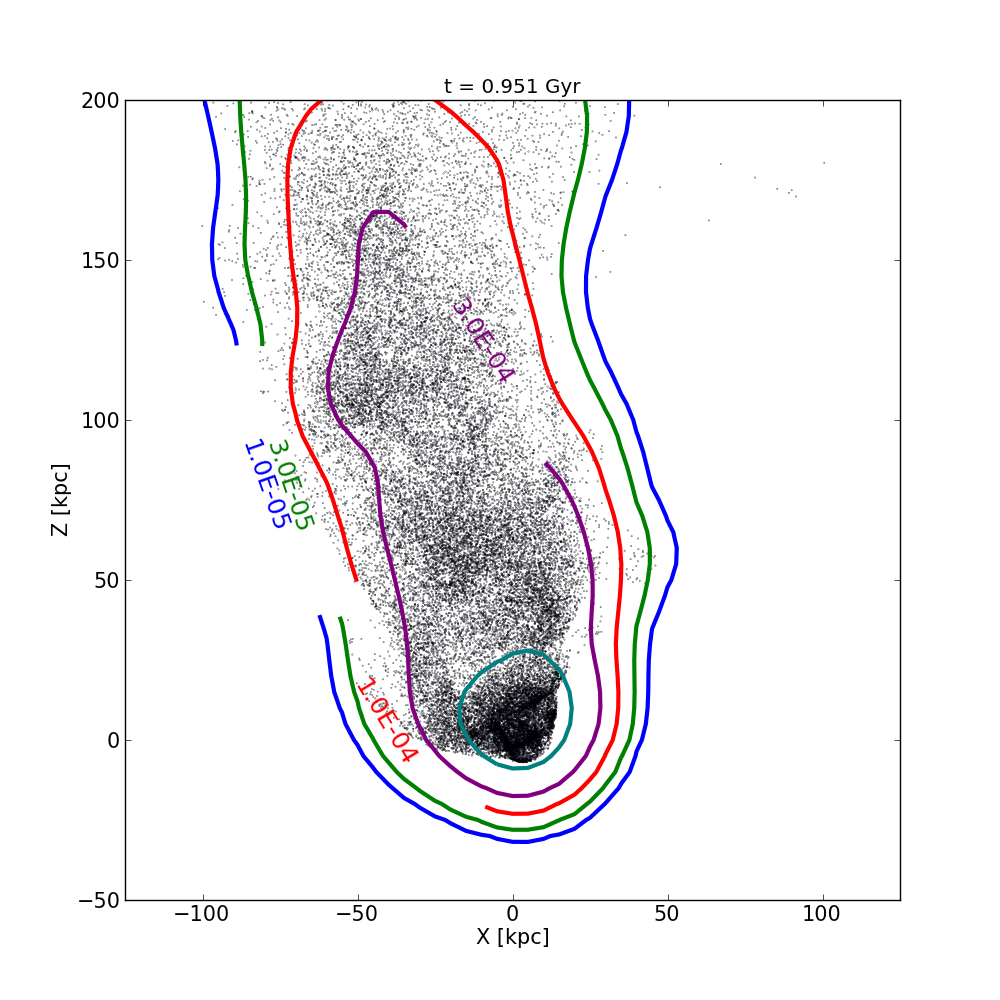} \label{fig:galcontour}}    
     \caption{Left: Scatter plot of tracer particles for gas originally bound to an arbitrary group galaxy at $t = 0.952$ Gyr. The tracer particles are colored by the ram pressure experienced by the gas parcel they trace. The solid red circle shows the surface at which the initial thermal pressure within this galaxy balances the thermal pressure of the ICM. The solid blue line shows the surface where the net thermal plus ram pressure of the ICM balance the galaxy's initial internal thermal pressure. Right: Scatter plot of passive tracer particles for the same group galaxy with contours of constant surface mass density (in units of $\mbox{g} \, \mbox{cm}^{-2}$) overlaid. }
  \end{center}  
\end{figure*}

Here we briefly illustrate the effect of ram pressure in comparison to the ICM's thermal pressure. Figure~\ref{fig:prampthermillus} shows the relationship between the stripped surface of galactic gas for a sample group galaxy at $t = 0.952$ Gyr and the strength of ram pressure and thermal pressure to which the galaxy is subjected. In Figure~\ref{fig:prampthermillus}, the red circle is a projection of the surface that defines the region within which this galaxy's initial thermal pressure balances the group ICM's initial thermal pressure, i.e. where 
$ P_{\rm therm, galaxy}(\mathbf{r_{\rm gal}}) \geq P_{\rm therm, ICM}(\mathbf{r_{\rm group}} + \mathbf{r_{\rm gal}})$. Here, $\mathbf{r_{\rm gal}}$ is the galaxy-centric position, and $\mathbf{r_{\rm group}}$ is the galaxy's position vector in the group's frame. Clearly, this surface is not a good tracer of the stripped leading edge of the galaxy.

The blue curve in Figure~\ref{fig:prampthermillus} defines a pressure-balanced surface that includes the contribution due to ram pressure. To estimate this surface and compare our prediction to the actual simulation, we solve the following pressure balance equation for ${\bf r}_{\rm conf}$:
\begin{align}
  P_{\rm therm, galaxy}(\mathbf{r_{\rm conf}}) & =  P_{\rm therm, ICM}(\mathbf{r_{\rm group}} + \mathbf{r_{\rm conf}}) + \nonumber \\
   &\qquad P_{\rm ram, ICM}(\mathbf{r_{\rm group}} + \mathbf{r_{\rm conf}}) \hat{\mathbf{v}}\cdot\hat{\mathbf{r}}_{\rm conf}.
   \label{eqn:prconf}
\end{align}
The contribution due to ram pressure ($P_{\rm ram, ICM}(\mathbf{r_{\rm group}} + \mathbf{r_{\rm conf}}) \hat{\mathbf{v}}\cdot\hat{\mathbf{r}}_{\rm conf}$) depends on the relative velocity vector between the galaxy and the ICM: $\hat{\mathbf{v}}$ is the unit relative velocity vector ($\mathbf{v_{\rm rel,gal}}/|\mathbf{v_{\rm rel,gal}}|$), so $\hat{\mathbf{v}}\cdot\hat{\mathbf{r}}_{\rm conf}$ is the direction cosine of a given gas parcel in its galaxy's frame of reference. Since the contribution due to ram pressure can be positive or negative, the RHS of equation~\ref{eqn:prconf} is allowed to be negative, while the LHS is always positive. Consequently, at certain galaxy-centric angles, there is no solution to equation~\ref{eqn:prconf}. Therefore, the confinement surface defined by the blue curve in Figure~\ref{fig:prampthermillus} is not a closed surface. This makes intuitive sense as well, as the trailing edge of galaxy does not experience strong ram pressure, and this where the stripped gas is initially deposited in the form of a tail that trails its host galaxy.

We therefore see, analytically, that incorporating the direction-dependent contribution of ram pressure in the pressure balance equation gives a confinement surface solution that is good estimator of the leading surface of galactic gas. This predictor does not work as well at $t \gtrsim 1.5 $ Gyr, since it only accounts for the instantaneous thermal and ram pressure that a galaxy experiences. In real simulated galaxies, the confinement surface of stripped gas is correlated with the highest ram pressure that a galaxy has experienced during its orbit, since gas once stripped cannot be recaptured by a galaxy.

\subsubsection{Stripped tails and ICM correlations}
\label{sec:wakecorr}
The properties of observed galactic tails and wakes depend on a number of factors. The ram pressure, which depends on the density of the surrounding medium and relative velocity, correlates with the confinement surface radius or the size of the leading edge. The size of the trailing edge, or the length of a galactic tail, correlates with the galaxy's orbital properties: faster galaxies should have longer, narrower tails. \citet{Roediger08} measured the width of ram pressure-stripped tails of cold disk gas and showed that galaxies with wider tails have lower velocities than galaxies with wide tails.

\begin{figure*}
  \begin{center}
    \subfigure[Group]
    {\includegraphics[width=3.5in]{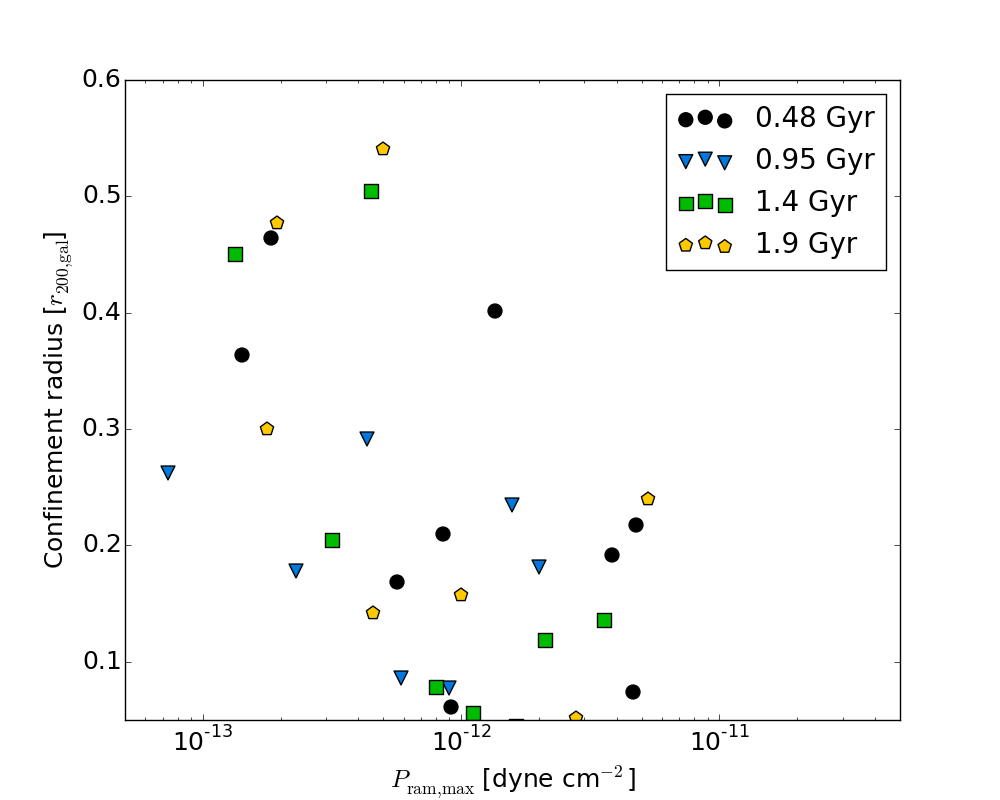} \label{fig:grouprminpram}}
    \subfigure[Cluster]
    {\includegraphics[width=3.5in]{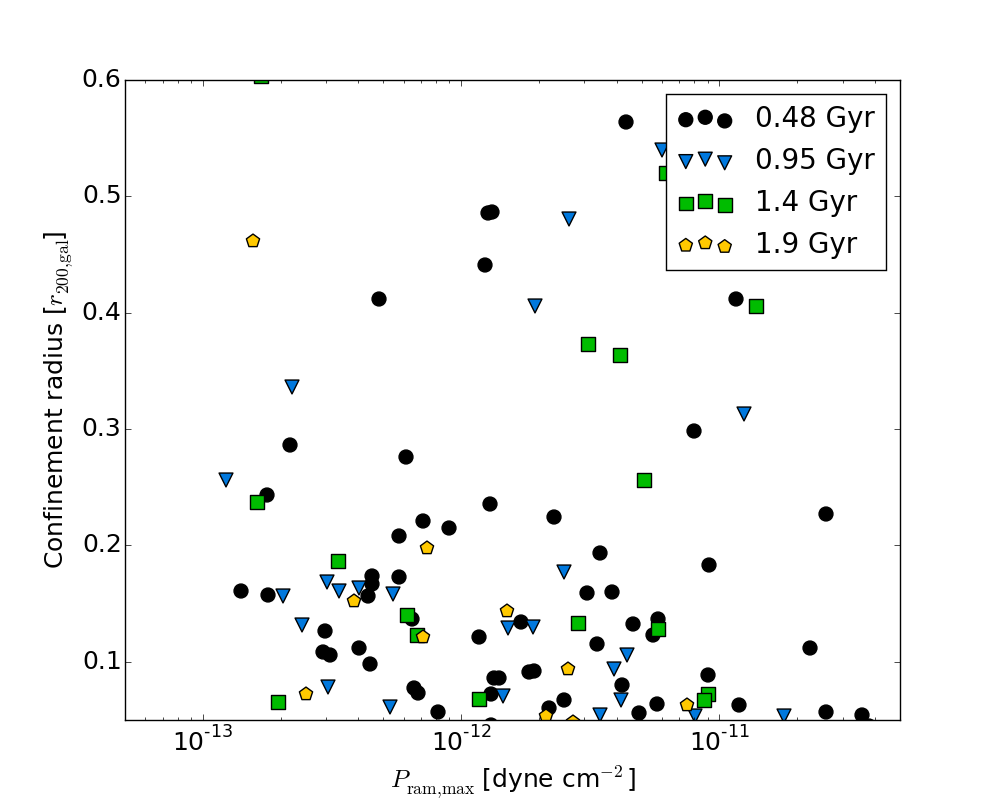} \label{fig:clusterrminpram}}    
    \caption{Projected confinement radius at the leading edge vs. ram pressure, for group (left) and cluster (right) galaxies. The confinement radius is normalized to the galaxies' intial $R_{200}$ radii. The colors and symbols correspond to different simulation times. \label{fig:rminpram} }
  \end{center}  
\end{figure*}

To observationally characterize tail dynamics, we only have access to 1D radial velocity information and 2D density and temperature distributions. We might ask whether we can use the spatial information to infer, for example, a galaxy's velocity components in the plane of the sky. To investigate this question, we have studied the correlation between projected tail properties and ram pressure or transverse velocity. Figure~\ref{fig:rminpram} shows the correlation between the measured confinement radius, $r_{\rm conf}$ (in projection) and the maximum ram pressure, $P_{\rm ram, max}$, that galaxies have been subjected to on their leading edges.  Figure~\ref{fig:grouprminpram} shows these correlations for group galaxies, and Figure~\ref{fig:clusterrminpram} for cluster galaxies.  We only consider those galaxies more massive than $10^{11}\ \mbox{M}_{\odot}$ at the beginning of the simulation in this correlation, since lower-mass galaxies lose most of their gas by $\sim 1$ Gyr and do not have measurable confinement radii. In general, group galaxies that are subject to lower values of $P_{\rm ram, max}$ have larger $r_{\rm conf}$, and galaxies subject to higher ram pressure have smaller values of $r_{\rm conf}$. 

There is no corresponding correlation between $r_{\rm conf}$ and $P_{\rm ram, max}$ for cluster galaxies. We see in Figure~\ref{fig:clusterrminpram} that more than half the galaxies with measurable confinement radii have $r_{\rm conf} \lesssim 0.1 R_{200}$, while fewer than a quarter of group galaxies have $r_{\rm conf} < 0.1 R_{200}$. This is a consequence of stronger ram pressure in the cluster, which results in more efficient gas removal and smaller confinement radii (previously seen in the form of more rapid gas removal in \S~\ref{sec:massloss}).

\begin{figure*}
  \begin{center}
  	 \subfigure[Group galaxies]
    {\includegraphics[width=3.5in]{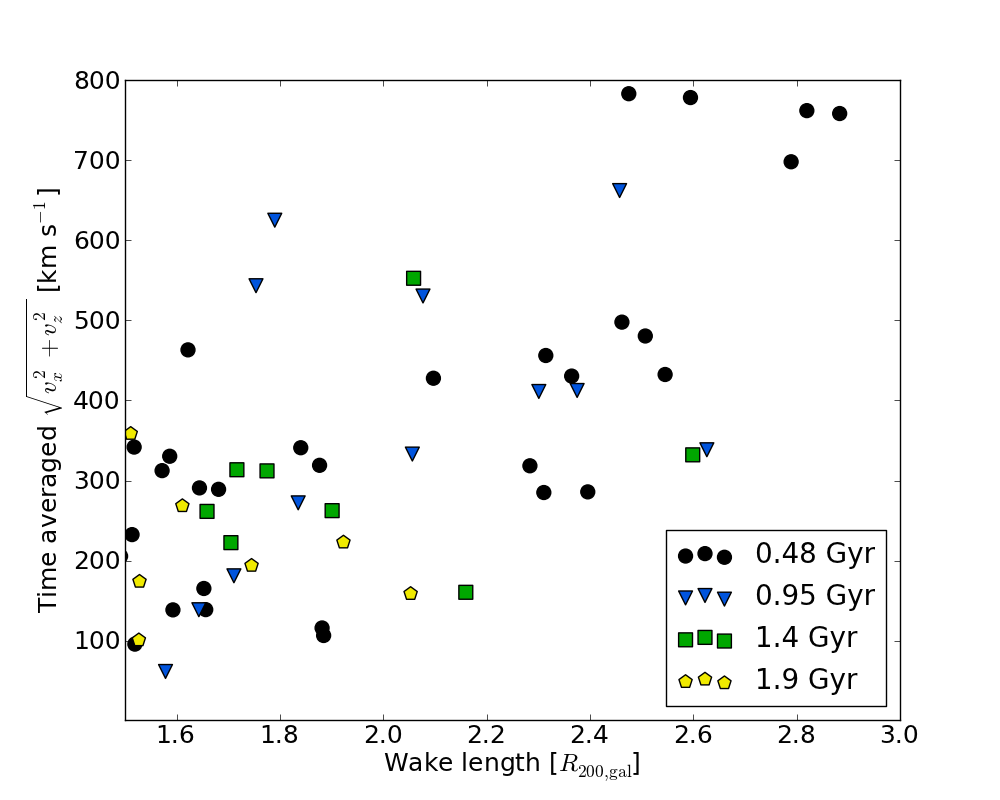} \label{fig:grouplongaxisvtrans}}  
    \subfigure[Cluster galaxies]
    {\includegraphics[width=3.5in]{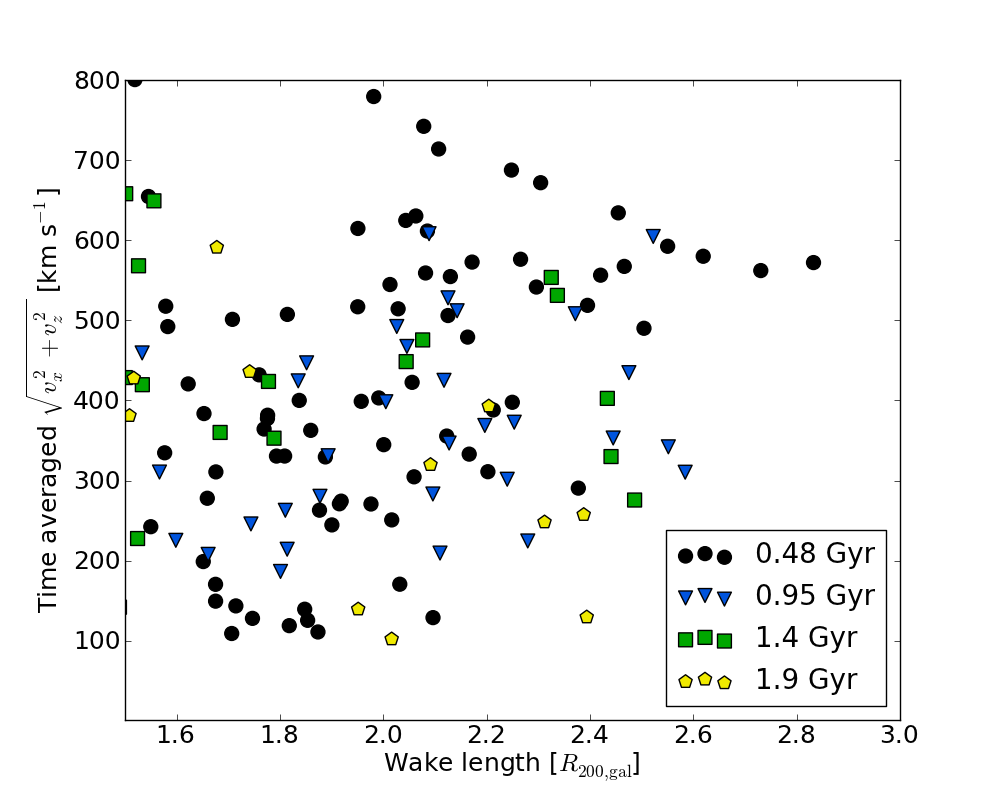} \label{fig:clusterlongaxisvtrans}}      
    \caption{The time averaged transverse velocity (averaged over the previous $\sim 0.2$ Gyr) vs. the length of the $10^{-4} \, \mbox{g} \, \mbox{cm}^{-2}$ surface density contour (normalized to the galaxies' initial $R_{200}$ values) for group and cluster galaxies. We include three different projections of the surface density contour in this plot. \label{fig:longaxisvtrans}}
  \end{center}  
\end{figure*}

We can also compare the length of the trailing edge of the galaxies' tails with their transverse velocities. To calculate the lengths of these tails, we plot contours of constant surface mass density on two-dimensional projections of the galactic gas distribution, as illustrated in Figure~\ref{fig:galcontour}, then define the length of the tails as the long axis of the ellipse that best fits the $10^{-4} \, \mbox{g} \, \mbox{cm}^{-2}$ contour. We find that instantaneous galaxy velocities are not well-correlated with the lengths of galaxy tails, since tails are better tracers of galaxies' velocity histories. We therefore calculate the time-averaged galaxy velocity over the previous five simulation snapshots, corresponding to $\sim 0.25$ Gyr. For typical galaxy velocities of $\sim 500 \, \mbox{km} \, \mbox{s}^{-1}$, this corresponds to a distance traversed of $\sim 125$ kpc,  the approximate length of a typical galaxy tail. 

Figure~\ref{fig:longaxisvtrans} shows the correlation between galaxy tail lengths and transverse velocities for group and cluster galaxies. For the group galaxies (Figure~\ref{fig:grouplongaxisvtrans}), although there is a large scatter, as with the $r_{\rm conf} $ -- $ P_{\rm ram, max}$ relationship, overall galaxies with longer tails have higher transverse velocities. The scatter in transverse velocity vs. tail length is even larger for the cluster galaxies (Figure~\ref{fig:clusterlongaxisvtrans}), although the trend is the same as for the group. For both the group and the cluster, more galaxies have detectable tails at earlier simulation times, i.e., at $0.48$ Gyr and $0.95$ Gyr, compared to $1.4 - 1.9$ Gyr. At late times, particularly at $1.9$ Gyr, there are very few surviving tails with surface densities of at least $10^{-4} \, \mbox{g} \, \mbox{cm}^{-2}$. By this time, most surviving tails have been disrupted or detached from their original host galaxies. After $2$ Gyr, most galaxies do not have detectable tails. However, a few galaxies still have distinct, concentrated coronae.

\subsection{Synthetic X-ray images and stacked profiles}
\label{sec:xraycoronae}
While tails have been detected in X-rays within group and cluster environments for many types of galaxies (\citealt{Forman79}, \citealt{Irwin96}), hot coronae have been only recently detected (\citet{Vikhlinin01}, \citealt{Yamasaki02}). In this section, we generate synthetic \textit{Chandra} X-ray images of the group, cluster, and their galaxies to determine how these coronae should appear. Because we expected the galaxy emission to be faint, we also calculate stacked radial surface brightness profiles and evaluate the detectability of coronae. 

The synthetic X-ray observations are generated using the \texttt{photon\char`_simulator} module of the \texttt{yt} analysis package. The algorithm for generating the synthetic X-ray observations is described in detail in \citet{ZuHone14}. Briefly, the \texttt{photon\char`_simulator} module generates a photon sample for each zone at a given temperature, density, and metallicity, and additionally, an assumed spectral model, cosmological redshift, angular diameter distance, exposure time, and detector area. These photon samples are then convolved with the instrument response to generate a realistic observation. We assumed a fixed source redshift $z = 0.05$, corresponding to an angular diameter distance of $200$ Mpc, a constant metallicity $Z = 0.3 \, \mbox{Z}_{\odot}$, and a spectral model based on the APEC model for a thermal plasma from the AtomDB database\footnote{http://www.atomdb.org/}. While using $z = 0.05$ as the source redshift in our mock-image generation is not consistent with the lookback times to the observed simulation snapshots, placing the images at the correct redshifts can be trivially accomplished using the angular diameter and luminosity distance-redshift relations for our chosen cosmology.

\begin{figure*}
  \begin{center}
  \subfigure
    {\includegraphics[width=2.35in]{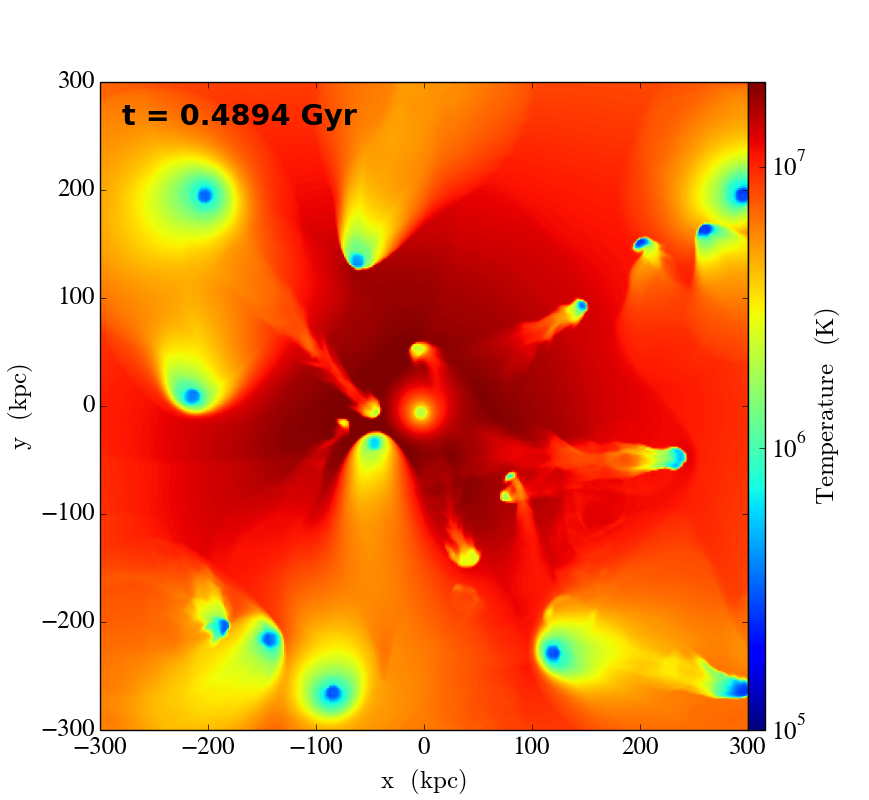} \label{fig:TEMsim30}}
    \subfigure
    {\includegraphics[width=2.2in]{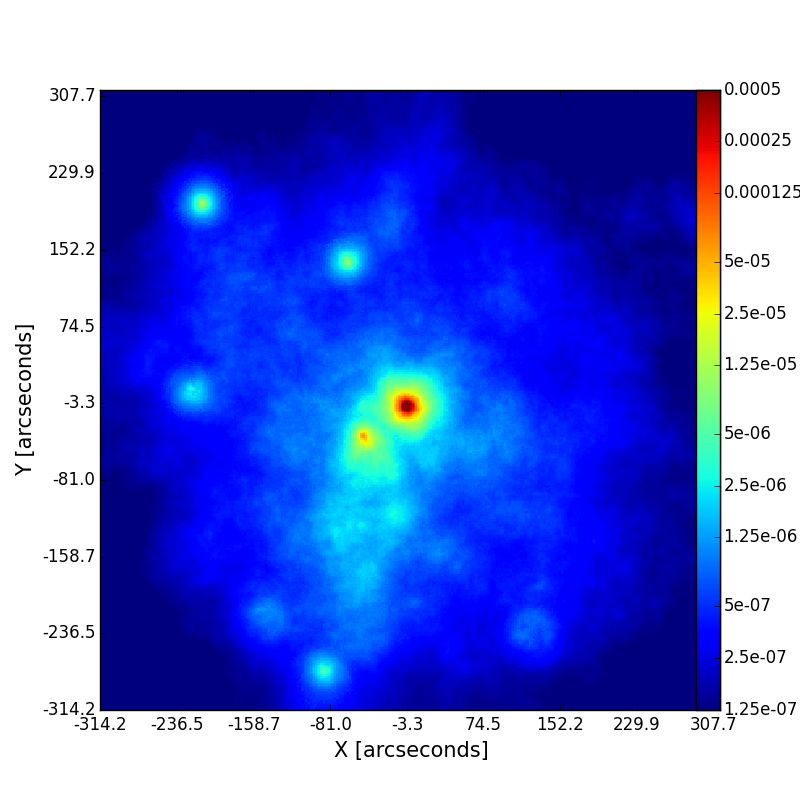} \label{fig:xray40k30}}
  	 \subfigure
    {\includegraphics[width=2.2in]{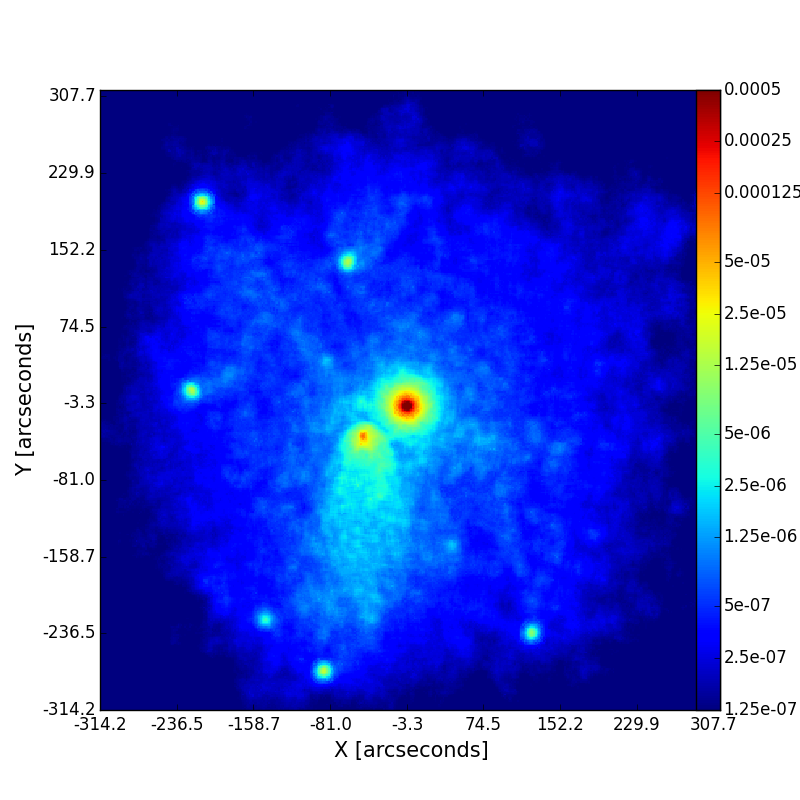} \label{fig:xray400k30}}
  
  \subfigure
    {\includegraphics[width=2.35in]{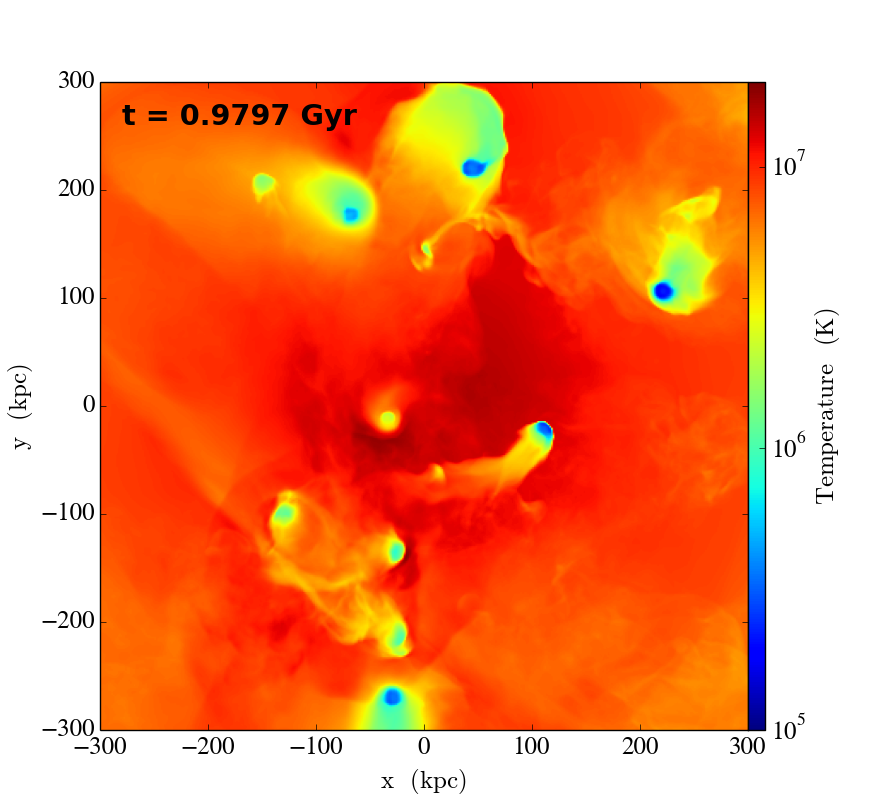} \label{fig:TEMsim60}}
    \subfigure
     {\includegraphics[width=2.2in]{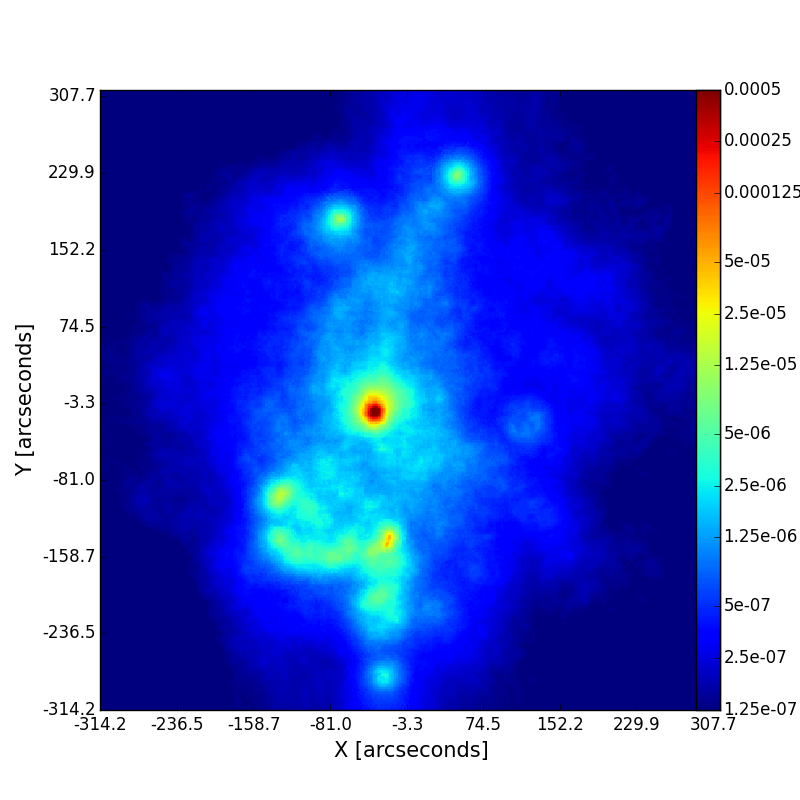} \label{fig:xray40k60}}
  	 \subfigure
    {\includegraphics[width=2.2in]{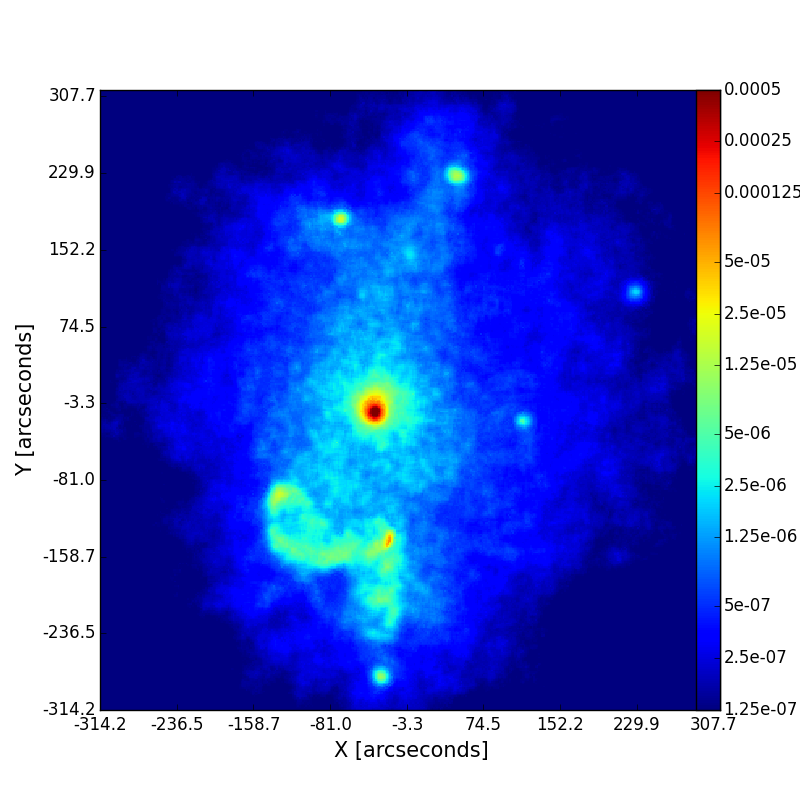} \label{fig:xray400k60}}
  
  \subfigure
    {\includegraphics[width=2.35in]{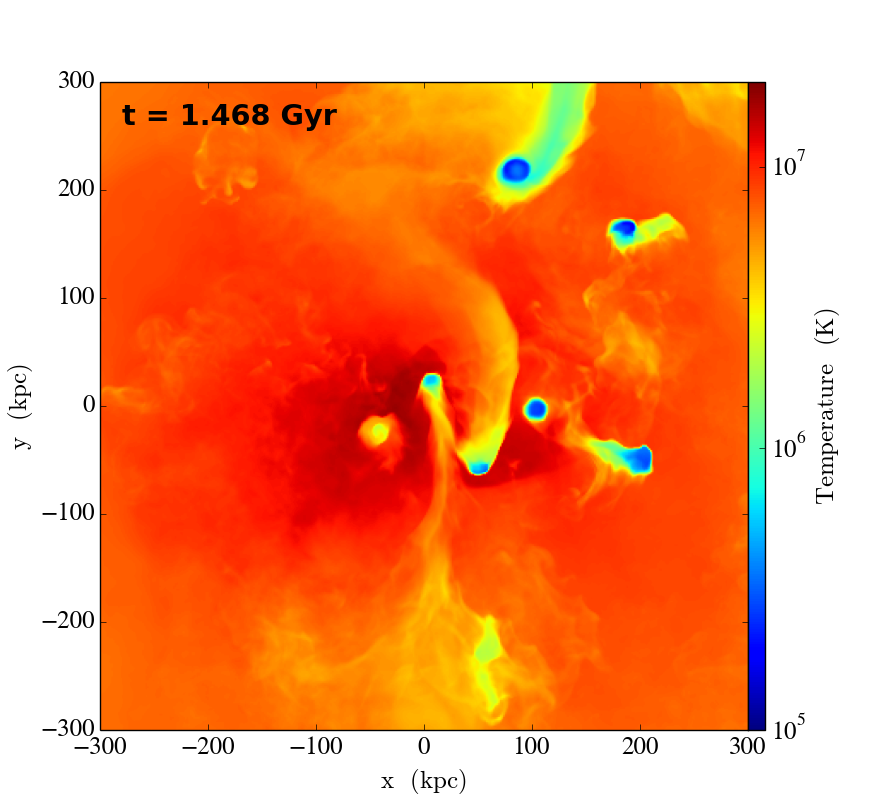} \label{fig:TEMsim90}}
    \subfigure
    {\includegraphics[width=2.2in]{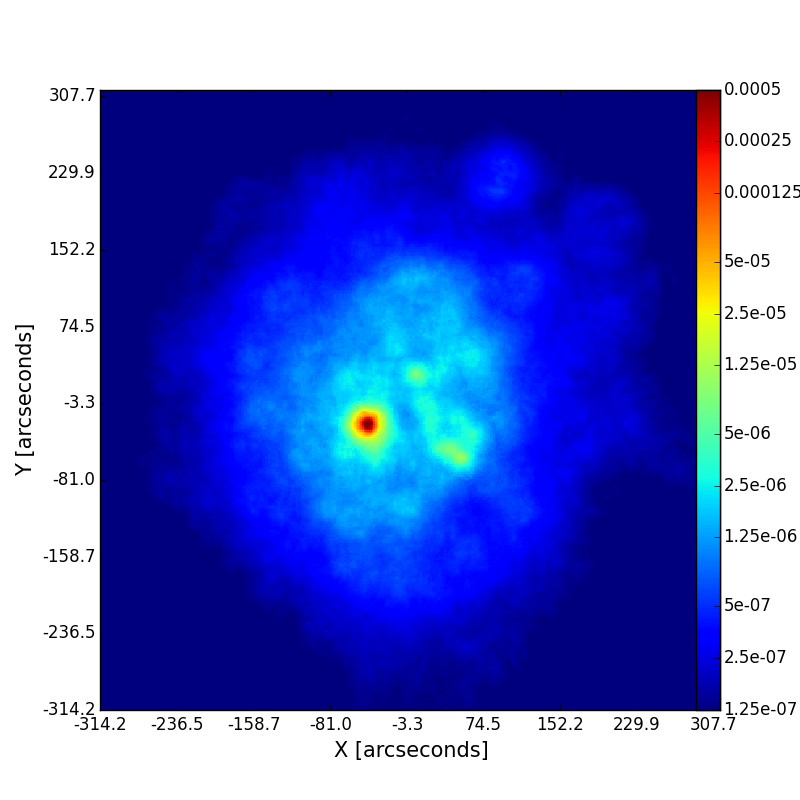} \label{fig:xray40k90}}
  	 \subfigure
    {\includegraphics[width=2.2in]{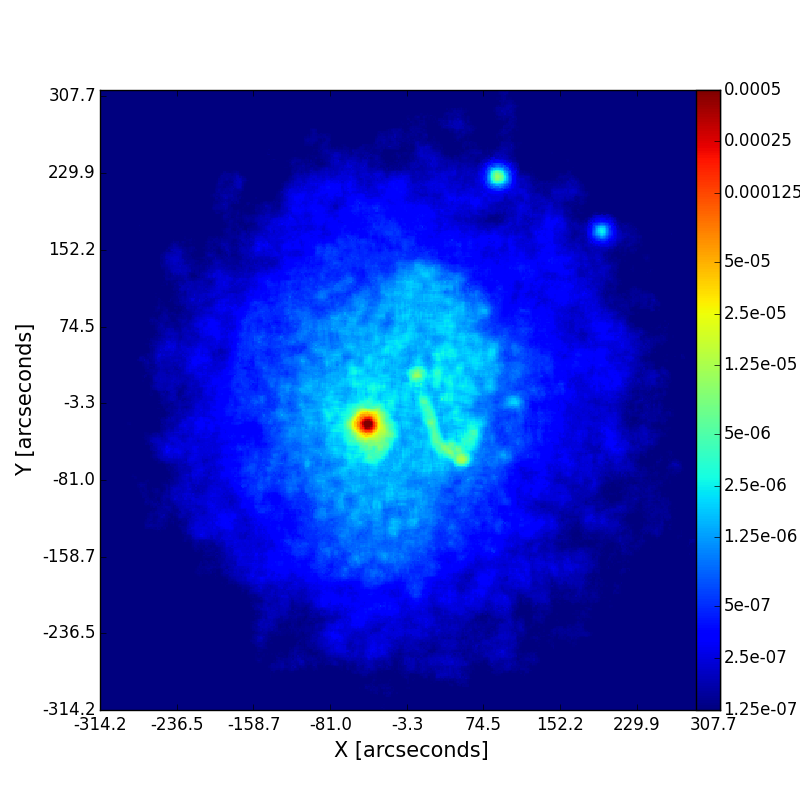} \label{fig:xray400k90}}
 
  \subfigure
    {\includegraphics[width=2.35in]{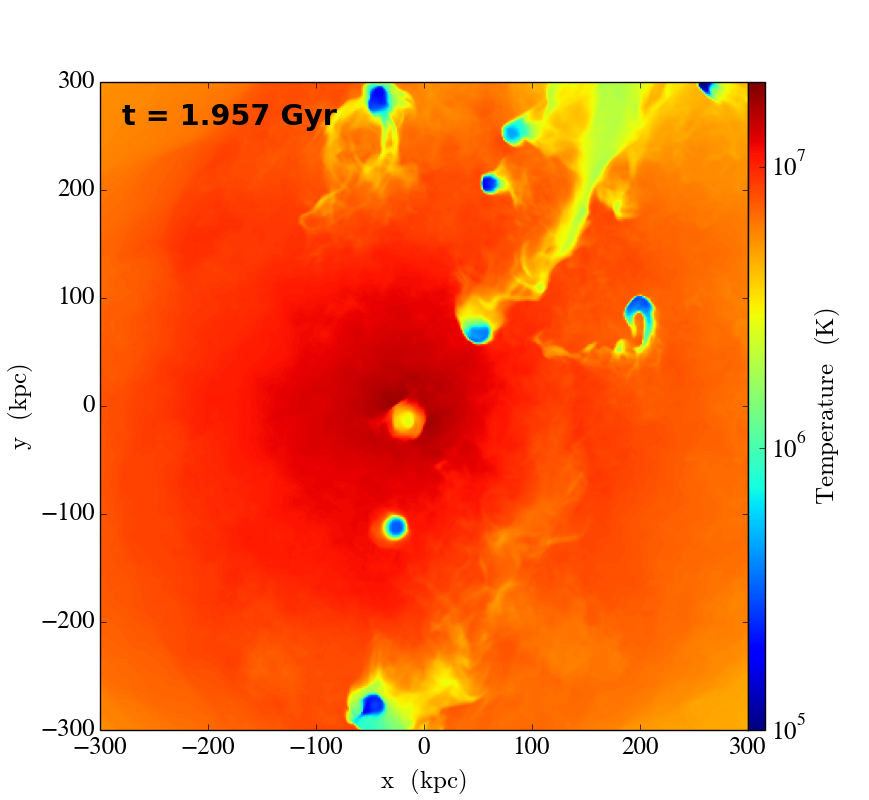} \label{fig:TEMsim120}}
    \subfigure
    {\includegraphics[width=2.2in]{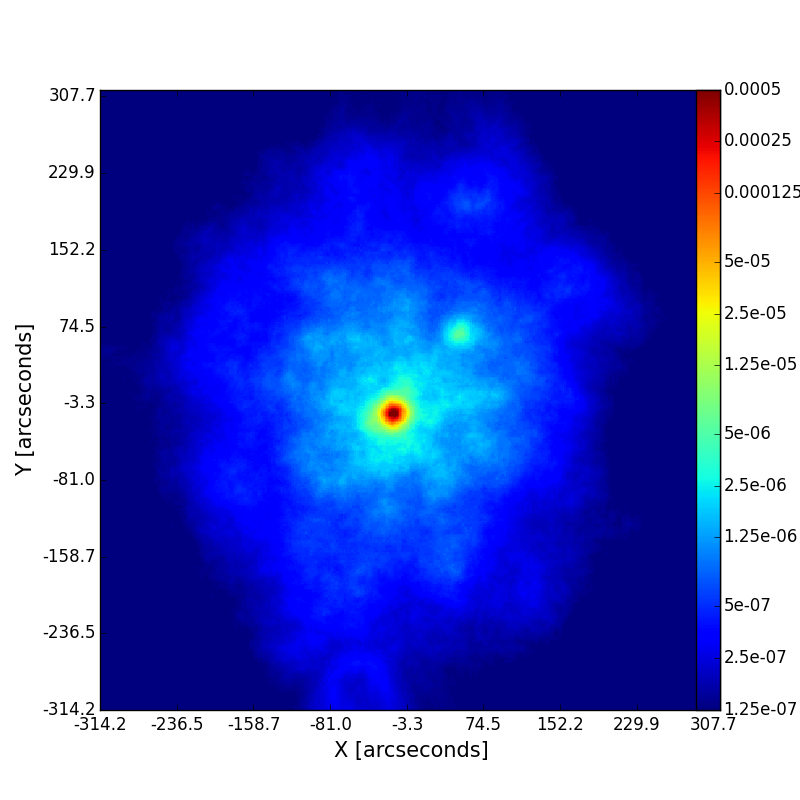} \label{fig:xray40k120}}
  	 \subfigure
    {\includegraphics[width=2.2in]{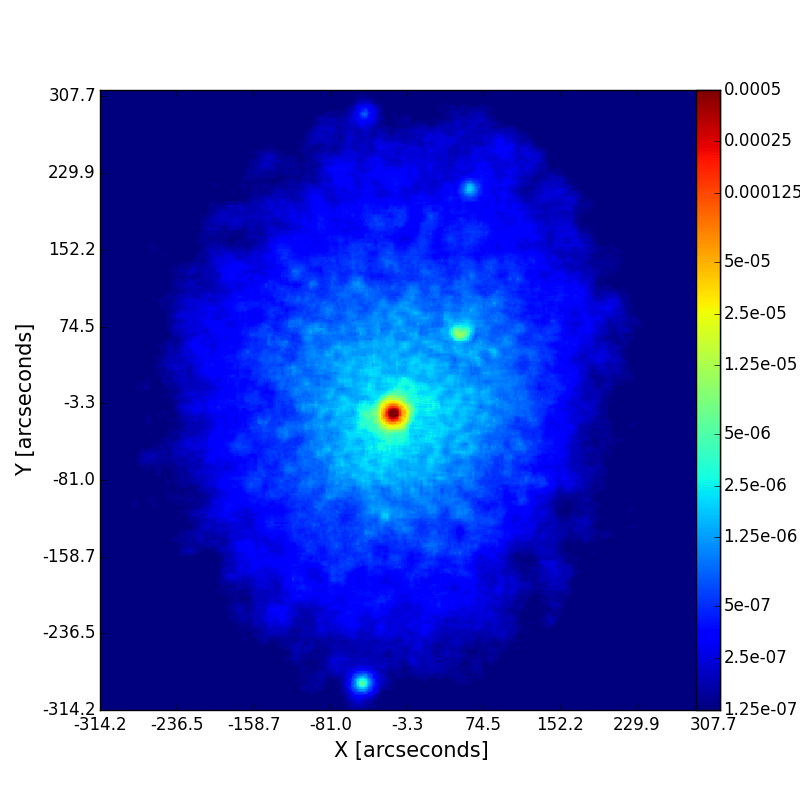} \label{fig:xray400k120}}
  \end{center}  
    \caption{Left: Projections of the emission measure weighted temperature of the group's central $600 \, \mbox{kpc}^2 $ region at $t = 0.47, 0.98, 1.47, 1.96$ Gyr. Center and right: Mock 40 ks and 400 ks images of the central region, after accumulative smoothing. The colors correspond to the photon flux in units of counts second$^{-1}$ arcsecond$^{-2}$.  \label{fig:xray_sample} }
   
\end{figure*}

\begin{figure*}
  \begin{center}
  \subfigure
    {\includegraphics[width=2.35in]{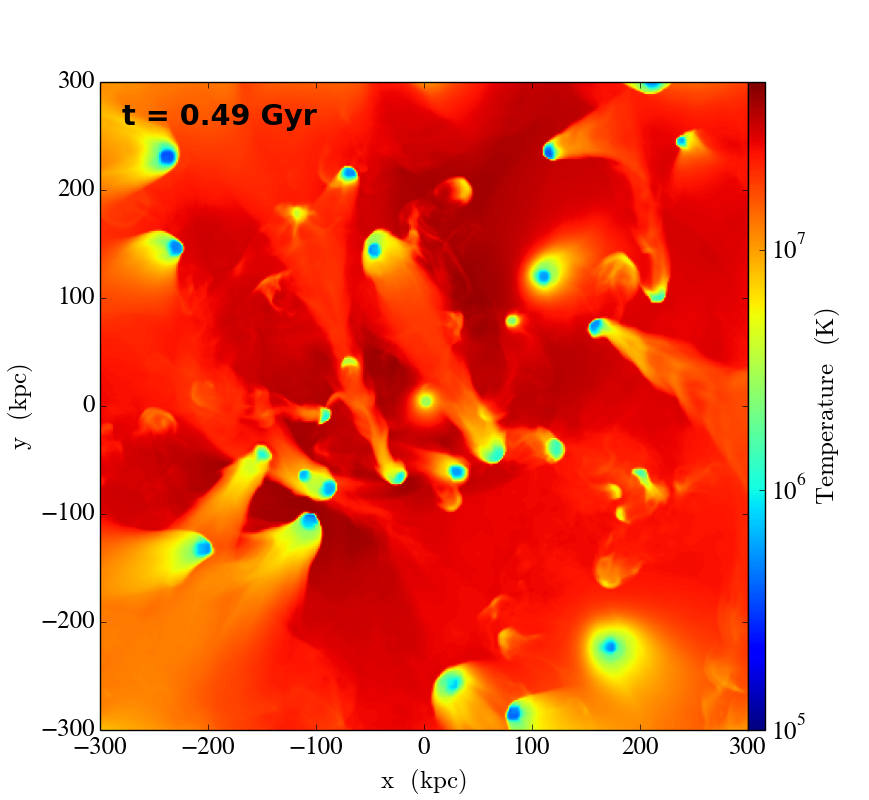} \label{fig:cTEMsim30}}
    \subfigure
    {\includegraphics[width=2.2in]{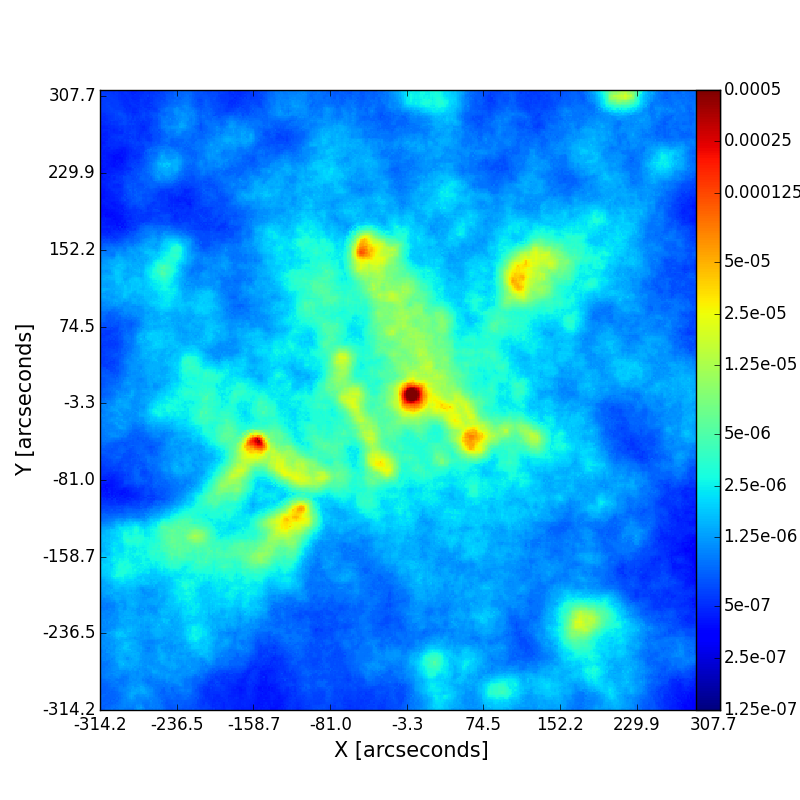} \label{fig:cxray40k30}}
  	 \subfigure
    {\includegraphics[width=2.2in]{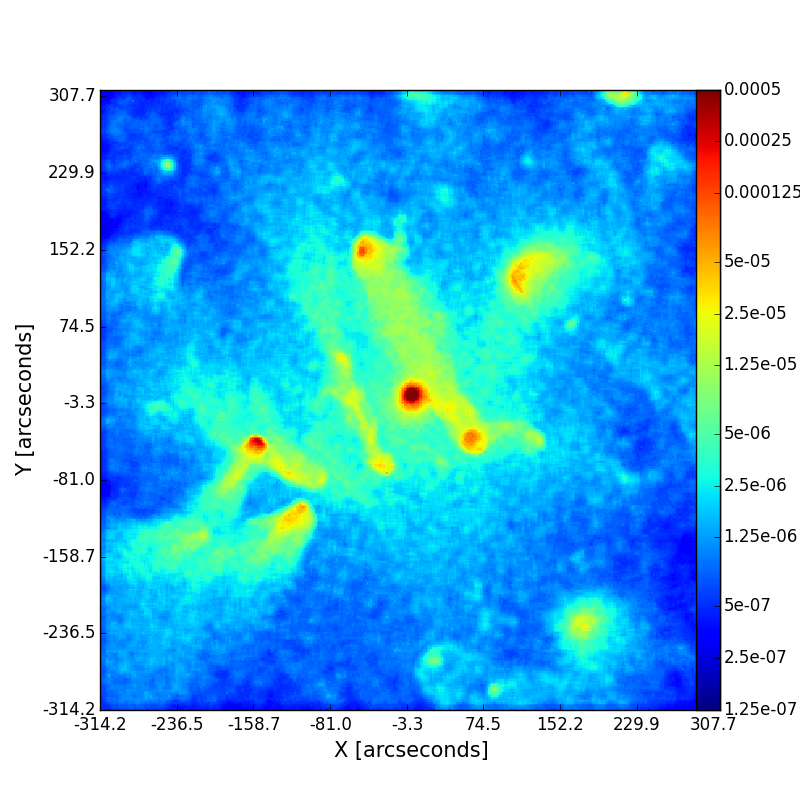} \label{fig:cxray400k30}}
 
 \subfigure
    {\includegraphics[width=2.35in]{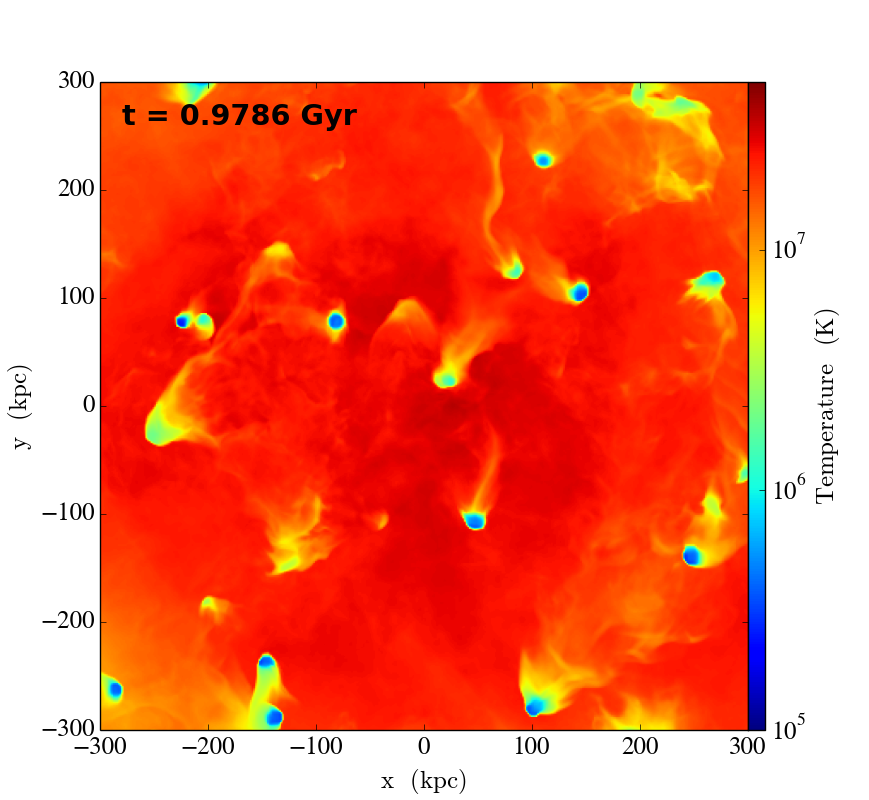} \label{fig:cTEMsim60}}
    \subfigure
    {\includegraphics[width=2.2in]{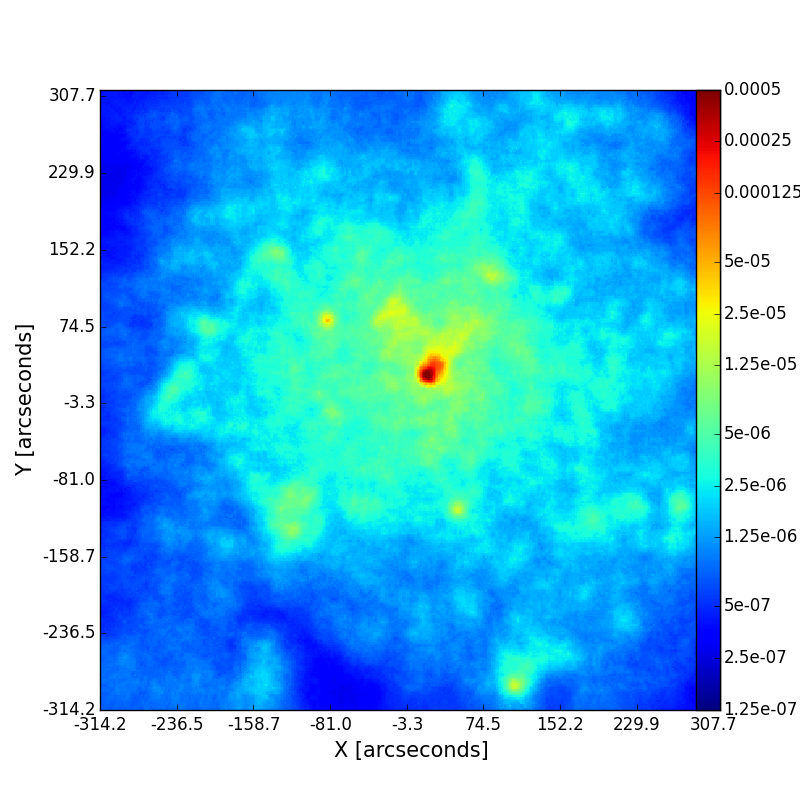} \label{fig:cxray40k60}}
  	 \subfigure
    {\includegraphics[width=2.2in]{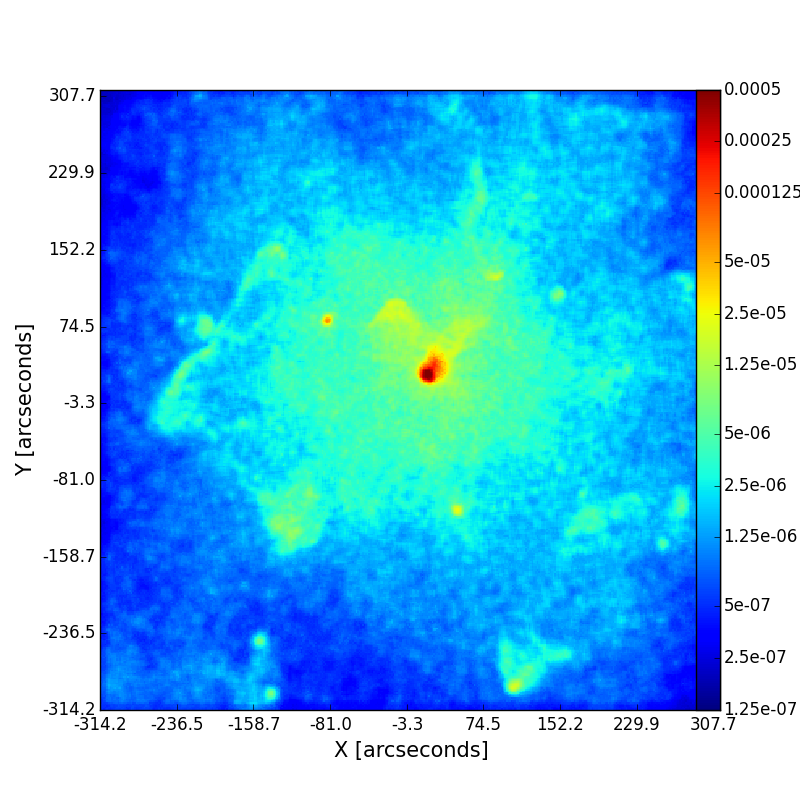} \label{fig:cxray400k60}}
  
  \subfigure
    {\includegraphics[width=2.35in]{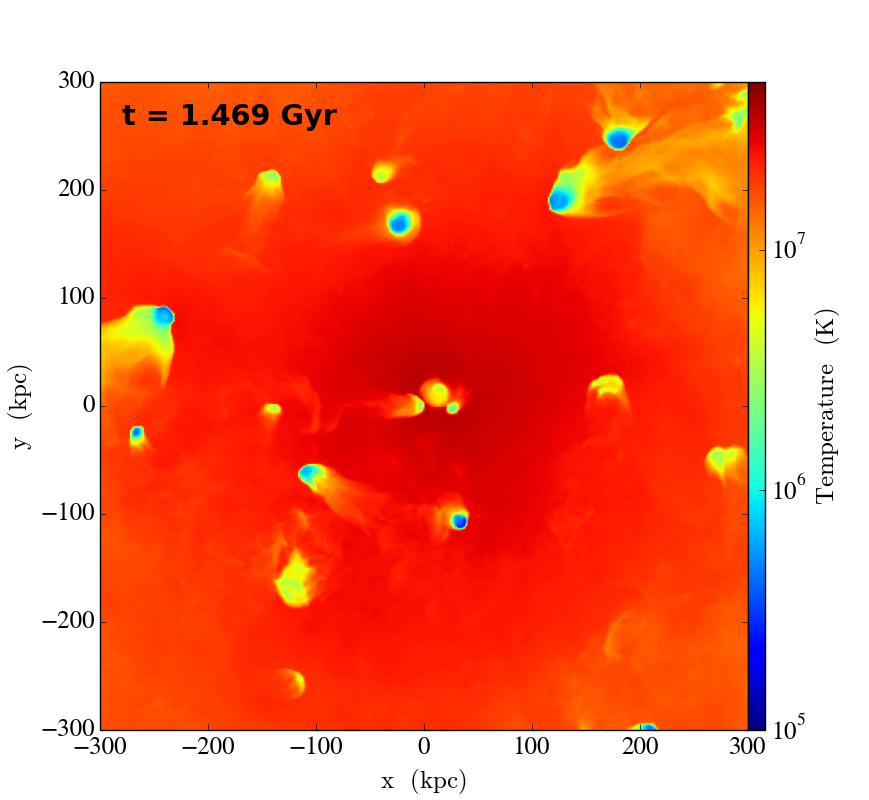} \label{fig:cTEMsim90}}
    \subfigure
    {\includegraphics[width=2.2in]{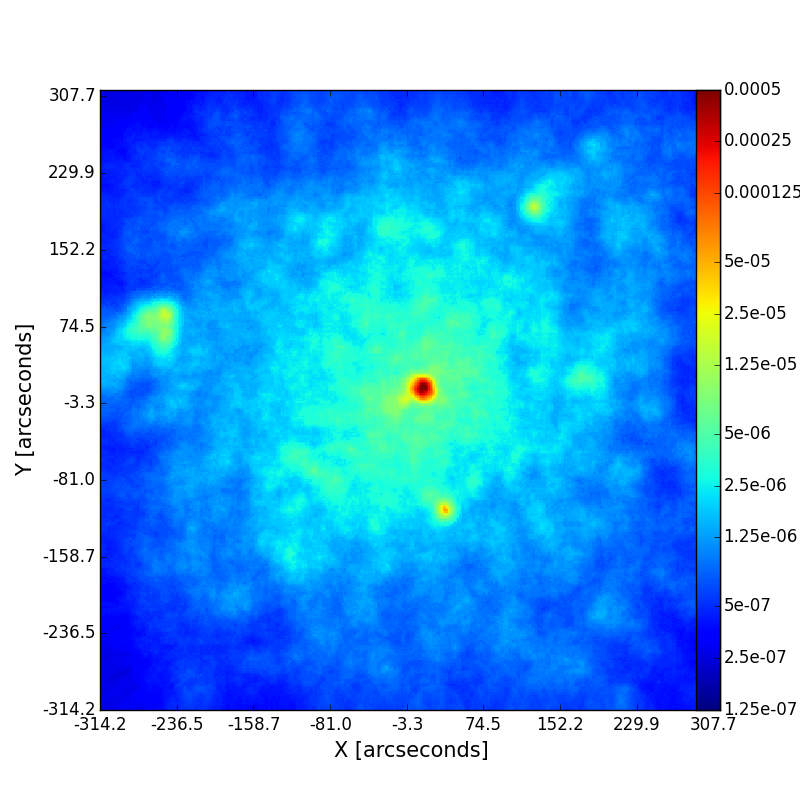} \label{fig:cxray40k90}}
  	 \subfigure
    {\includegraphics[width=2.2in]{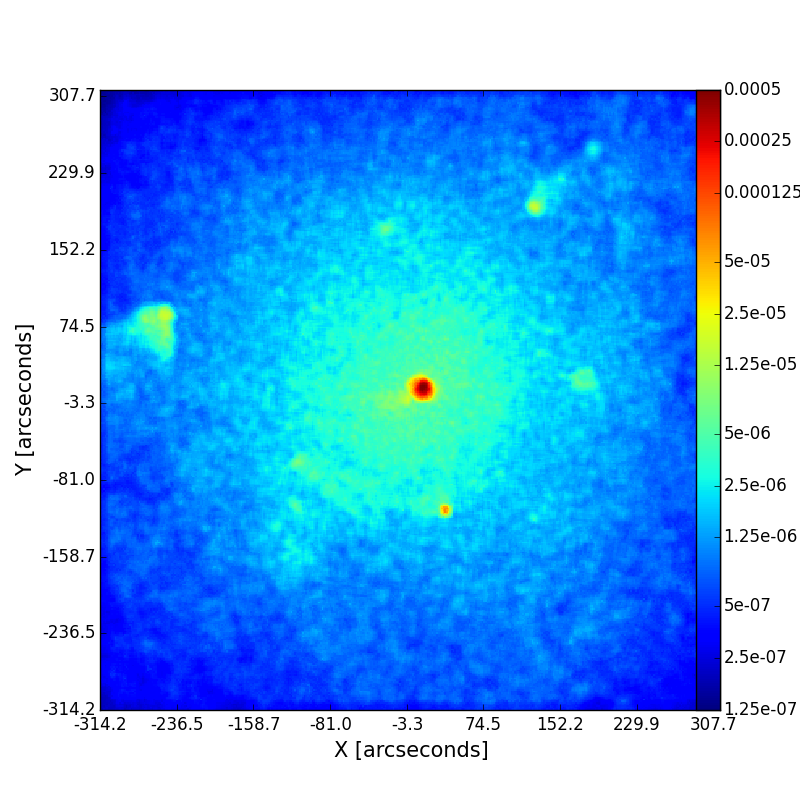} \label{fig:cxray400k90}}
  
  \subfigure
    {\includegraphics[width=2.35in]{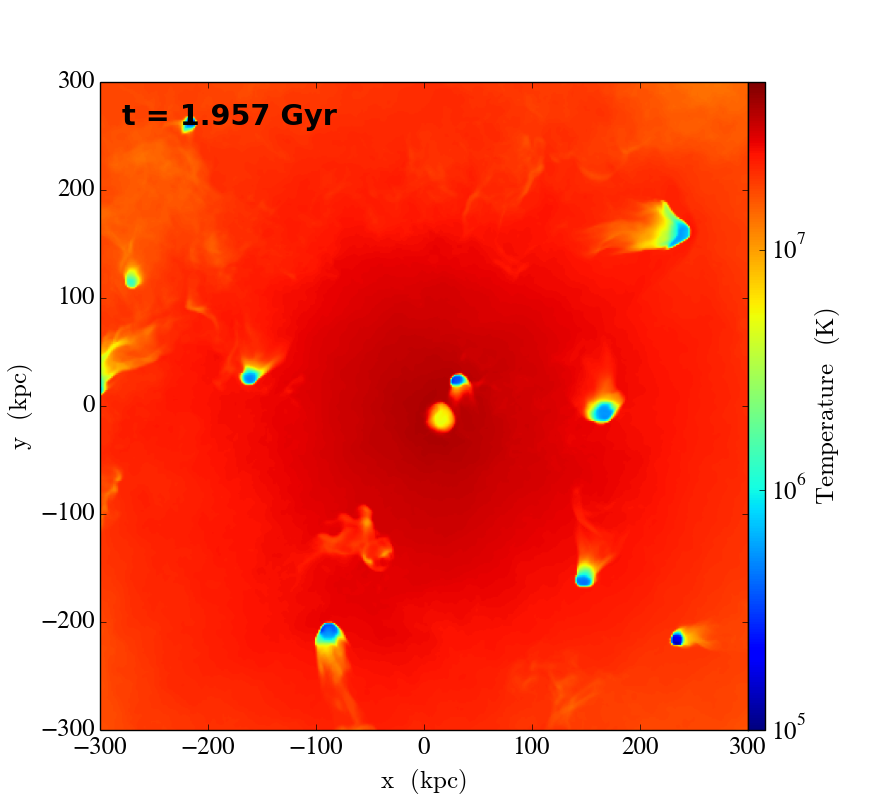} \label{fig:cTEMsim120}}
    \subfigure
    {\includegraphics[width=2.2in]{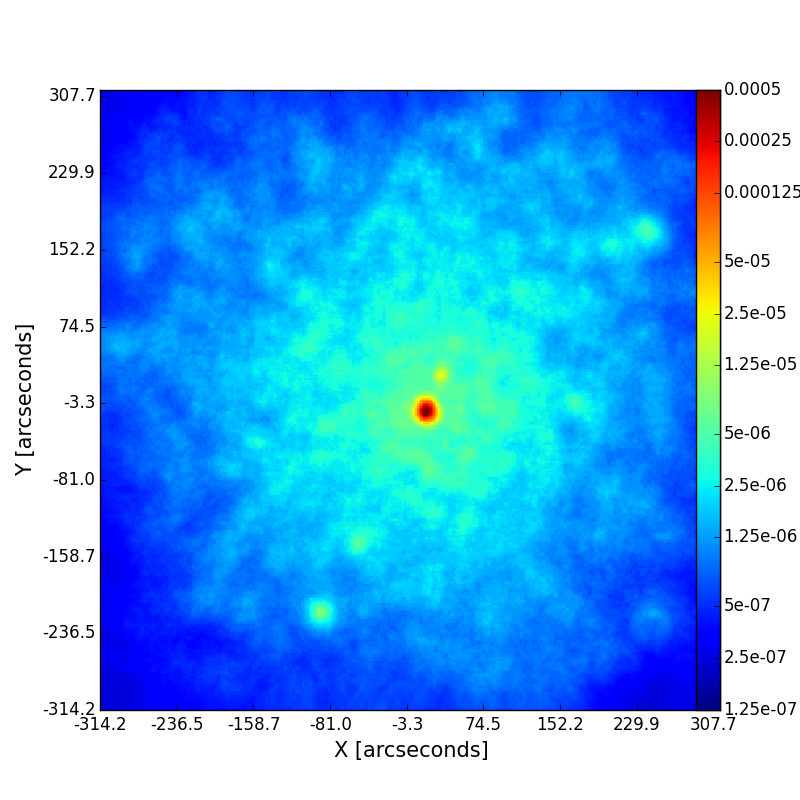} \label{fig:cxray40k120}}
  	 \subfigure
    {\includegraphics[width=2.2in]{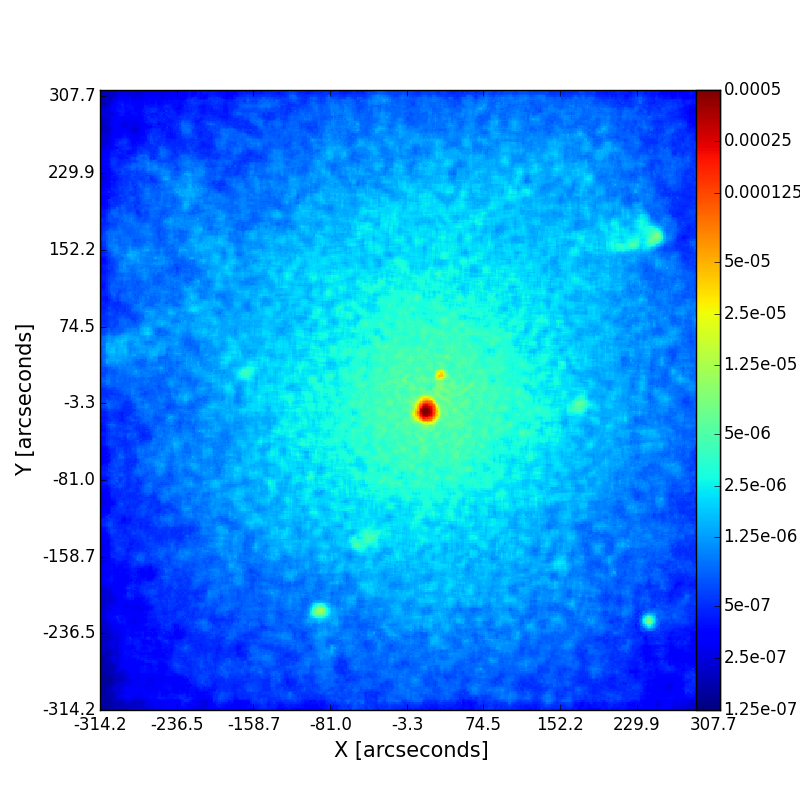} \label{fig:cxray400k120}}
  \end{center}  
    \caption{Left: Projections of the emission measure weighted temperature of the cluster's central $600 \, \mbox{kpc}^2 $ region at $t = 0.47, 0.98, 1.47, 1.96$ Gyr. Center and right: Mock 40 ks and 400 ks images of the central region, after accumulative smoothing. The colors correspond to the photon flux in units of counts second$^{-1}$ arcsecond$^{-2}$. \label{fig:xray_sample_c}}
   
\end{figure*}

Figures~\ref{fig:xray_sample} and ~\ref{fig:xray_sample_c} show three-dimensional snapshots projected in two dimensions from the group and cluster simulations along with the corresponding mock observations. The images in the left column are the emission measure-weighted temperature maps of the central $600 \, \mbox{kpc}^2 $ of the group and its galaxies, projected through the whole group, at $t = 0.49$, $0.98$, $1.47$, and $1.96$ Gyr. The images in the central and right columns are mock 40 ks and 400 ks observations respectively of the region in the left column. These mock \textit{Chandra} observations have been reblocked by a factor or 4, corresponding to the maximum resolution of the simulation. These images have been smoothed using the \texttt{accumulative smoothing} method in \citet{Sanders06}, part of the \texttt{contour binning} package\footnote{\url{http://www-xray.ast.cam.ac.uk/papers/contbin/}}. In this algorithm, the image is adaptively smoothed with a top-hat kernel, and the size of the kernel is varied so that the minimum signal to noise ratio in the kernel is 5. The colors in the images correspond to photon counts per pixel. For our simulation resolution and assumed redshift, each pixel width is 1.63 arcseconds. 

We see in Figures~\ref{fig:xray_sample} and \ref{fig:xray_sample_c} that most of the observed X-ray emission is associated with the ICM, particularly the central core. The surviving X-ray coronae and stripped tails and wakes that are distinctly visible in the temperature maps cannot be as easily distinguished in the 40 ks images. As expected, tails are more prominent in the 400 ks observations; for instance, the 40 ks group image at $t = 0.49$ Gyr (top row, \autoref{fig:xray_sample}) has only one distinct stripped tail, but at least 5--6 galaxies' associated tails are visible in the 400 ks observation at the same time. Additionally, multiple tails may appear as just one tail, as seen at $t = 0.98$ Gyr (second row, Figure~\ref{fig:xray_sample}) when the tails associated with the three galaxies at $[X, Y] = [-100, -100]$ kpc appear to be blended. Qualitatively, the mock observations of the cluster and its galaxies show characteristics similar to the group's. We see in the top row of Figure~\ref{fig:xray_sample_c} that more tails are detected in the 400 ks image than in the 40 ks image. The 40 ks observations in the second and third rows of Figure~\ref{fig:xray_sample_c} do not show any distinct tails, while these tails are clearly visible in the 400 ks images. The late-time images for both the group and cluster ($t = 1.96$ Gyr, bottom rows in Figures~\ref{fig:xray_sample} and \ref{fig:xray_sample_c}) do not show any tail features. However, a few distinct galactic coronae are visible in the 400 ks observations.

While stripped tails dissipate within $\sim 1.5 - 2$ Gyr and are too diffuse to be detected at $t \gtrsim 1.5$ Gyr, the denser central gas associated with galactic coronae is visible in the mock images, particularly in the 400 ks observations. However, as galaxies continue to be stripped and their coronae diminish, their individual signal to noise declines from $1.5 - 2 \sigma$ at  $t = 0.49$ Gyr to being undetectable at $t = 1.97$ Gyr in the 40 ks observations. While these coronae can be detected in the 400 ks observations at $2 - 3 \sigma$ significance at $t = 1.97$ Gyr, such observations are impractical for a large sample of galaxy clusters and groups. More typical are X-ray observations of $10 - 100$ ks. Therefore, to quantify the effectiveness of strangulation in these environments using existing and future short-duration X-ray observations, we consider stacking X-ray images centered on known optical galaxy centers. The X-ray signal from these stacked galaxy images will be at a higher significance level. Below, we describe the properties of stacked mock observations and relate them to the underlying physical processes. 

To stack the images, we first calculate the locations of the density peaks of the particles initially bound to each galaxy using a cloud-in-cell (CIC) technique, as proxies for the observed surface density peaks of optical galaxies; these are the galaxy centers. Using the \texttt{photon\char`_simulator} module, we generate a photon sample for the $400 \, \mbox{kpc}^2$ region centered on each galaxy, integrating through the whole group or cluster, then stack these mock observations for all the galaxies at different times separated by a 0.48 Gyr interval. Each galaxy's exposure time is 40 ks. At each mock observation time, we calculate the radial profile of the stacked photons. We also bin the photons in our sample in three different energy bins: $E_{\rm soft} = 0.1 - 1.2 \, \mbox{keV}$, $E_{\rm medium} = 1.2 - 2.0 \, \mbox{keV}$, and $E_{\rm hard} = 2.0 - 10.0 \, \mbox{keV}$. We only stack those galaxies that are at least $200$ kpc in projection from the group and cluster centers to minimize contamination from the group and cluster's cores. 

\begin{figure*}
  \begin{center}
  	 \subfigure[Stacked surface brightness]
    {\includegraphics[width=3.5in]{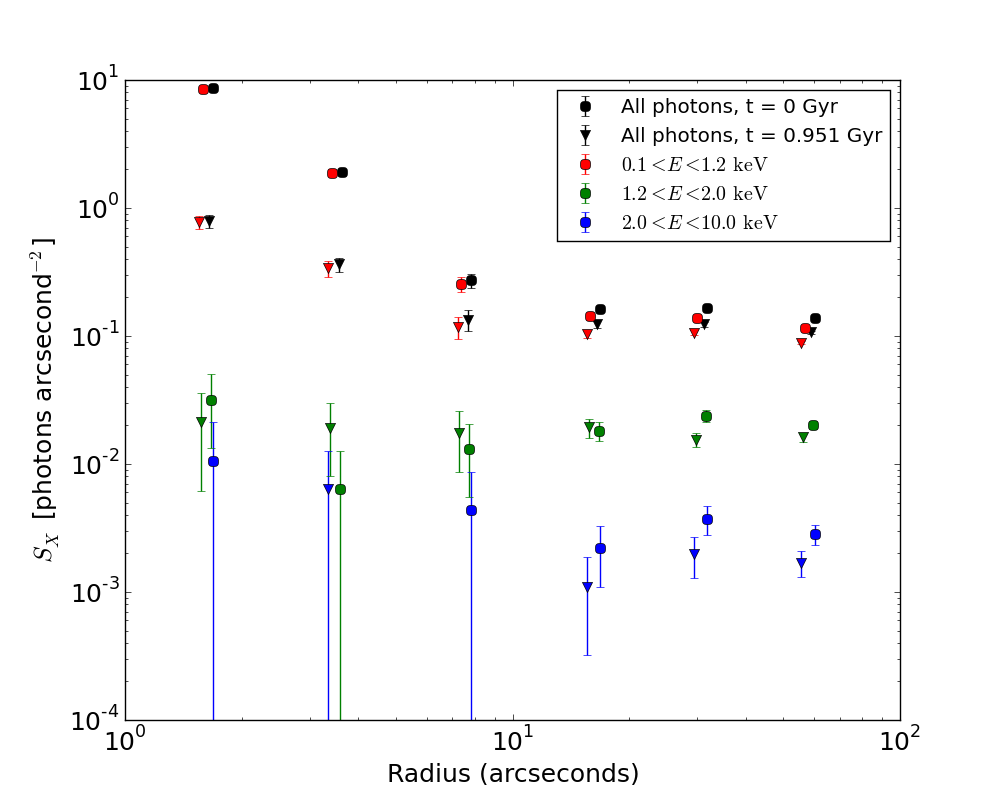} \label{fig:photonrc_groupearlytime}}  
    \subfigure[Opposite-subtracted surface brightness]
    {\includegraphics[width=3.5in]{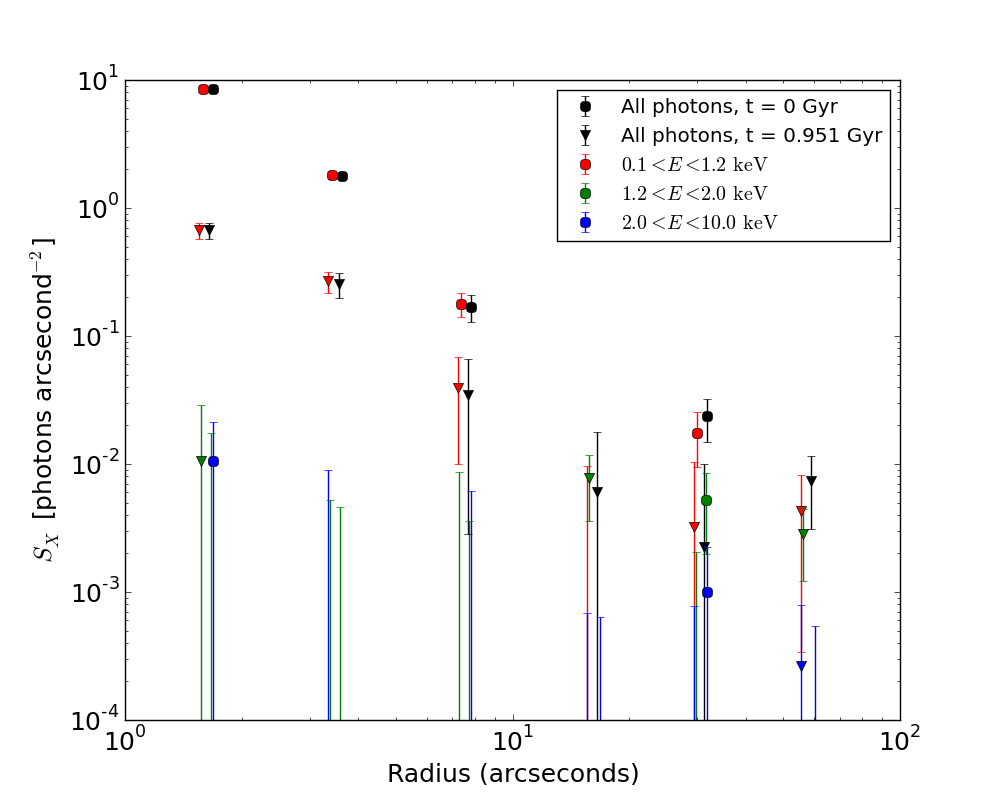} \label{fig:photonrc_opp_groupearlytime}}      
    \caption{Stacked surface brightness profiles of the group galaxies at early times. The circles correspond to the surface brightness at the beginning of the simulation, and the triangles to $t = 0.95$  Gyr. The colors correspond to different energy bins: red to the lowest energy bin ($0.1 - 1.2$ keV), green to the medium energy bin ($1.2 - 2$ keV), blue to the highest energy bin ($2  - 10$ keV), and black to the total count. We calculate the errors by assuming Poisson statistics; the error bars are $1 \sigma$ limits. The data points  in each radial bin are slightly offset for clarity. Left: Stacked radial profile for group galaxies (that are at least 200 kpc from the group center in projection). Right: Opposite-subtracted radial profile, as described in the text, where errors are calculated using error propagation. \label{fig:photongroupearlytime}}
  \end{center}  
  
   \begin{center}
  	 \subfigure[Stacked surface brightness]
    {\includegraphics[width=3.5in]{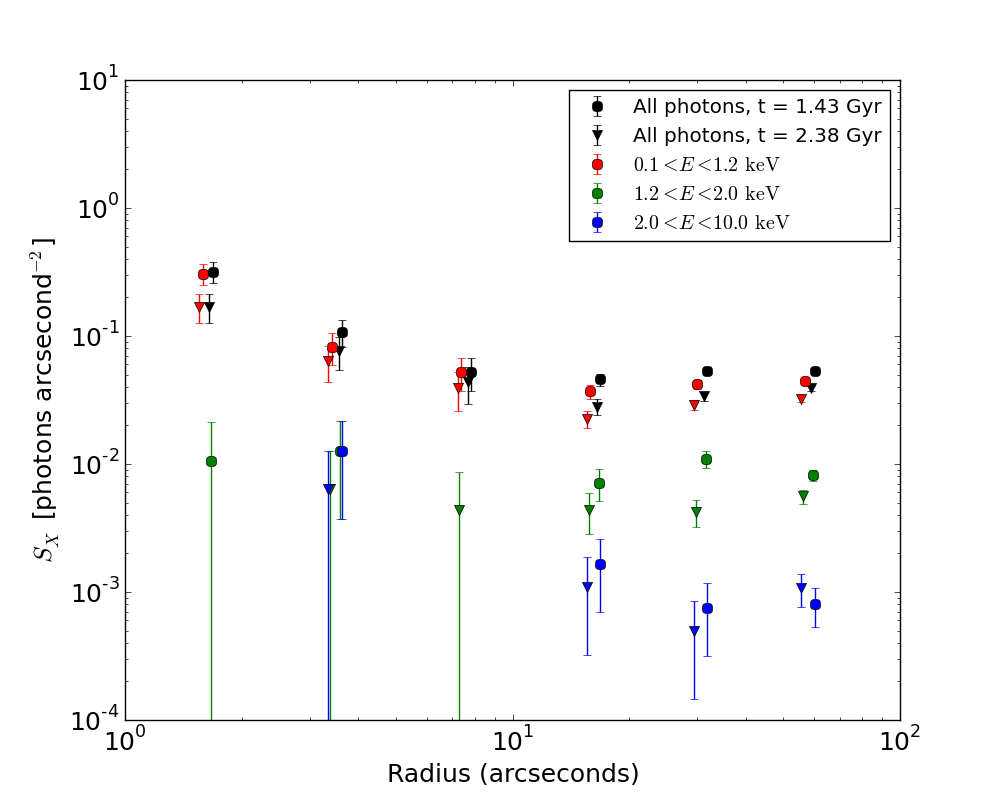} \label{fig:photonrc_grouplatetime}}  
    \subfigure[Opposite-subtracted surface brightness]
    {\includegraphics[width=3.5in]{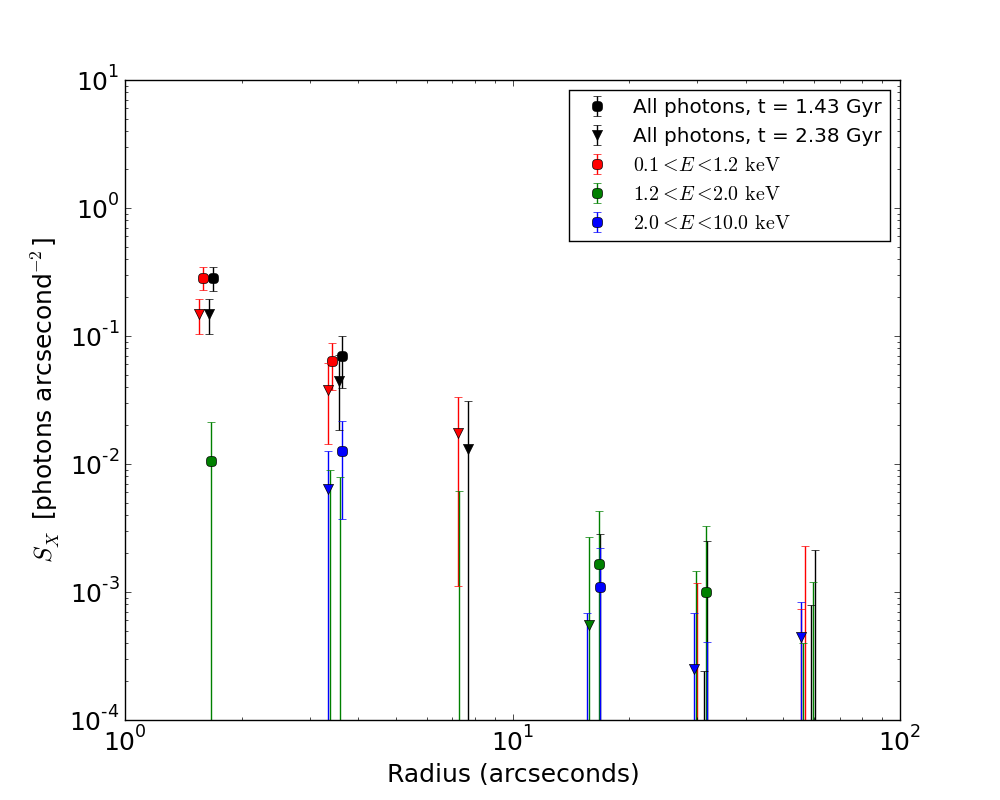} \label{fig:photonrc_opp_grouplatetime}}      
    \caption{Stacked surface brightness profiles of group galaxies at late times. The circles correspond to the surface brightness at $t = 1.43$ Gyr, and the triangles to $t = 2.38 $ Gyr. The colors and error bars are as in Figure~\ref{fig:photongroupearlytime}. Left: Stacked radial profile for group galaxies (that are at least 200 kpc from the group center in projection). Right: Opposite-subtracted radial profile.   \label{fig:photongrouplatetime}}
  \end{center}  
  
\end{figure*}

\begin{figure*}
  \begin{center}
  	 \subfigure[Stacked surface brightness]
    {\includegraphics[width=3.5in]{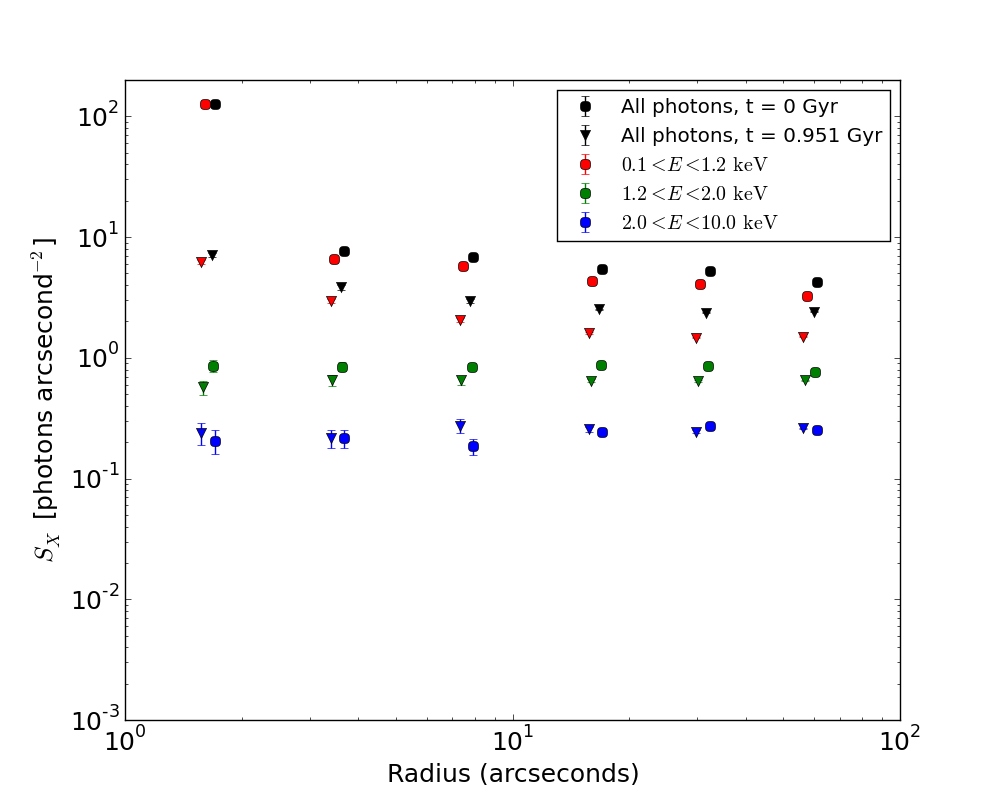} \label{fig:photonrc_clusterearlytime}}  
    \subfigure[Opposite-subtracted surface brightness]
    {\includegraphics[width=3.5in]{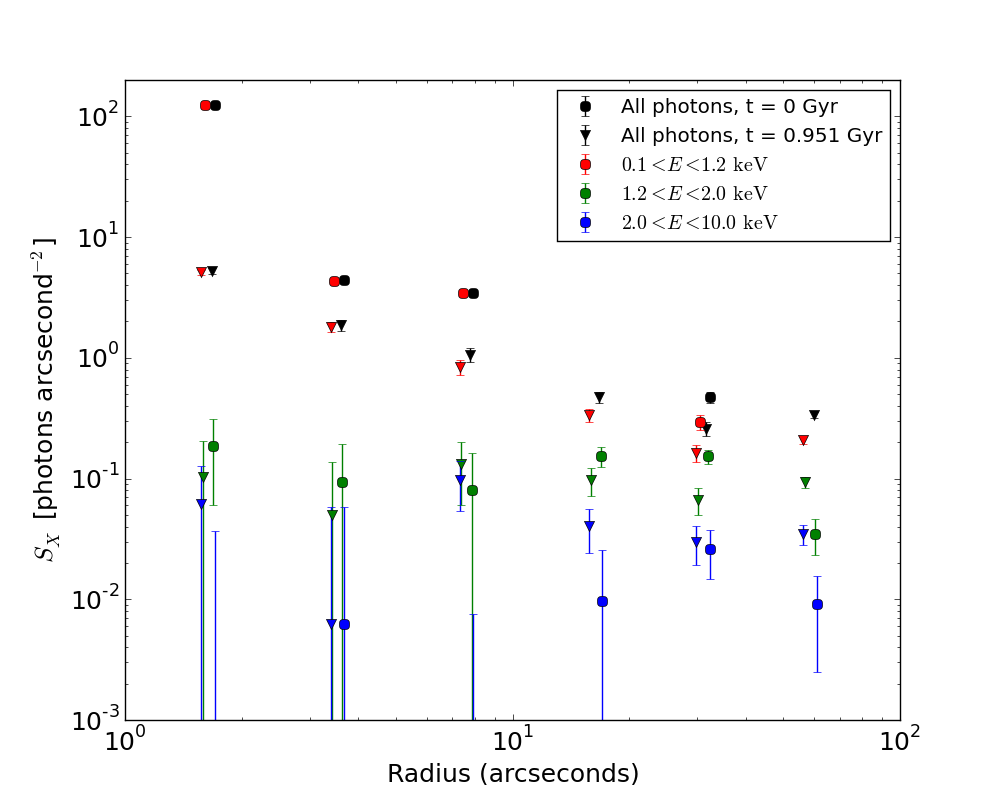} \label{fig:photonrc_opp_clusterearlytime}}      
    \caption{Stacked surface brightness profiles of the cluster galaxies at early times. The colors and symbols are as in Figure~\ref{fig:photongroupearlytime}, and the data points in each radial bin are slightly offset for clarity. Left: Stacked radial profile for cluster galaxies (that are at least 200 kpc from the cluster center in projection). Right: Opposite-subtracted radial profile. \label{fig:photonclusterearlytime}}
  \end{center}  
  
   \begin{center}
  	 \subfigure[Stacked surface brightness]
    {\includegraphics[width=3.5in]{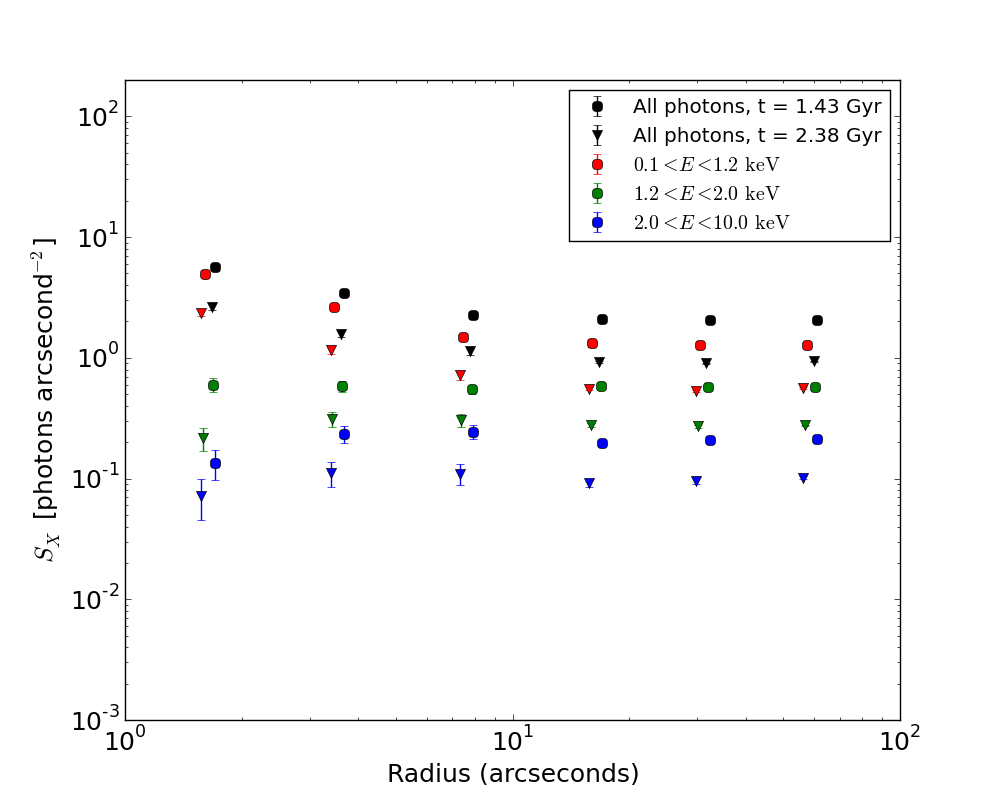} \label{fig:photonrc_clusterlatetime}}  
    \subfigure[Opposite-subtracted surface brightness]
    {\includegraphics[width=3.5in]{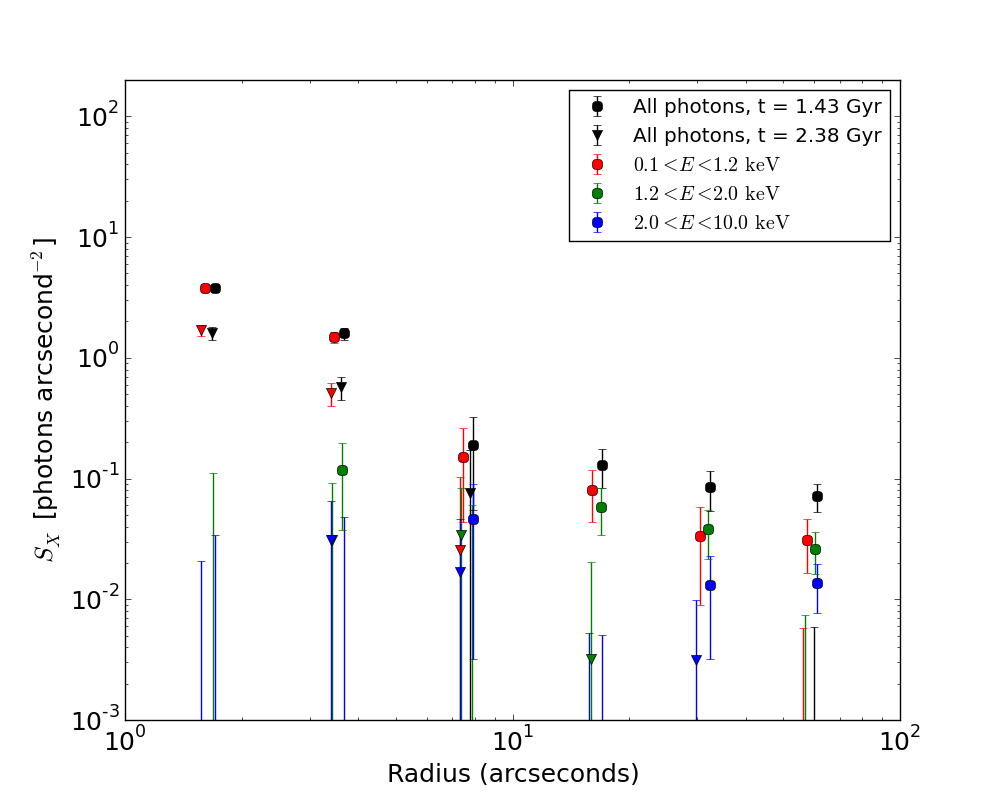} \label{fig:photonrc_opp_clusterlatetime}}      
    \caption{Stacked surface brightness profiles of cluster galaxies at late times. The colors and symbols are as in Figure~\ref{fig:photongrouplatetime}. Left: Stacked radial profile for cluster galaxies (that are at least 200 kpc from the cluster center in projection). Right: Opposite-subtracted radial profile.   \label{fig:photonclusterlatetime}}
  \end{center}  
  
\end{figure*}

\begin{figure*}
  \begin{center}
  	 \subfigure[$t = 0 - 0.951$ Gyr]
    {\includegraphics[width=3.5in]{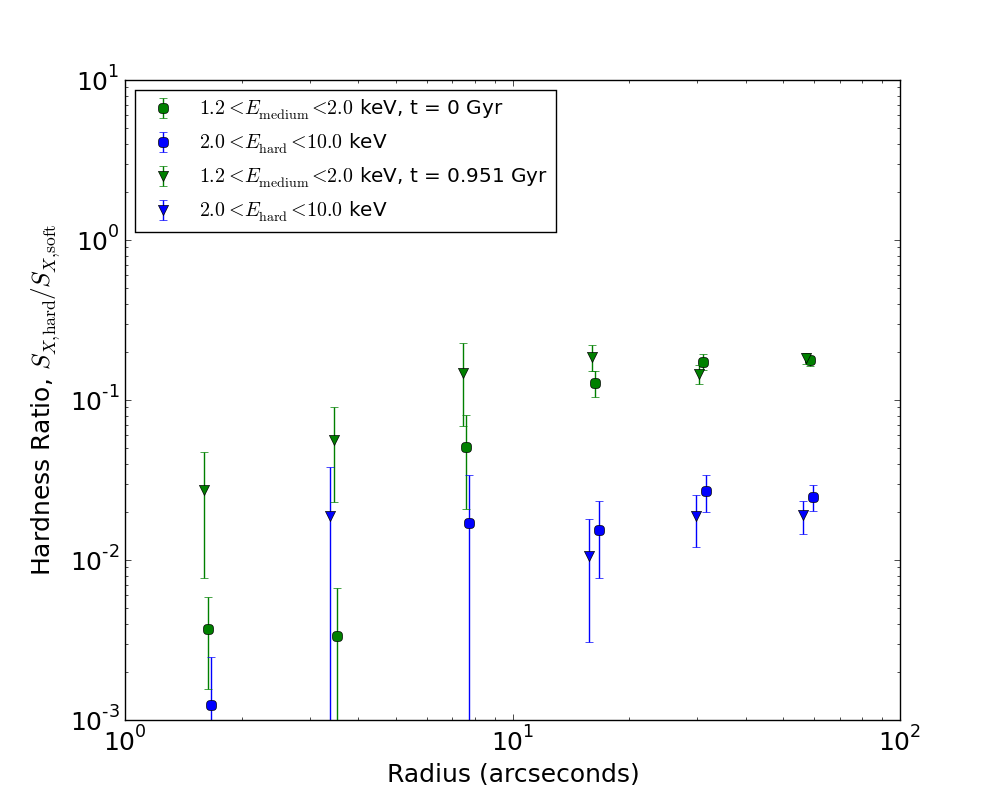} \label{fig:hrgroupearlytime}}  
    \subfigure[$t = 1.43 - 2.38$ Gyr]
    {\includegraphics[width=3.5in]{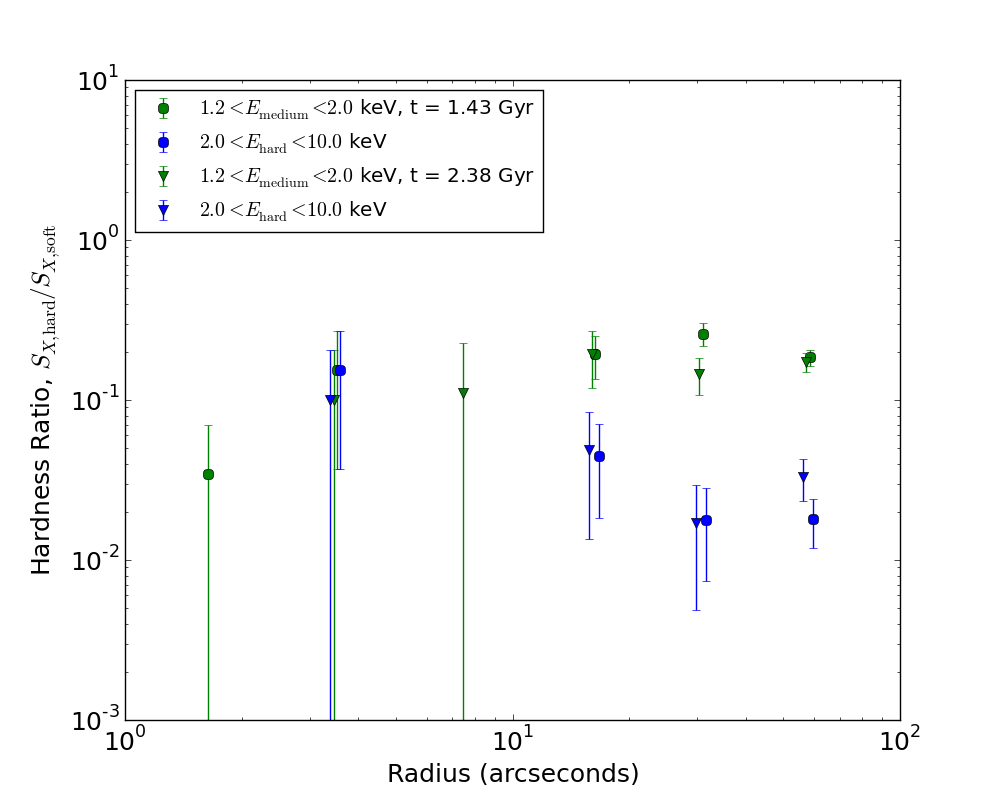} \label{fig:hrgrouplatetime}}      
    \caption{The hardness ratio, defined as the ratio of surface brightness in the medium and hard bins to the surface brightness in the soft or lowest energy radial bin, for group galaxies. The colors and symbols are as in Figures~\ref{fig:photongroupearlytime} and \ref{fig:photongrouplatetime}, and the error bars are calculated using error propagation. \label{fig:hrgroup}}
  \end{center} 
\end{figure*}

\begin{figure*}
  \begin{center}
  	 \subfigure[$t = 0 - 0.951$ Gyr]
    {\includegraphics[width=3.5in]{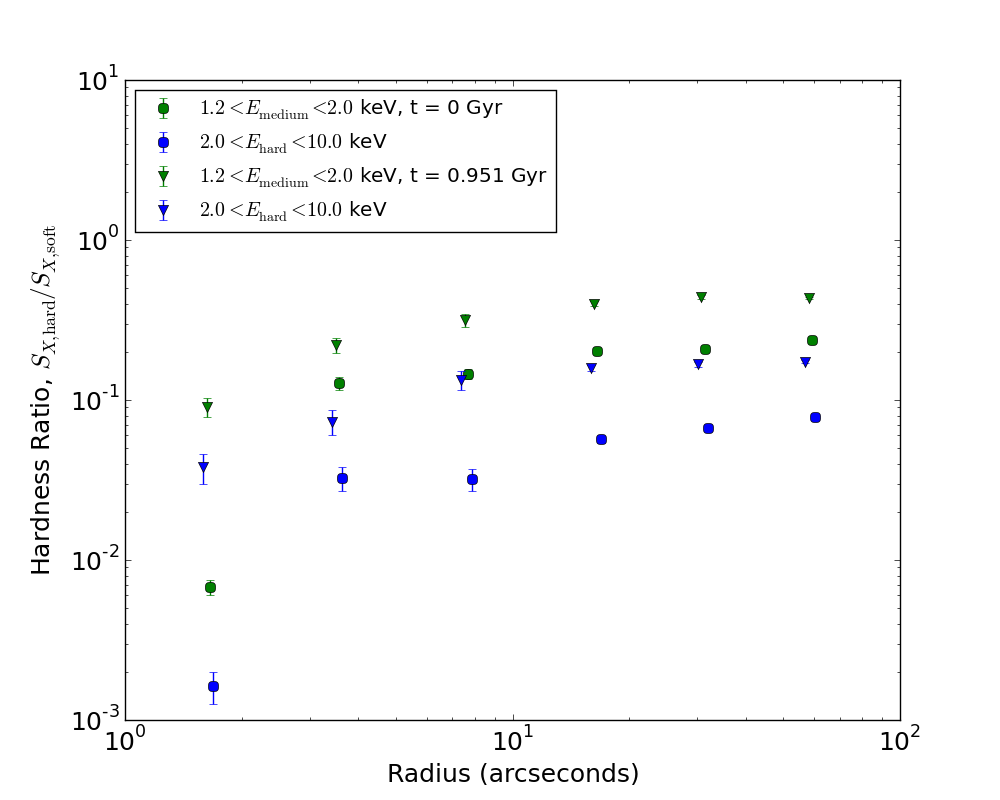} \label{fig:hrclusterearlytime}}  
    \subfigure[$t = 1.43 - 2.38$ Gyr]
    {\includegraphics[width=3.5in]{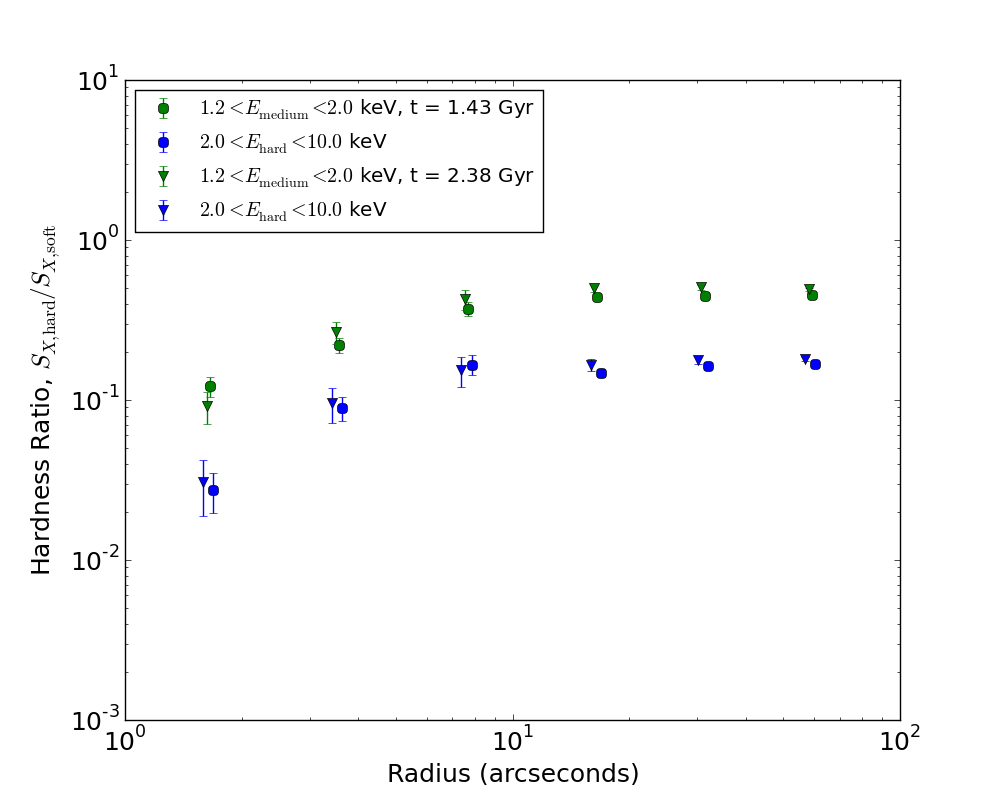} \label{fig:hrclusterlatetime}}      
    \caption{ The hardness ratios of stacked cluster galaxies. The colors and symbols are as in Figure~\ref{fig:hrgroup}, and the error bars are calculated using error propagation. \label{fig:hrcluster}}
  \end{center} 
\end{figure*}

The stacked radial profiles of group galaxies early in the simulation ($t = 0 - 0.95$ Gyr) and at late times when most of the galaxies' gas has been stripped ($t = 1.43 - 2.38 $ Gyr) are plotted in Figures~\ref{fig:photongroupearlytime} and \ref{fig:photongrouplatetime}. Each plot shows the surface brightness in three different energy bins at two early and two late timesteps. The plots on the left (Figures~\ref{fig:photonrc_groupearlytime} and \ref{fig:photonrc_grouplatetime} show the stacked galaxy radial profiles only, while the plots on the right (Figures~\ref{fig:photonrc_opp_groupearlytime} and \ref{fig:photonrc_opp_grouplatetime}) show the radial surface brightness with the emission from the stacked opposite-point radial profiles subtracted. To calculate the stacked opposite-point profiles, we generate a photon sample and mock observations for regions centered on points diametrically opposite the galaxies' density peaks, in 2D projection. This is done with the assumption that the X-ray emission centered on these opposite points will be uncorrelated with the galaxies' emission. These opposite-point mock observations are stacked in the same fashion as the galaxy-centered observations, and their radial surface brightness profiles are subtracted from the galaxy-centered profiles to generate Figures~\ref{fig:photonrc_opp_groupearlytime} and \ref{fig:photonrc_opp_grouplatetime} with appropriately propagated error bars. The above stacking and opposite-point subtraction analysis is repeated for the cluster's galaxies, and Figures~\ref{fig:photonclusterearlytime} and \ref{fig:photonclusterlatetime} show the corresponding stacked surface brightness profiles for cluster galaxies. 

The surface brightness profiles of the initial conditions correspond to the circles in Figures~\ref{fig:photongroupearlytime} and \ref{fig:photonclusterearlytime}. The triangle symbols in these plots correspond to a more realistic emission profile at $t = 0.95$ Gyr. The peak of the emission is at $r \simeq 1.6 \arcsec$. Studying the surface brightness profiles of the group's galaxies, we see on comparing Figures~\ref{fig:photonrc_groupearlytime} and \ref{fig:photonrc_opp_groupearlytime} that the stacked emission in the softest energy band ($0.1 < E < 1.2$ keV) for $r \lesssim 10 \arcsec$ at $t = 0$ Gyr and $r \lesssim 5\arcsec$ at $t = 0.95$ Gyr is robust to opposite-point radial profile subtraction. In contrast, the emission from larger galaxy-centric radii, particularly at $t = 0.95$  Gyr, is consistent with zero after subtraction. The total stacked  emission from the galaxies' centers decreases by an order of magnitude at $r \simeq 1.6 \arcsec$  from $t = 0$ Gyr to $t = 0.95$ Gyr due to efficient gas stripping by the ICM. On average, $\sim 70\%$ of the gas within $R_{200}$ has been stripped by this time. However, the emission in the harder energy bands ($E > 1.2$ keV) remains unaffected by the stripping of cooler (relative to the ICM) gas. Additionally, the emission in the harder energy bands remains relatively flat before subtraction and is close to zero after subtraction. We further elaborate on the hard energy band emission later in this section in the discussion of hardness ratios. 

Figure~\ref{fig:photongrouplatetime} shows the radial surface brightness profiles at $t = 1.43 - 2.38 $ Gyr. The central surface brightness at $t = 2.38$ Gyr is $\sim 0.5 \times$ the central surface brightness at $t = 1.43$ Gyr (Figure~\ref{fig:photonrc_grouplatetime}), compared to the factor of 10 decrease during the same time interval from $t = 0 - 0.95$ Gyr, since the denser coronal gas responsible for this emission is disrupted on a longer timescale than the diffuse gas at larger galactic radii. The central surface brightness after opposite-point subtraction (Figure~\ref{fig:photonrc_opp_grouplatetime}) at $r \lesssim 5 \arcsec$ is also robust to opposite-point subtraction, unlike the emission at $r \gtrsim 10 \arcsec$. Therefore, the stacked coronae that are the source of this emission can be reliably detected even after $\sim 1$ dynamical time within the group. Note that no astrophysical background or projected emission has been included.

The stacked surface brightness profiles of the cluster galaxies (Figures~\ref{fig:photonclusterearlytime} and \ref{fig:photonclusterlatetime}) are qualitatively similar to those of the group galaxies. The cluster's galaxies are subject to stronger ram pressure than the group's galaxies (since $P_{\rm ram} \propto v_{\rm gal}^2$, and the more massive cluster has a higher velocity dispersion), so the central surface brightness decreases by a factor of $\sim 25$ in the first 0.95 Gyr (Figure~\ref{fig:photonclusterearlytime}) compared to the factor of $\sim 11$ decrease seen in the group. The cluster's emission at $r \lesssim 5 \arcsec$ is robust to opposite-point subtraction (Figure~\ref{fig:photonrc_opp_clusterearlytime}). At late times, the central  surface brightness further declines as expected, but persists after opposite-point subtraction at $r \lesssim 5 \arcsec$. This expected emission from highly stripped cluster galaxies after more than one dynamical time is an optimistic sign for future observational studies.  As seen in the group, the emission in the harder energy bands remains relatively flat at all times. There is, however, an increase in the emission in the harder energy bands at late times, particularly from 1.43 Gyr to 2.38 Gyr at large galaxy-centric radii ($r > 10 \arcsec$). This is because the emission from these regions is increasingly dominated by the ICM, and stripped galactic gas is additionally heated to the temperature of the ICM. 

The coronal gas bound to galaxies is cooler than the hot ICM because of the galaxies' lower virial temperatures. Therefore, we expect the emission in the $0.1 - 1.2$ keV energy band within $\sim 5 \arcsec$ to be significantly higher relative to $r \gtrsim 10 \arcsec$. We quantify this effect using the hardness ratio $S_{X, {\rm hard}}/S_{X, {\rm soft}}$, where $S_{X, {\rm hard}}$ is the total photon flux in the $1.2 < E < 2$ keV band or the $2 < E < 10$ keV band. Figure~\ref{fig:hrgroup} shows the hardness ratio for the group at early (Figure~\ref{fig:hrgroupearlytime} and late (Figure~\ref{fig:hrgrouplatetime} times, and Figure~\ref{fig:hrcluster} similarly shows the cluster's stacked hardness ratio profiles. 

The stacked emission from the group's galaxies at early times lowers the hardness ratios ($t = 0 - 0.95$ Gyr, Figure~\ref{fig:hrgroupearlytime}) at $r \lesssim 10 \arcsec$ relative to large galaxy-centric radii. The hardness ratios increase with radius up to $10 \arcsec$ and then flatten out. There is, however, a large scatter in the monotonically increasing hardness ratios at $r \lesssim 10 \arcsec$ due to the low photon counts in the hard bands. The hardness ratios do not vary significantly with time within each hard energy band. At late times (Figure~\ref{fig:hrgrouplatetime}), the group galaxies' measured hardness ratio is consistent with being constant with radius.

The cluster galaxies' hardness ratios also increase monotonically within $r \lesssim 10 \arcsec$, and this increase is more significant than that of the group's galaxies. The slope of the hardness ratio profile decreases from $t = 0$ to $t = 0.95 $ Gyr (Figure~\ref{fig:hrclusterearlytime}). However, the trend in increasing hardness ratio up to $r \simeq 10 \arcsec$ and the flattening out beyond this radius are significant. At late times (Figure~\ref{fig:hrclusterlatetime}), the cluster galaxies' hardness ratio profiles flatten out, but if a sufficiently large number of galaxies is stacked, the hardness ratio within $10 \arcsec$  is still significantly lower than at $r \gtrsim 10 \arcsec$. Additionally, as seen in Figure~\ref{fig:photonclusterlatetime}, the ICM at $r \gtrsim 10 \arcsec$ heats up, and the increase in temperature is reflected in the hardness ratio. The overall hardness ratio increases steadily in both high energy bands from $t = 0$ Gyr to $t = 2.38$ Gyr. 
 
\section{Discussion}
\label{sec:discussion}

\subsection{Ram pressure stripping and gas mass loss rates}

The results discussed in \S~\ref{sec:massloss} show that galaxy strangulation rates depend strongly on both galaxy mass and parent halo mass, in the sense that less massive galaxies within more massive parent halos have higher rates of mass loss. For example, group galaxies on average lose 90\% of the gas within $R_{200,{\rm gal}}$ within 2.4~Gyr, while the cluster galaxies require only 1.7~Gyr on average. When we consider gas loss as a function of galaxy mass, group galaxies with $M_{\rm init} > 10^{11} \ \mbox{M}_{\odot}$ lose 80\% of their gas by 2.4~Gyr, while those with smaller initial masses lose 95\% by this time. In the cluster, the higher-mass galaxies lose 90\% of their gas by 2.4~Gyr, while the lower-mass galaxies lose 100\%.

Qualitatively we agree with other idealized simulation results \citep{McCarthy08,Roediger14a} and with X-ray observations of galactic coronae \citep{Sun07,Jeltema08}. A potential caveat in comparing our results to other studies is the dependence of gas mass loss rate on the assumed galactic gas density profile. The exact density profiles of the hot gas in galaxies, particularly in group and cluster environments, are not well constrained observationally. As noted by \citet{McCarthy08}, non-gravitational processes like radiative cooling, thermal conduction, and feedback from starbursts, supernovae, and AGN can destroy or replenish the hot coronal component of galactic gas, changing the profile shape. We do not account for these processes in our current simulations.

\begin{figure}
  \begin{center}
    {\includegraphics[width=0.48\textwidth]{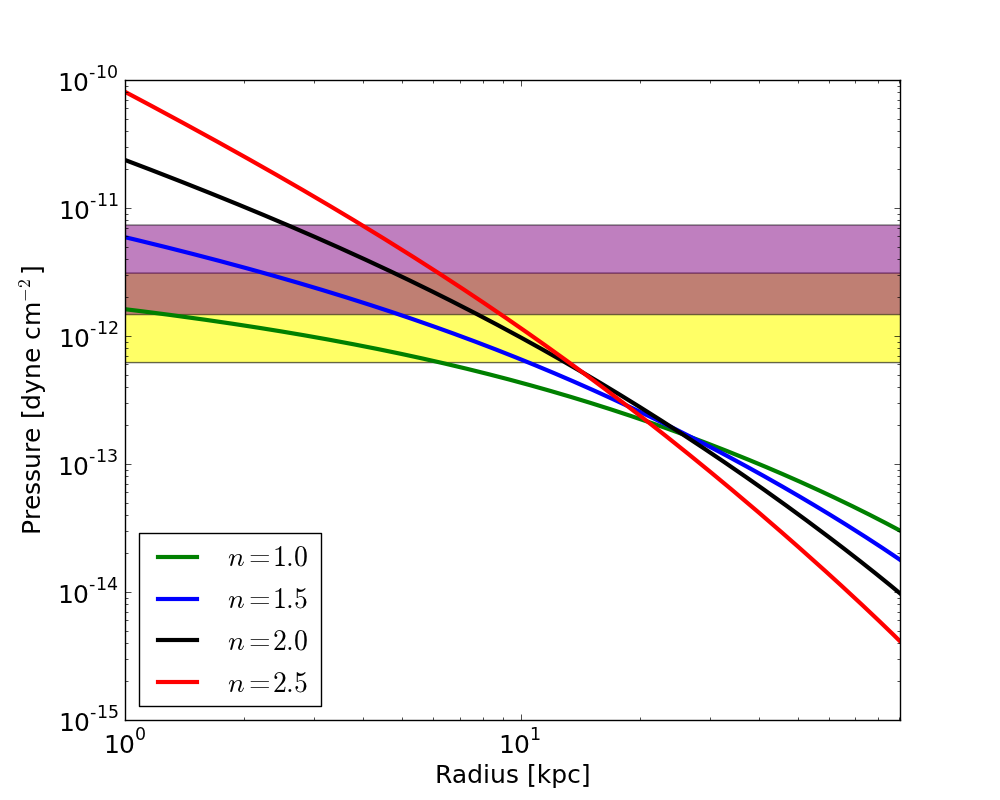}}   
    \caption{Initial $P(r)$ profiles for a $2.69 \times 10^{11} \ \mbox{M}_{\odot}$ group galaxy, calculated assuming hydrostatic equilibrium for varying values of $n$ in $\rho_{\rm gas} = k_n r^{-n}$. $k_n$ is calculated from $M_{\rm gas} = 0.1 M_{\rm 200, gal} = 4\pi \int_{0}^{r_{200}}{ r^2 \rho_{\rm gas}(r) dr}$. The black line in this figure corresponds to the $n = 2$ profile used in this study. The shaded regions correspond to typical ranges of ram pressure in the group (yellow) and cluster (purple), for typical velocities of 557 km s$^{-1}$ (group) and 858 km s$^{-1}$ (cluster) and densities from $2 \times 10^{-28}$\ g cm$^{-3}$ to $10^{-27}$\ g cm$^{-3}$.  \label{fig:pcheck}}
  \end{center} 
\end{figure}

To investigate the effect of the gas density profile, we parametrize it locally using $\rho_{\rm gas}(r) \propto r^{-n}$.  Previous simulations have assumed initial NFW density profiles (\citealt{McCarthy08}; $n \sim 1$ for $r \lesssim r_{\rm s}$ and $n \sim 3$ for $r \gtrsim r_{\rm s}$) or $\beta$ model density profiles (\citealt{Roediger14a}; $\beta = 0.4$ and 0.5 corresponding asymptotically to $n = 1.2$ and $n = 1.5$) for the gas in individual galaxies. We use $n = 2$, while our total density profile is NFW. Therefore, at small radii the initial hydrostatic pressure $P(r)$ satisfies  $dP / dr \propto \rho_{\rm gas} (r)$. Figure~\ref{fig:pcheck} illustrates the dependence of the galactic $P(r)$ profile on $n$ for a fixed mass and gas fraction and the expected range in $P_{\rm ram}$ in the group and cluster for a typical galaxy of mass $2.69 \times 10^{11} \ \mbox{M}_{\odot}$. The pressure profile steepens with increasing $n$. If $n < 2$, the pressure is higher at larger galaxy-centric radii and lower at smaller radii relative to our assumed profile. Given the ram pressures observed in our simulations ($P_{\rm ram} \sim 10^{-12}$ to $10^{-11}$~dyne~cm$^{-2}$), for a $2.69 \times 10^{11} \ \mbox{M}_{\odot}$ galaxy the flatter profiles characteristic of other work would result in complete stripping of the gas. For steeper profiles the remnant corona size is larger for the group than for the cluster.

Although the flatter gas density profile in \citet{McCarthy08} works against the retention of gas in their simulations, their galaxies are significantly more massive than ours, putting more of the core pressure profile above the level of ram pressure and allowing them to retain more gas than we observe. They found, using simulations of individual galaxies orbiting within $10^{14}\ \mbox{M}_{\odot}$ clusters, that a $2 \times 10^{12}\ \mbox{M}_{\odot}$ galaxy loses 75\% of its gas within 2 Gyr to strangulation, while a $10^{13}\  \mbox{M}_{\odot}$ galaxy loses 50\% of its gas by the same time. 

In principle the external thermal pressure due to the ICM could act to confine galactic coronae, inhibiting strangulation (\citealt{Mulchaey10}). Clearly this does not occur in our simulations. Including more complete physics does not appear to alter this conclusion. For example, using cosmological hydrodynamics simulations including radiative cooling, star formation, and stellar feedback, \citet{Bahe12} found that ram pressure dominates over thermal pressure in 84\% of galaxies in parent halos with $10^{13}\ \mbox{M}_{\odot} < M_{200} < 10^{15.1}\ \mbox{M}_{\odot}$. Even galaxies for which thermal pressure dominates showed evidence of strangulation in their simulations, because those galaxies were found to have lost gas during earlier pericentric passages. The fraction of galaxies with any hot gas was at best weakly dependent on the ratio of thermal to ram pressure (their Figure~4).

Early X-ray observations of galaxies inconsistently supported the idea that environmental influences like strangulation are important. For example, using {\it Einstein} data, \citet{White91} showed that early-type galaxies with low X-ray luminosities for their optical luminosities tend to be found in denser environments. Using a larger sample of early-type galaxies observed with {\it ROSAT}, \citet{OSullivan01} found no evidence for environmental dependence on X-ray-to-optical ratio. However, the low spatial resolution of the {\it ROSAT} observations did not allow for accurate subtraction of the ICM and point source contributions \citep{Sun07}.

More recently, systematic \textit{Chandra}-based studies of galactic coronae in group and cluster environments by \citet{Sun07} and \citet{Jeltema08} have produced strong evidence of coronal gas and probed its dependence on galaxy and parent halo mass. Using 179 galaxies in archival \textit{Chandra} observations of 25 nearby clusters, \citet{Sun07} found that 60\% of early-type galaxies with 2MASS $K_s$-band luminosities $L_{K_s} > 2L_{*}$ have detected X-ray coronae with radii of $1.5 - 4$~kpc. Although detections of fainter coronae were likely not complete, only 40\% of galaxies with $2L_{*} > L_{K_s} > L_{*}$ and 15\% with $L_{K_s} < L_{*}$ had detectable coronae. \citet{Jeltema08} observed 13 groups with \textit{Chandra} and found that $\sim 80\%$ of $L_{K_s} > L_{*}$ galaxies in poor group environments have detectable coronae. Taken together, these observations show evidence for ram pressure stripping and agree with our result that coronae should last longer in group environments and for larger galaxies. \citet{Bahe12} also make this point, noting that while the X-ray luminosities of X-ray-detected field galaxies are similar to those of group and cluster galaxies at a given stellar mass, the detection fraction is significantly lower in denser environments and increases with galaxy stellar mass.

\subsection{Confinement surfaces and stripped tails}
\label{sec:disc_conf}

In \S~\ref{sec:confsurface} we saw that galaxies with ram pressure-stripped coronae have well-defined confinement surfaces, where the galaxies' internal thermal pressure (or equivalently, the gravitational restoring force per unit area assuming hydrostatic equilibrium) balances the ICM ram pressure. These surfaces appear as temperature and surface brightness jumps in the synthetic observations; Figures~\ref{fig:xray_sample} and \ref{fig:xray_sample_c} show these effects in the 400 ks images at all times and in the 40 ks images at $t = 0.49$ Gyr. The jumps correspond to contact discontinuities or cold fronts, across which the pressure is constant.

X-ray observations of real galaxies also display cold fronts. NGC 4472 (M49), an elliptical galaxy falling into the Virgo cluster, has a distinct bow-shaped contact discontinuity at its leading edge (\citealt{Irwin96}, \citealt{Kraft11}) in addition to a ram pressure-stripped tail. \citet{Irwin96} showed with \textit{ROSAT} observations that this edge is consistent with being the surface where the galaxy's internal potential gradient is equal to the ram pressure. \citet{Machacek06} showed using \textit{Chandra} observations that NGC 4552 (M89), an elliptical galaxy in the Virgo cluster, has a sharp surface brightness jump and gas tail extending in the direction opposite to the surface brightness discontinuity. \citet{Kim08} used \textit{Chandra} and \textit{XMM-Newton} observations to show that NGC 7619, an elliptical galaxy in the Pegasus group, has a sharp discontinuity and an X-ray tail on the opposite side of the discontinuity, consistent with being a ram pressure-stripped structure. 

In principle, the size of a galaxy's confinement surface should correlate with the ICM ram pressure, assuming steady-state ICM flow past the galaxy. However, as seen in Figure~\ref{fig:rminpram}, the confinement radii of galaxies in realistic orbits are at best weakly correlated with the maximum ram pressure experienced by the galaxies over their orbits. Galaxies with supersonic velocities also form bow shocks ahead of their orbital locations, but these shocks are not prominent enough to be seen in projection. 

\begin{figure*}
  \begin{center}
  	 \subfigure[]
    {\includegraphics[width=3.2in]{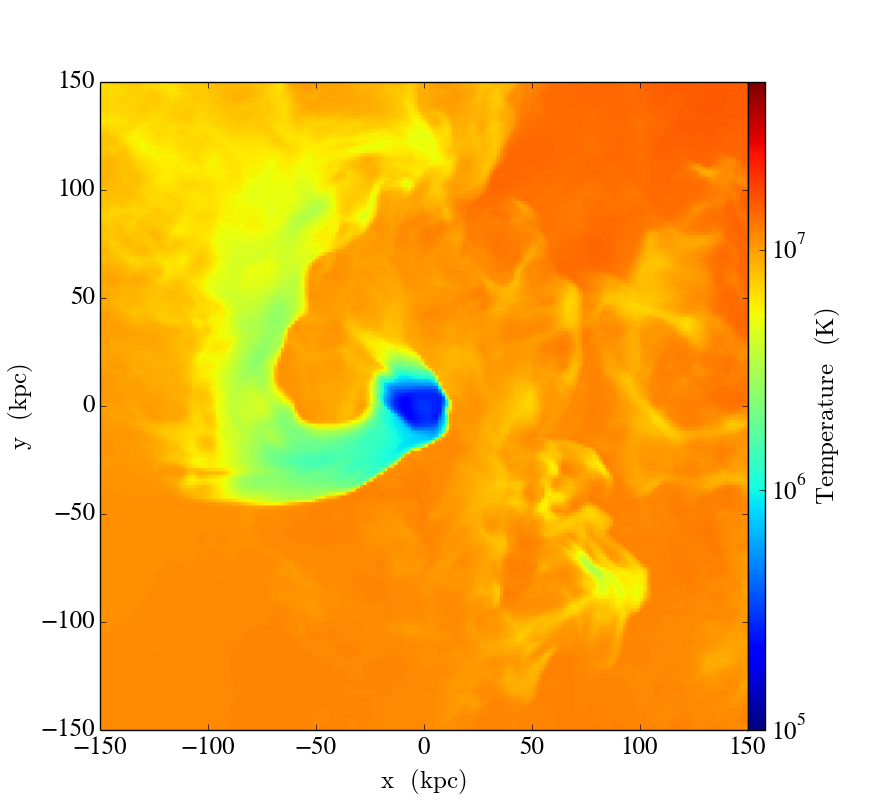} \label{fig:tail_090_21}}  
     \subfigure[]
    {\includegraphics[width=3.2in]{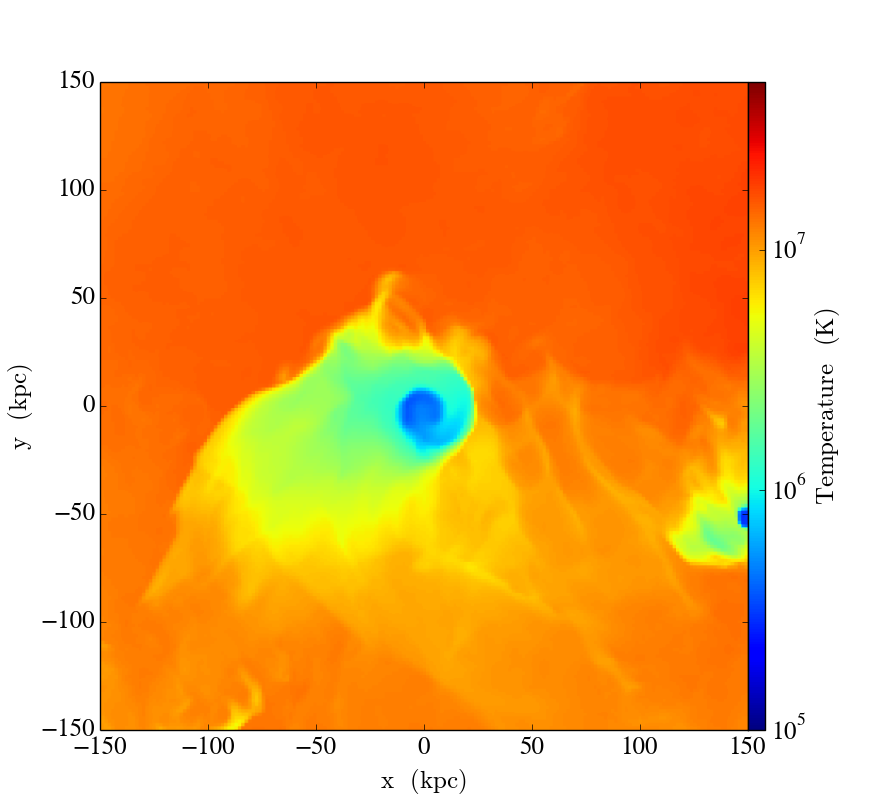} \label{fig:tail_090_63}}  
    \subfigure[]
    {\includegraphics[width=3.2in]{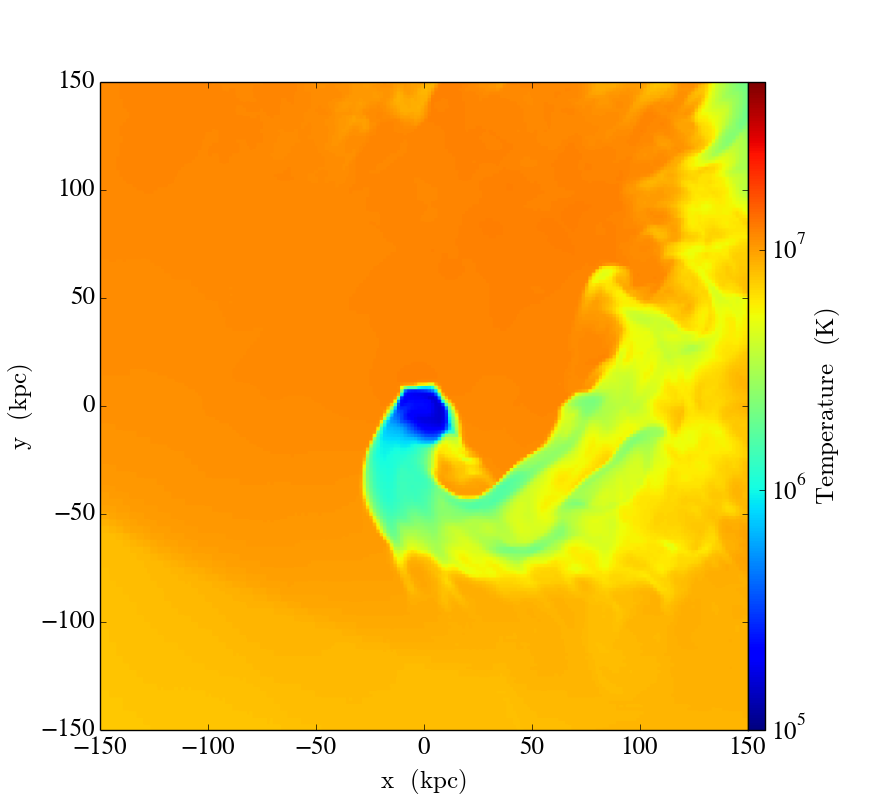} \label{fig:tail_090_94}}  
    \subfigure[]
    {\includegraphics[width=3.2in]{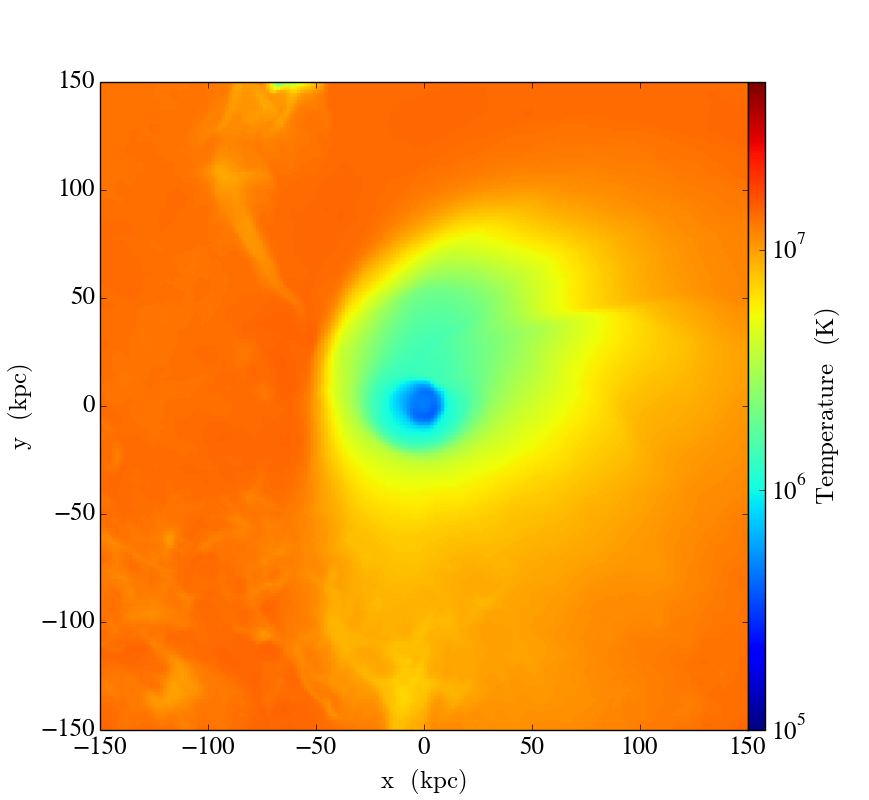} \label{fig:tail_090_143}}  
    \caption{Emission measure-weighted temperature maps of stripped galaxy tails at  $t = 1.47$ Gyr in the $1.2 \times 10^{14}\ \mbox{M}_{\rm \odot}$ cluster. \label{fig:galtail}}
  \end{center} 
\end{figure*}

Our simulations demonstrate the complexity in the structure of stripped tails over a range of galaxy masses. Galaxy tails can bend as galaxies turn in their orbits, and stripped tails and wakes can be narrow or broad depending on the galaxies' velocities. Figure~\ref{fig:galtail} shows a variety of galaxies being stripped and the structure of their tails. Figure~\ref{fig:tail_090_21} shows a galaxy with a bent tail, almost at $90^{\circ}$ to the main galaxy corona. Kelvin-Helmholtz rolls form at the interface between the cooler, denser tail and the ICM. The corona in Figure~\ref{fig:tail_090_63} appears to have a spiral-shaped plume. Figure~\ref{fig:tail_090_94} shows a galaxy whose tail is split into two distinct structures beyond the turning point. The galaxy in Figure~\ref{fig:tail_090_143} is less stripped than the other three galaxies and has a prominent leading edge. Tides and ram pressure can also detach tails from their host galaxies (as seen for the long vertically tailed galaxy at $t = 1.523$ and $2.03$ Gyr in Figure~\ref{fig:groupgasT_EM}) and briefly survive for $\sim 500$ Myr before dissipating within the ICM . 

The stripped tails of real galaxies also often have complex morphologies. The X-ray emitting tail of M86 was first detected using the \textit{Einstein} X-ray Observatory  by \citet{Forman79}. Later high-resolution \textit{Chandra}  observations by \citet{Randall08} show that the stripped hot gas of M86 has a plume-like structure offset from the galaxy's central emission in addition to a bifurcated and bent tail which most likely traces the galaxy's orbit. This interpretation is consistent with the structure of the tails of some of the massive galaxies in our cluster. \citet{Sivakoff04} showed with \textit{Chandra} observations that NGC 1603, a group galaxy, has an extended tail, and the central peak of this galaxy's emission is slightly bent with respect to its tail. \citet{Machacek06} showed using \textit{Chandra} observations that the hot gas of M89 has two horn-like structures at its leading edge in addition to a bent tail.

Late-type galaxies can also have prominent tails. \citet{Wang04} showed using \textit{Chandra} observations that C153, a late-type galaxy in Abell 2124, has a  distinct X-ray tail pointing away from its direction of motion. \citet{Machacek04}, also using \textit{Chandra}, showed that NGC 4438 in the Virgo cluster has a network of 4 -- 10 kpc-long X-ray filaments extending out from the galaxy disk, caused by ram pressure and tidal stripping in addition to a collision with the galaxy NGC 4435. One of the most dramatic examples of a ram pressure-stripped late-type galaxy is ESO 137-001 in Abell 3627, which has a 70 kpc-long bifurcated X-ray tail (\citealt{Sun06}, \citealt{Sun10}). The ESO 137-001 tail also has actively star forming knots and shows emission from molecular gas (\citealt{Jachym14}). \citet{Sun10} also showed that ESO 137-002, another late-type galaxy in Abell 3627, also has a 40 kpc-long X-ray tail. 
 
\subsection{The properties of X-ray coronae in the presence of additional physical processes}
\label{sec:physlimit}
The X-ray properties of simulated galactic coronae, discussed in the following section, can be modified with the inclusion of additional physics and consequent changes in gas loss and stripping rates. Physical properties and processes that influence gas loss rates and survival timescales of coronae include viscosity in the ICM, magnetic fields, thermal conduction, radiative cooling, and feedback from AGN and stellar outflows. A viscous ICM will suppress the formation of Kelvin-Helmholtz instabilities in stripped gas tails and the consequent mixing of stripped gas with the ICM, as shown in simulations by \citet{Roediger13, Roediger14a, Roediger14b}. Stripped tails therefore survive longer in a viscous ICM, and can be observed in X-rays for $\sim 300$ Myr longer than in an inviscid ICM (\citealt{Roediger14b}). However, the \citet{Roediger13, Roediger14a, Roediger14b} simulations show that the properties of gas within the central bound corona of a galaxy remain relatively unaffected in the presence of viscosity. 

The presence of $\mu$G magnetic fields in the ICM can also affect the survival of galactic coronae. Considering cluster-subcluster mergers, \citet{Asai07} showed that magnetic fields in the ICM suppress thermal conduction between cold dense subcluster gas and the hot diffuse ICM. \citet{Dursi07} and \citet{Dursi08} further showed that magnetic field draping over a moving subcluster can suppress hydrodynamic instabilities and thermal conduction on the leading edge of the moving subcluster. Magnetic fields in the ICM as well as within galaxies also affect the structure of stripped galactic gas tails. \citet{Ruszkowski14} showed that while the presence of ICM magnetic fields can lead to longer lived and more filamentary tails, the amount of gas lost from a galaxy does not vary significantly between a magnetized and an unmagnetized ICM. \citet{Tonnesen14} compared the amount of gas lost from magnetized and unmagnetized galaxies. They showed that the total amount of gas stripped did not vary significantly between the two cases, although the velocities of stripped gas in magnetized disks were slower than those in unmagnetized disks. Based on these results, one can conclude that while galactic and ICM magnetic fields play an important role in the structure and survival timescales of stripped tails as well as in suppressing thermal conduction, their effect on the amount of gas lost, and consequently the appearance of X-ray coronae, is not likely to be significant.

The sizes of X-ray coronae will be reduced in the presence of radiative cooling. \citet{Tonnesen09} quantified the effect of cooling in stripped galactic disks, and showed that the formation of dense clumps in the multiphase ISM allowed a larger fraction of gas to be stripped. However, the \citet{Tonnesen09} simulations do not account for heating from AGN and stellar outflows, or thermal conduction; the balance between these processes is uncertain. Observational evidence in \citet{Vikhlinin01} and \citet{Sun07} suggests that the conductive heat flux between galactic coronae and the ISM generally exceeds the X-ray luminosity of coronae. The survival of coronae in the presence of thermal conduction with the ICM therefore implies that conductivity between galactic coronae and the ICM should be suppressed, possibly by magnetic fields. 

Gas loss due to radiative cooling and stripping can be offset by heating and outflows from stellar outflows, supernovae, and AGN. An observational analysis by \citet{Sun07} shows that the kinetic energy released by stellar mass loss is $\sim 2 - 3.5$ times lower than the X-ray luminosity of cluster galaxies' coronae, so stellar outflows alone cannot reheat radiatively cooled gas. In addition, stellar outflows can only partially replace stripped coronal gas. Our simulations show that $\sim 80\%$ of the coronal gas bound to a $10^{11} \, \mbox{M}_{\odot}$ group galaxy and $\sim 90 - 95\%$ of gas in a cluster galaxy of the same mass is stripped within 2.4 Gyr. With our assumed gas mass fraction of $10 \%$, this corresponds to a gas mass loss rate, due to ram pressure and tidal stripping alone, of $3 - 4 \, \mbox{M}_{\odot} \, \mbox{yr}^{-1}$. Stellar mass loss rates, on the other hand, are at least an order of magnitude lower (based on the generally used \citet{Faber76} value of $\dot{M}_{*} = 1.5 \times 10^{-11} \, \mbox{M}_{\odot} \, \mbox{yr}^{-1} \, \mbox{L}_{\odot}^{-1} $ for early-type galaxies). Stellar outflows are therefore unlikely to replenish stripped gas or significantly modify the X-ray emission from stripped coronae.

The effect of supernova heating on galactic coronae is less clear. \citet{Vikhlinin01} argue that the $\sim 0.6$ keV supernova ejecta cannot heat the $1 - 1.8$ keV coronae in their central galaxies. For cooler satellite galaxy coronae, \citet{Sun07} show that the kinetic energy released by supernovae inside observed galactic coronae can balance energy losses due to cooling, assuming energy coupling efficiencies of $\sim 20 \%$ for low luminosity galaxies and $\sim 100 \%$ for luminous galaxies. The effects of AGN on coronae are even more uncertain. Cooling coronae can fuel the central supermassive black holes in cluster galaxies triggering AGN. Observationally, \citet{Sun07} find a correlation between the radio luminosity and X-ray luminosity of the galaxies in their sample, and also find instances of AGN radio jets outside galactic coronae. AGN jets can also be powerful enough to destroy coronae. If coronae are destroyed by AGN, our analysis might overestimate the evolution of coronal emission. In the absence of significant stellar replenishment, if a combination of supernova and AGN heating balance energy losses due to radiative cooling, coronae can remain in approximate energy balance, and the environmental effects should dominate their overall evolution. 

Therefore, although our simulations do not account for the full complexity in physical processes that affect the survival and detectability of galactic coronae, we expect that the general trends observed in our synthetic images and stacked profiles will not be significantly altered. These include the environmental dependence of coronal emission where cluster galaxies are stripped faster than group galaxies, the decrement in hardness ratio towards the central regions of coronae, and the overall decrease in emission and increase in hardness ratio with time spent in group and cluster environments. We will investigate the effects of additional physical processes in a future paper.

\subsection{Detectability of stripped X-ray coronae and tails}
\label{sec:disc_xray}

In galaxy preprocessing scenarios (e.g.\ \citealt{Rasmussen12}, \citealt{Lu12}, \citealt{Bahe12}, \citealt{VR13}), much of the evolution of cluster galaxies occurs in groups or other dense environments before they join their current parent halos. Detecting ongoing ram pressure stripping within clusters, on the other hand, should support a picture in which cluster galaxies continue to evolve within their current hosts. In this context, if galactic coronae are found to be common in clusters, we can infer that preprocessing plays a minor role or that gas replenishment is efficient. If coronae are not found to be common, then either preprocessing is efficient or galaxies do not have coronae to begin with.

Our simulations show that galactic wakes, stripped tails, and remnant coronae can be detected in long-exposure X-ray images of individual group and cluster galaxies (\S~\ref{sec:xraycoronae}). These structures are detectable for more than a Gyr after infall and last longer in groups than in clusters. The significance at which the coronal emission is detected for individual galaxies in the group at $t = 0.98$ Gyr is $\sim 5\sigma$ in the 400 ks image. However, the significance is only $\sim 1.5\sigma$ in the 40 ks image, and this significance level decreases as galaxies are further stripped. Given that most X-ray observations of clusters are $\mathcal{O} (10 \, \mbox{ks})$, the X-ray emission from individual galaxies in these systems often cannot be detected, even for relatively nearby clusters ($z \simeq 0.05$). Stacking the X-ray emission centered on known optical centers of cluster galaxies improves the significance. We show in \S~\ref{sec:xraycoronae} that this stacked emission should be visible for at least 2.38 Gyr, longer than the dynamical time ($t_{\rm dyn} \simeq 1.61$ Gyr). Galactic coronae are cooler than the surrounding ICM, and most of their emission is in the $0.5 < E < 1.2$ keV band. The stacked profiles in harder bands ($1.2 < E < 2$ keV and $2 < E < 10$ keV) are flat compared to the low-energy emission. The hardness ratio ($S_{X, {\rm hard}} / S_{X, {\rm soft}}$) of the stacked emission thus increases with increasing galaxy-centric radius.

Existing and future cluster catalogs can be used to detect stacked galactic X-ray emission. \citet{Anderson13} and \citet{Anderson14} used a stacking procedure on the ROSAT All-Sky Survey (RASS) data to study the extended X-ray emission around isolated galaxies. We suggest a similar analysis, but performed using group and cluster galaxies. In this analysis, one would add together $100 \arcsec \times 100 \arcsec$ regions of X-ray images centered on the centers of optical cluster members, correspondingly stack the regions diametrically opposite the stacked galaxies, and subtract the opposite stacked image from the galaxy stacked image. Emission from galactic coronae will be visible in the lowest-energy band at small galaxy-centric radii ($r \lesssim 10 \arcsec$ at $z = 0.05$). The hardness ratio should also correspondingly decrease. 

Several low-redshift X-ray cluster catalogs exist and could potentially be used to look for coronae via stacked observations. The \textit{ACCEPT} (``Archive of \textit{Chandra} Cluster Entropy Profile Tables'') cluster sample\footnote{\url{http://www.pa.msu.edu/astro/MC2/accept}} compiled by \citet{Cavagnolo09} is a catalog of 241 clusters with redshifts $z < 0.89$ from the \textit{Chandra} Data Archive. The typical exposure times for clusters in the \textit{ACCEPT} catalog are $\sim 10 - 100$ ks, comparable to the exposure time of 40 ks assumed in our stacking analysis. This catalog has a total of 56 clusters at $0 < z < 0.05$, 54 clusters at $0.05 < z < 0.1$, 14 clusters at $0.1 < z < 0.15$, 25 clusters at $0.15 < z < 0.2$, 30 clusters at $0.2 < z < 0.25$, and 17 clusters at $0.25 < z < 0.3$. The \textit{XMM} Cluster Survey (XCS; \citealt{Romer01,Lloyd11,Mehrtens11}) is a compilation of $\sim 500$ galaxy clusters serendipitously detected in the \textit{XMM-Newton} science archive. The typical exposure times for clusters in this catalog range from 10 -- 50 ks, and the redshifts of the clusters are $z \sim $  0.05 -- 0.6. \citet{Clerc12} compiled a similar \textit{XMM} cluster catalog (X-CLASS) of 850 clusters of 10 -- 20 ks exposures at $z \sim $  0.05 -- 0.5. A caveat to using \textit{XMM-Newton} observations in detecting stacked emission is the relatively low spatial resolution, $5\arcsec$, which is comparable to the size of a 5 kpc galaxy corona at $z = 0.05$. \textit{Chandra}, in contrast, has a much higher spatial resolution ($0.4\arcsec$), allowing the resolved detection of galactic coronae even at $z \sim 0.2$. eROSITA (\citealt{Merloni12}) will perform an all-sky X-ray survey and is expected to detect $\sim 10^5$ galaxy clusters. However, its low spatial resolution ($16 \arcsec$) will make the detection of kpc-scale stacked coronae difficult.

In addition to exposure time and spatial resolution, field of view (FOV) must be considered when choosing a cluster X-ray catalog to stack. Although the X-ray flux and spatial resolution are higher for low-redshift clusters, a single exposure often can only cover the core of such a cluster, so multiple pointings must be used. The \textit{Chandra} ACIS-I instrument has an FOV of $16.9\arcmin = 1014\arcsec$. For our chosen cosmological parameter values (\S~\ref{sec:ic}), this corresponds to 0.98~Mpc at $z = 0.05$ and 3.3~Mpc at $z = 0.2$. Since the flux from an individual galaxy at $z = 0.2$ is approximately 1/20 the flux from the same galaxy at $z = 0.05$, the `sweet spot' at which FOV, spatial resolution, exposure time, and number of clusters available are optimized probably lies between $z = 0.05$ and $0.2$.

In performing an observational stacking analysis we might ask whether it is better to stack many galaxies in a small number of clusters or a few galaxies in a large number of clusters. Although lower-mass galaxies dominate cluster galaxy populations in terms of numbers, our simulations show that they lose their coronal gas most rapidly. We should thus expect a point of diminishing returns when stacking galaxies with lower and lower masses in a given cluster. When reaching this point, stacking additional clusters is the only way to improve the signal to noise ratio.

\begin{figure}
  \begin{center}
    {\includegraphics[width=0.48\textwidth]{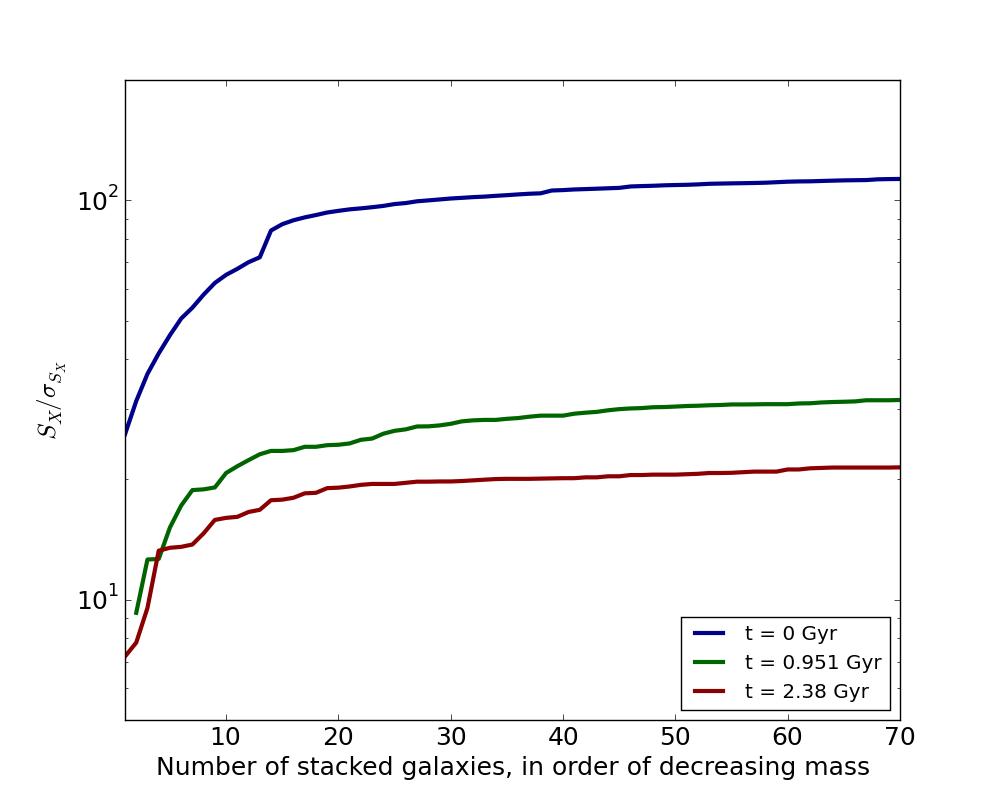}}   
    \caption{Signal to noise ratio in the stacked surface brightness within $5 \arcsec$ for galaxies in the $1.2 \times 10^{14} \, \mbox{M}_{\odot}$ cluster, versus increasing number of stacked galaxies. In this calculation, galaxies at projected radii $r > 200$ kpc are rank-ordered in decreasing order of their initial mass, and the stacked, opposite-subtracted surface brightness and its corresponding Poisson noise are calculated for each additional galaxy stacked. For instance, $S_X / \sigma_{S_X}$ for 5 galaxies is the value of $S_X / \sigma_{S_X}$ on stacking the five most massive galaxies in the cluster at $r > 200$ kpc. \label{fig:snrgalaxy}}
  \end{center} 
\end{figure}

To determine how far down in the cluster galaxy mass distribution we should go, we rank-ordered the galaxies in our simulated cluster outside a projected radius of 200~kpc by decreasing initial mass, taking this as a proxy for the galaxies' stellar masses. (The group and cluster galaxies in our simulations have initial masses greater than $10^9 \, \mbox{M}_{\odot}$, or luminosities greater than $10^8 \, \mbox{L}_{\odot}$ for a mass-to-light ratio of $10 \, \mbox{M}_{\odot} / \mbox{L}_{\odot}$. In comparison, large optical cluster surveys like the Dark Energy Survey have limiting apparent magnitudes $ m \simeq 24.0$, corresponding to galaxy luminosities of $\sim 10^7 \, \mbox{L}_{\odot}$ at $z = 0.05$ and $\sim 10^8 \, \mbox{L}_{\odot}$ at $z = 0.2$.) The 200~kpc radius cutoff is used to minimize confusion due to non-axisymmetric physics (e.g. AGN bubbles) in the cluster core; this is not present in our simulation but could be expected in real clusters. We then computed the signal to noise ratio for the X-ray surface brightness within $5\arcsec$ when stacking galaxies up to increasing ranks. The 40~ks synthetic {\it Chandra} observation was used, and the cluster was taken to be at $z = 0.05$. The results appear in Figure~\ref{fig:snrgalaxy} for three different simulation times. It is clear from the figure that it is profitable to stack at most the first 20 galaxies outside 200~kpc. Beyond this point the signal to noise ratio changes minimally. Moreover, for a single cluster, the achievable signal to noise ratio even at late times is close to 20. At $z = 0.2$, for a fixed exposure time and angular bin size, one would need to stack galaxies from 20 clusters of similar mass to obtain the same signal to noise ratio. This may be feasible with the {\it ACCEPT} catalog, although a systematic study of different parent halo masses would require more clusters.

Our simulations show that all galaxies with coronae are stripped by the host ICM, forming characteristic X-ray tails that survive for up to $\sim 1.5$ Gyr in the cluster and up to $\sim 2 - 2.5$ Gyr in the group (Figures~\ref{fig:groupgasT_EM} and \ref{fig:clustergasT_EM}). These tails are, however, not prominent in the 40 ks X-ray images at $t \gtrsim 1$ Gyr, and stacked observations should not significantly improve prospects for detecting them since the tails should be randomly oriented within the cluster. However, stripped tails are detectable even in the massive cluster at $t =  0.5$ Gyr, so the observed frequency of individual stripped tails in clusters should provide an estimate of the amount of galactic stripping over the last $\sim 0.5 - 1$ Gyr. 

Using optical cluster catalogs in which cluster membership is determined using photometric redshifts can potentially degrade the stacked X-ray signal. Projected interlopers that are not cluster members are unlikely to have their gas stripped and therefore introduce additional X-ray emission. However, estimates of the impact of non-cluster galaxies in photometric redshifts of clusters by \citet{Rozo11} show that the presence of these interlopers does not significantly affect galaxy membership properties of $\gtrsim 95\%$ of clusters. Additional projection effects from nearby clusters are also low ($\lesssim 5\%$) in recent sophisticated cluster-finding algorithms like the redMaPPer algorithm (\citealt{Rykoff14}).  

A potential caveat in interpreting the observed stacked X-ray emission from cluster galaxies is the contribution from low-mass X-ray binaries (LMXB). Previous spectroscopic studies of observed galactic coronae (\citealt{Sun07}, \citealt{Jeltema08}) model the contribution from LMXB and AGN point sources as power law sources in the spectra of galactic coronal emission to estimate the temperatures of coronae. The contribution of LMXB's to the overall X-ray luminosity is less well-known; in general, it should trace the stellar light distribution (\citealt{Sarazin01}. \citet{Vikhlinin01}, in their study of galactic coronae in the Coma cluster, ruled out any contribution from LMXB's based on their spectral analysis and the fact that the observed X-ray emission does not trace the galactic stellar light. 

A significant limitation in comparing our results to observations of galactic coronae and tails is that the cluster galaxies in our idealized simulations are initialized with all their hot ISM. In a real cosmological scenario, galaxies can be `pre-processed' (\citealt{Bahe13}, \citealt{VR13}) and be stripped of their gas in a group environment or cosmological filaments before cluster infall. Additionally, we do not account for non-adiabatic physical processes that can remove or replenish galactic gas in cluster environments. We discuss the implications of accounting for these processes in \S~\ref{sec:physlimit}. We will address these processes in a future paper.

\section{Conclusions}
\label{sec:conclusions}

We have simulated the evolution of a cosmologically motivated population of galaxies with hot coronal gas and collisionless dark matter in group and cluster environments within isolated boxes. Snapshots from these simulations are seen in Figures~\ref{fig:groupgasT_EM}, ~\ref{fig:groupgasT_EM2}, and ~\ref{fig:clustergasT_EM}. With these simulations, we have studied the effect of ram pressure stripping on the retention of galactic gas and the observational consequences of gas loss in these environments. We showed that ram pressure and tidal stripping can remove on average $\sim 90\%$ of the gas bound to galaxies within 2.4 Gyr. The amount of gas removed depends on the mass of the galaxy and the host. Galaxies in the less massive group have smaller velocities and experience weaker ram pressure compared to galaxies in the massive, high velocity dispersion cluster. Group galaxies therefore lose gas at a slower rate than cluster galaxies. In a given environment, more massive galaxies, with larger gravitational restoring forces, are more resistant to ram pressure stripping. 

We also studied the effect of ram pressure stripping on individual galaxies. We showed that ram pressure stripping produces a well-defined confinement surface at each galaxy's leading edge, defined by the surface where the external ICM ram pressure plus thermal pressure balances the galaxy's internal thermal confinement pressure. The stripped gas is deposited in the form of a tail which trails the galaxy on its orbit. This tail can bend or be distorted and become bifurcated. The location of the confinement surface is correlated with the ram pressure experienced by a galaxy over its entire orbit; galaxies that experience stronger ram pressure on average have smaller confinement surfaces. This correlation is weaker in the cluster and for later times.

We generated synthetic \textit{Chandra} X-ray observations with 40 ks and 400 ks exposure times of the simulated group and cluster, including their galaxies. We found that galaxy wakes and tails are visible up to $\sim 1$ Gyr in the 40 ks image, and their surviving central coronae up to $\sim 2$ Gyr, albeit at low significance levels above the cluster background. Galactic tails are visible up to $2$ Gyr in the 400 ks images. Practical constraints imply that most cluster X-ray observations are $\mathcal{O}(10)$ ks. We therefore evaluated the possibility that galactic coronal emission can be detected observationally by stacking regions around individual cluster galaxies identified in other wavebands. We found that there is an excess in stacked galactic surface brightness profiles at $r \lesssim 10 \arcsec$ in group and cluster galaxies up to $2.38$ Gyr in the low energy $0.1 < E < 1.2$ keV band. This excess persists on subtracting the correspondingly stacked emission centered on points diametrically opposite known galaxy centers. We also found that the X-ray emission from cluster galaxies declines faster than that of group galaxies, since galaxies in massive clusters experience stronger ram pressure. Additionally, the emission from galaxies at small galaxy-centric radii manifests itself in measurements of the hardness ratio ($E_{\rm hard} / E_{\rm soft}$), as a noticeable decrease in hardness ratio in the regions with significant galactic emission. 

We evaluated the suitability of existing and future X-ray catalogs of clusters for performing such a stacking analysis. The \textit{ACCEPT} sample (\citet{Cavagnolo09} of 241 clusters can possibly be used, since the clusters in this sample have an appropriate redshift distribution and exposure times, field of view, and spatial resolution adequate to detect coronal emission. We performed all our mock X-ray analyses at $z = 0.05$;  to extend this analysis to higher redshifts, one should stack galaxies from multiple clusters rather than more galaxies from the same clusters.  Stacking galaxies rank-ordered by mass reaches a point of diminishing returns, as the signal-to-noise ratio does not significantly improve on stacking galaxies beyond the first 20 most massive galaxies.  Other cluster catalogs, like the \textit{XMM} Cluster Survey and future eROSITA cluster catalogs, also have exposure times and sensitivities suitable for stacking cluster galaxies. However, these catalogs have a lower spatial resolution than \textit{Chandra} and will therefore have a harder time resolving galactic emission.  Systematic studies of stacked galactic emission in clusters over a range of masses, as functions of cluster-centric distance, and as a function of galaxy mass and morphology, would be useful in understanding the effect of ram pressure in different environments.

The survival of galactic coronae also depends on other physical processes as described in \S~\ref{sec:physlimit}. 
In particular, their survival for timescales on the order of the Hubble time implies that there exists a balance between radiative cooling, AGN activity, and magnetic field draping that suppresses thermal conduction. In subsequent papers we will address these effects in detail, although we expect the general environmental and evolutionary trends observed in our synthetic observations to persist in the presence of additional physical processes. 

\section*{Acknowledgments}

The simulations presented here were carried out using the NSF XSEDE Kraken system at the National Institute for Computational Sciences and the Stampede system at the Texas Advanced Computing Center under allocation TG-AST040034N. FLASH was developed largely by the DOE-supported ASC/Alliances Center for Astrophysical Thermonuclear Flashes at the University of Chicago. Support for this work was provided by the National Aeronautics and Space Administration through Chandra Award Number TM5-16008X issued by the Chandra X-ray Observatory Center, which is operated by the Smithsonian Astrophysical Observatory for and on behalf of the National Aeronautics Space Administration under contract NAS8-03060. This work was partially supported by the Graduate College Stutzke Dissertation Completion Fellowship at the University of Illinois at Urbana-Champaign. We thank John ZuHone for useful discussions and for assistance with the \texttt{yt} package and X-ray analysis. We also thank Yen-Ting Lin, Chris Miller, and Neal Dalal for useful discussions and the anonymous referee for useful comments and suggestions that improved this paper. 

\bibliography{ms}

\end{document}